\documentclass[12pt]{article}

\usepackage{latexsym,amsmath,amscd,amssymb,graphics,cite,svg}
\usepackage{authblk}
\usepackage{enumerate}
\usepackage{cancel} 
\usepackage{graphicx}
\usepackage{framed}
\usepackage[colorlinks]{hyperref}
\usepackage{url}
\usepackage{cite}
\usepackage[margin=1in]{geometry}% Just for this example
\usepackage{lipsum}% Just for this example
\usepackage{graphicx,caption}

\usepackage[all]{xy}

\makeatletter

\@addtoreset{figure}{section}
\def\thefigure{\thesection.\@arabic\c@figure}
\def\fps@figure{h, t}
\@addtoreset{table}{bsection}
\def\thetable{\thesection.\@arabic\c@table}
\def\fps@table{h, t}
\@addtoreset{equation}{section}

\makeatother

\newtheorem{theorem}{Theorem}

\newtheorem{conjecture}[theorem]{Conjecture}

\newtheorem{remark}[theorem]{Remark}

\numberwithin{theorem}{section}

\def\be{\begin{equation}}
\def\ee{\end{equation}}
\def\bea{\begin{eqnarray}}
\def\eea{\end{eqnarray}}
\def\ba{\begin{array}}
\def\ea{\end{array}}

\def\boldeta{\boldsymbol{\eta}}
\def\brho{\boldsymbol{\rho}}

\def\bx{{\mathbf {x} }}

\def\div{\mbox{div}\,}

\let\<\langle
\let \>\rangle
\newcommand{\rem}[1]{}

\newcommand{\de}{\delta}

\newcommand{\bvarphi}{\boldsymbol{\varphi}}

\newcommand{\bm}{\boldsymbol{m}}

\newcommand{\bu}{\boldsymbol{u}}
\newcommand{\bPsi}{\boldsymbol{\Psi}}
\newcommand{\bv}{\boldsymbol{v}}
\newcommand{\bX}{\boldsymbol{X}}

\newcommand{\bpsi}{\boldsymbol{\Psi}}

\newcommand{\bF}{\boldsymbol{F}}

\newcommand{\bU}{\boldsymbol{U}}

\newcommand{\bY}{\boldsymbol{Y}}

\newcommand{\bn}{\boldsymbol{n}}

\newcommand{\pp}[2]{\frac{\partial #1}{\partial #2}}

\newcommand{\dede}[2]{\frac{\delta #1}{\delta #2}}

\newcommand{\id}{\,\mathrm{Id}\,}

\newcommand{\todo}[1]{\vspace{5 mm}\par \noindent
\framebox{\begin{minipage}[c]{0.95 \textwidth}
\tt #1 \end{minipage}}\vspace{5 mm}\par}
\newcommand{\todoFGB}[1]{\vspace{5 mm}\par \noindent
\framebox{\begin{minipage}[c]{0.95 \textwidth} \color{red}FGB: \tt #1
\end{minipage}}\vspace{5 mm}\par}
\newcommand{\revision}[2]{#2}

\begin{document}
\title{Actively deforming porous media in an incompressible fluid: a variational approach}
\author{Tagir Farkhutdinov$^{1,2}$, Fran\c{c}ois Gay-Balmaz$^{3}$, Vakhtang Putkaradze$^{1,2}$}
\addtocounter{footnote}{1}
\footnotetext{Department of mathematical and statistical sciences, University of Alberta, 
Edmonton, AB, T6G 2G1 Canada. (TF) \texttt{farkhutd@ualberta.ca}, (VP) \texttt{putkarad@ualberta.ca}}
\addtocounter{footnote}{1}
\footnotetext{ATCO SpaceLab, 5302 Forand St SW, Calgary, AB, T3E 8B4, Canada.}
\addtocounter{footnote}{1}
\footnotetext{LMD - Ecole Normale Sup\'erieure de Paris - CNRS, 75005, Paris.
\texttt{francois.gay-balmaz@lmd.ens.fr}}

%\author[1,2]{ Tagir Farkhutdinov} 
%\author[3]{Fran{\c{c}}ois Gay-Balmaz}
%\author[1,2]{Vakhtang Putkaradze} 
%\affil[1]{\small Department of mathematical and statistical sciences, University of Alberta, 
%Edmonton, AB, T6G 2G1 Canada; email: (TF) farkhutd@ualberta.ca, (VP) putkarad@ualberta.ca
%} 
%\affil[2]{ATCO SpaceLab, 5302 Forand St SW, Calgary, AB, T3E 8B4, Canada
%}
%\affil[3]{
%LMD - Ecole Normale Sup\'erieure de Paris - CNRS, 75005, Paris; email: gaybalma@lmd.ens.fr
%} 
%, email: farkhutd@ualberta.ca
%, email: gaybalma@lmd.ens.fr
%, email: putkarad@ualberta.ca

\date{}

\maketitle

\abstract{ 
Many parts of biological organisms are comprised of deformable porous media. The biological media is both pliable enough to deform in response to an outside force and can deform by itself using the work of an embedded muscle. For example, the recent work \cite{ludeman2014evolutionary} has demonstrated interesting 'sneezing' dynamics of a freshwater sponge, when the sponge contracts and expands to clear itself from surrounding polluted water. 
We derive the equations of motion for the dynamics of such an active porous media (\emph{i.e.}, a deformable porous media that is capable of applying a force to itself with internal muscles), filled with an incompressible fluid. These equations of motion extend the earlier derived equation for a passive porous media filled with an incompressible fluid. We use a variational approach with a Lagrangian written as the sum of terms representing the kinetic and potential energy of the elastic matrix, and the kinetic energy of the fluid, coupled through the constraint of incompressibility. We then proceed to extend this theory by computing the case when both the active porous media and the fluid are incompressible, with the porous media still being deformable, which is often the case for biological applications. For the particular case of a uniform initial state, we rewrite the equations of motion in terms of two coupled telegraph-like equations for the material (Lagrangian) particles expressed in the Eulerian frame of reference, particularly suitable for numerical simulations, formulated for both the compressible media/incompressible fluid case and the doubly incompressible case. We derive interesting conservation laws for the motion, perform numerical simulations in both cases and show the possibility of self-propulsion of a biological organism due to particular running wave-like application of the muscle stress.
}

\section{Introduction} 
The mechanics of deformable porous media filled with fluid, also known as  \emph{poromechanics},  plays an important role in understanding the dynamics of biological organisms. From arterial walls to fibrous spider silk and to sea sponges, multi-layer materials with the complex internal structure are ubiquitous in nature \cite{meyers2008biological,naleway2015structural}. When such porous materials are immersed in fluids, for example, as is the case for underwater organisms or various internal organs of the human body, the porous media interacts with the fluid permeating it \cite{khaled2003role}. Thus the understanding of the dynamics of such a system is important for our comprehension of the science of biological tissue. In many biological applications, there is an additional complication due to an internal muscle acting on the material and deforming it. In this case, the effect of the muscle stress should be combined with the dynamics of the elastic matrix (we call it 'solid' for briefness in this article) and of the fluid permeating the matrix. Finally, an essential feature of many biological materials is the large percentage of water in the elastic matrix itself, leading to virtual incompressibility of the elastic material. This paper addresses these challenges using a variational approach to the motion of active porous media.   

It is useful to start with a short review of earlier works in poromechanics. Due to the large amount of work in the area, our description must necessarily be brief and only focus on the works essential to this discussion. 
 The earlier developments in the field of poromechanics are due to K. von Terzaghi \cite{Te1943} and M. Biot \cite{biot1941general,biot1955theory,biot1957elastic} in the consolidation of porous media, and subsequent works by M. Biot which derived the time-dependent equations of motion for poromechanics, based on certain assumptions on the media. M. Biot also considered the wave propagation in both low and high wavenumber regime  \cite{biot1962mechanics,biot1962generalized,biot1963theory,biot1972theory}. The amount of recent work in the field of porous media is vast, both in the field of model development  \cite{joseph1982nonlinear,detournay1993fundamentals,dell1998micro,brovko2007continuum,carcione2010computational,grillo2014darcy} and their subsequent mathematical analysis   \cite{showalter2000diffusion,bociu2016analysis,bastide2018penalization}. We refer the reader interested in the history of the field to the review \cite{rajagopal2007hierarchy} for a more detailed exposition of the literature. 

 Biot's work still remains highly influential today, especially in the field of acoustic propagation of waves through porous media. However, subsequent careful investigations have revealed difficulties in the interpretation of various terms through the general principles of mechanics, such as material objectivity, frequency-dependent permeability and changes of porosity in the model, as well as the need to describe large deformations of the model \cite{wilmanski2006few}. Based on this criticism, \cite{wilmanski2006few} 
develops an alternative approach to saturated porous media equations which does not have the limitations of the Biot's model. Alternatively, \cite{ChMo2010,ChMo2014} develop equations for saturated porous media based on the general thermodynamics principles of mechanics. 

The mainstream approach to the porous media has been to treat the dynamics as being friction-dominated by dropping the inertial terms from the equations. The seminal book of Coussy \cite{coussy1995mechanics} contains a lot of background information and analysis on that approach. For more recent work, we will refer the reader to, for example, the studies of multi-component porous media flow  \cite{seguin2019multi}, 
as well as the gradient approach to the thermo-poro-visco-elastic processes \cite{both2019gradient}. Our work, in contrast, is dedicated to the development of modes using variational principles of mechanics, which is a sub-field of all the approaches to porous media. The equations we will derive here, without the viscous terms, will be of an infinite-dimensional Hamiltonian type and approximate the inertia terms and large deformations consistently. On the other hand, the friction-dominated approach gives equations of motion that are of gradient flow type. \revision{R1Q1}{However, these approaches are not mutually exclusive, as we demonstrate in Sec.~\ref{sec:previous_models}, and in particular in Sec.~\ref{sec:PME}}.

Fluid-filled elastic porous media, by its very nature, is a highly complex object involving both the individual dynamics of the fluid and the media and highly nontrivial interactions between them. It is not realistic to assume complete knowledge of the micro-structured geometry of pores in the elastic matrix and the details of fluid motion inside the pores. Hence, models of porous media must include interactions between the macroscopic dynamics and an accurate, and yet treatable, description of \emph{relevant aspects} of the micro-structures. Because of the large variations of the geometry and dynamics of micro-structures between different porous media (e.g. biological materials vs geophysical applications), the task of deriving a detailed, unified theory of porous media suitable for all applications is most likely not possible. However, once a detailed set of assumptions is provided and their limitation is understood, deriving a consistent theory is possible. In such a framework, we believe, the variational theory is advantageous since it can develop a consistent mathematical model satisfying the physical assumptions on the system. 
Variational methods are developed by first describing the Lagrangian of the system on an appropriate configuration manifold, and then computing the critical curves of the associated action functional to obtain the equations of motion in a systematic way. The advantage of the variational methods is their consistency, as opposed to approaches based on balancing the conservation laws for a given point, or volume, of fluid. In a highly complex system like poromechanics, primarily when written in the non-inertial Lagrangian frame associated with the matrix, writing out all the forces and torques to obtain correct equations is very difficult. In contrast, the equations of motion follow from variational methods automatically, and the conservation laws are obtained in a general setting, \emph{i.e.} for arbitrary Lagrangians, as long as necessary symmetry arguments are satisfied.

 One of the earliest papers in the field of variational methods applied to the porous media was \cite{bedford1979variational} where the kinetic energy of microscopic expansion was incorporated into the Lagrangian to obtain the equations of motion. In that work, several Lagrange multipliers were introduced to enforce the continuity equation for both solid and fluid.
The works 
\cite{aulisa2007variational,aulisa2010geometric} use variational principles for explanation of the Darcy-Forchheimer law. Furthermore, 
\cite{lopatnikov2004macroscopic,lopatnikov2010poroelasticity} derive the equations of porous media using additional terms in the Lagrangian coming from the kinetic energy of the microscopic fluctuations. 
Of particular interest to us are the works on the Variational Macroscopic Theory of Porous Media (VMTPM) which was formulated in its present form in 
\cite{dell2000variational,sciarra2008variational,madeo2008variational,dell2009boundary,serpieri2011formulation,serpieri2015variationally,serpieri2016general,auffray2015analytical,serpieri2016variational,travascio2017analysis}, also summarized in a recent book \cite{serpieri2017variational}. In these works, the microscopic dynamics of capillary pores is modelled by a second-grade material, where the internal energy of the fluid depends on both the deformation gradient of the elastic media and the gradients of local fluid content.  The study of a pre-stressed system using variational principles and subsequent study of the propagation of sound waves was undertaken in \cite{placidi2008variational}.

One of the main assumptions of the VMTPM is the dependence of the internal energy of the fluid on the quantity measuring the micro-strain of the fluid, or, alternatively, the fluid content or local density of the fluid, including, in some works, the gradients of that quantity. This assumption is physically relevant for compressible fluid, but, in our view, for an incompressible fluid (which, undoubtedly, is a mathematical abstraction), such dependence is difficult to interpret. For example, for geophysical applications, fluids are usually considered compressible because of the high pressures involved. In the biological applications like the dynamics of highly porous sponges in the water we are interested in, the compressibility effects of the water, and, as we shall discuss here, of the sponge itself, can be neglected. For a truly incompressible fluid, it is difficult to assign physical meaning to the dependence of internal energy of the fluid on the parameters of the porous media. The paper  \cite{wilmanski2006few} points out the difficulties of developing the true variational principle for Biot's model because of the difficulty of interpreting the fluid content through a variational principle.
\rem{ %%%BEGIN REM 
by stating in the Conclusions: 
\\
\emph{It seems to be also clear that it is a waste of effort to try to construct a true variational principle as the Biot model contains a nonequilibrium variable, the increment of fluid contents which rules out the existence of such a principle.}
\\
} %%%END REM 
This difficulty in the interpretation of the fluid content was explained in \cite{FaFGBPu2020}, in terms of the fluid content being a constraint in the fluid incompressibility, and the fluid pressure being a Lagrange multiplier related to the incompressibility. Interestingly, Biot model for acoustic wave propagation in porous media was developed in that paper as the linearized case of wave propagation for certain Lagrangians. 

In this paper, we go one step beyond the initial variational approach developed in \cite{FaFGBPu2020} and show that the additional complexity of matrix incompressibility for biologically relevant materials can be treated similarly, as the solid's incompressibility constraint introduces another Lagrange multiplier related to the pressure inside the solid. Mathematically, we base our methods on the classical Arnold's description of incompressible fluid \cite{arnold1966geometrie} as geodesic motion on the group of volume-preserving diffeomorphisms of the fluid domain, in the absence of external forces. In Arnold's theory, the Lagrangian is simply the kinetic energy, as the potential energy of the fluid is absent, and the fluid pressure enters the equations from the incompressibility condition. This paper continues the initial derivation of \cite{FaFGBPu2020}, and achieves new results in the following directions: 
 \begin{enumerate} 
 \item We introduce incompressibility of both the fluid and the elastic matrix in the variational approach. 
 \item We also include the actions of the muscle by using a variational approach and show that the application of the muscle and its effect on the boundaries follows from a modified Lagrange-d'Alembert principle. 
 \item We show an exact reduction of the model for one-dimensional motion and derive integrals of motion, such as the net momenta and, in the case of double incompressibility, an affine relationship between the Lagrangian variables of the fluid and the solid.  
 \item We perform numerical simulations of the resulting reduced one-dimensional equations for incompressible fluid, for both compressible and incompressible matrix. We illustrate the difference between these cases and show the possibility of self-propulsion of the porous matrix (solid) while preserving the net-zero momentum of the fluid and solid.
\end{enumerate}

It is also useful to have a short discussion on the choice of coordinates and physics of what is commonly considered the saturated porous media. In most, if not all, previous works, the saturated porous media is a combined object consisting of an (elastic) dense matrix, and a network of small connected pores filled with fluid. The fluid encounters substantial resistance when moving through the pores due to viscosity and the no-slip condition on the boundary. In such a formulation, it is easier to consider the motion of the porous matrix to be 'primary', and the fluid motion to be computed relative to the porous matrix coordinates. Because the motion of the elastic matrix is 'primary', the equations are written in the system of coordinates consistent with the description of the elastic media, which is the material frame associated with the media. In this paper, we take an alternative view where we choose the same coordinate system of the stationary observer (Eulerian frame) for the description of both the fluid and the elastic media. Such a coordinate system is more frequently used in the classical fluid description but is less common in the description of elastic media. Ironically, the Eulerian frame is also frequently used in the description of wave propagation in the media, in particular, classical Biot's theory \cite{biot1962mechanics,biot1962generalized,biot1963theory,biot1972theory}. 
Physically, our description is more relevant for the case of a porous media consisting of a dense network of elastic 'threads' positioned inside the fluid, which is a case that has not been considered before apart from \cite{FaFGBPu2020}. It is worth noting that the combined Eulerian description that we use in this paper is also applicable to the traditional porous media with a 'dense' matrix, and is also well suited for the description of wave propagation in such media as shown in \cite{FaFGBPu2020} comparing the results of variational models with that of Biot. 
We shall also point out that our theory can be reformulated and is applicable for the familiar choice of the Lagrangian material description of the elastic porous matrix. These descriptions are completely equivalent from the mathematical point of view, and this is rigorously justified by using the process of Lagrangian reduction by symmetry in continuum mechanics \cite{GBMaRa12}.

\section{Equations of motion for porous media in spatial coordinates} 
\label{sec:3D_eqs}

In this section, we derive the equations of motion for a porous medium filled with an incompressible fluid, for the case of a solid elastic matrix, by using a variational formulation deduced from Hamilton's principle. For both the fluid and the elastic matrix, we shall follow the differential geometric description outlined in the book by Marsden and Hughes \cite{marsden1994mathematical}, where the reader can find the background and fill in technical details of the description of each media. This derivation closely follows the approach developed in \cite{FaFGBPu2020} to which the reader is referred for technical details. This derivation is presented here, first, to make this paper self-consistent, and second, in order to introduce the definitions of the variables. We start with some necessary background information on the description of the combined dynamics of the elastic media and fluid that is contained in it. 

\subsection{Definition of variables}

\paragraph{Configuration of the elastic body and the fluid.} The motion of the elastic body (indexed by $s$) and the fluid (indexed by $f$) is defined by two time dependent maps $\bpsi(t,\_\,): \mathcal{B} _s \rightarrow \mathbb{R} ^3  $ and $\bvarphi(t,\_\,): \mathcal{B} _f \rightarrow \mathbb{R} ^3$ with variables denoted as 
\[
\bx=\bpsi(t,\bX) \quad\text{and}\quad \bx=\bvarphi(t,\bY)\,.
\]
Here $ \mathcal{B} _s$ and $ \mathcal{B} _f$ denote the reference configurations containing the elastic and fluid labels $\bX$ and $\bY$.
We assume that there is no fusion of either fluid or elastic body particles, so the map $\bpsi$ and $\bvarphi$ are embeddings for all times $t$, defining uniquely the \textit{back-to-labels maps} $\bX=\bpsi^{-1}(t,\bx)$ and 
$\bY=\bvarphi^{-1}(t,\bx)$. 
We also assume, for now, that the fluid cannot escape the porous medium
or create voids, so at all times $t$, the domains occupied in space by the fluid  ${\mathcal B}_{t,f}=\bvarphi(t,\mathcal{B})$ and the elastic body ${\mathcal B}_{t,s}=\bpsi(t,\mathcal{B})$ coincide: 
${\mathcal B}_{t,f}={\mathcal B}_{t,s}={\mathcal B}_{t}$. Finally, we shall assume for simplicity that the domain ${\mathcal B}_t$ does not change with time, and will simply call it ${\mathcal B}$, hence both $\bvarphi: \mathcal{B} _f \rightarrow \mathcal{B} $ and $\bpsi: \mathcal{B} _s \rightarrow \mathcal{B}$ are diffeomorphisms for all time $t$. An extension to the case of the fluid escaping the boundary is possible, although it will require appropriate modifications in the variational principle and we shall not consider it in general for now, but only in a specific case later.

\paragraph{Velocities of the elastic body and the fluid.} 
The fluid velocity $\bu_f$ and elastic solid velocity $\bu_s$, measured relative to the fixed coordinate system, \emph{i.e.}, in the Eulerian representation, are given as usual by 
\begin{equation} 
\bu_f(t,\bx)=\partial_t  \bvarphi \big(t, \bvarphi^{-1}(t,\bx)\big) \, , \quad  \bu_s(t,\bx)=\partial_t \bpsi \big(t, \bpsi^{-1}(t,\bx)\big) \, ,
\label{vel_def} 
\end{equation} 
for all $\bx \in \mathcal{B}$.  Note that since $\bvarphi$ and $\bpsi$ restrict to the boundaries, the vector fields $\bu_f$ and $\bu_s$ are tangent to the boundary, \emph{i.e.},
\begin{equation} \label{free_slip} 
\bu_f\cdot\bn=0\, , \quad  \bu_s\cdot\bn=0 \, ,
\end{equation}
where $\bn$ is the unit normal vector field to the boundary.  One can alternatively impose that $\bvarphi$ and $\bpsi$ (or only $\bpsi$) are prescribed on the boundary. In this case, one gets no-slip boundary conditions
\begin{equation} \label{no_slip} 
\bu_f|_{ \partial \mathcal{B} }=0\, , \quad  \bu_s|_{ \partial \mathcal{B} }=0 \, , \quad \text{(or only $ \bu_s|_{ \partial \mathcal{B} }=0$)}.
\end{equation}

\paragraph{Elastic deformations of the dry media.} 
In order to incorporate the description of the elastic deformations of the media in the potential energy, we consider the deformation gradient of $\bpsi$ denoted
\begin{equation} 
\mathbb{F}(t,\bX)=\nabla\bpsi(t,\bX)\, ,
\label{F_def} 
\end{equation} 
where $ \nabla \bpsi$ denotes the derivative with respect to $\bX$.
In the spatial frame, we consider the Finger deformation tensor $b(t,\bx)$ defined by
\begin{equation} 
b(t,\bx)=\mathbb{F}(t,\bX)\, \mathbb{F}(t,\bX) ^\mathsf{T}\,,  
\label{b_def} 
\end{equation} 
where $\bx= \bpsi(t,\bX)$, see the paragraph below for the intrinsic geometric definition of $b$. In coordinates, we have
\[
\mathbb{F}^i_A= \frac{\partial\bpsi^i}{\partial X^A}\,,\qquad b^{ij}=  \frac{\partial\bpsi^i}{\partial X^A}\frac{\partial\bpsi^j}{\partial X^A}\,,
\]
with the summation over the material index $A$ is assumed.

From its definition \eqref{b_def}, the Finger deformation tensor $b$ is a symmetric 2-contravariant tensor field and satisfies the advection equation 
\begin{equation} 
\label{intrinsic_def_b}
\partial_t b + \pounds_{\bu_s} b=0 \, , 
\end{equation} 
where $\pounds_{\bu_s}$ is the Lie derivative of the corresponding tensor quantity. In coordinates, that Lie derivative is computed as
 \begin{equation}\label{Lie_der_b}
(\pounds_{\bu_s}b)^{ij}= \frac{\partial b^{ij}}{\partial x^k} u_s^k - b^{kj}\frac{\partial u_s^i }{\partial x^k} - b^{ik}\frac{\partial u_s^j }{\partial x^k}\,. 
\end{equation}

In general the deformation of an elastic media \emph{without fluid} leads to $b \neq \id$ (the unit tensor). The potential energy $V$  of deformation of the dry media thus depends on $b$. However, in our case, there is another part that leads to the elastic potential energy, namely, the microscopic deformations of the pores that we shall describe below.

\paragraph{Microscopic variables, pore size and free volume.} 
Out of many microscopic variables presented in the solid, the geometric shape of pores, and their connectivity are most important for computing the volume occupied by the fluid. In this paper, just like in \cite{FaFGBPu2020}, as well as several papers before us \cite{placidi2008variational,serpieri2016variational,serpieri2017variational} and others, the internal 'microscopic' volume of the pores is chosen as an important variable affecting the potential energy of the solid. 
This choice is true for the case when the pores' geometry will be roughly similar throughout the material. The model will need to be corrected when there is a drastic change of pores' geometry (e.g. from roughly spherical to elliptical pores, for merging of pores \emph{etc}). For now, we consider that the locally averaged internal volume of the pores is represented by the local variable $v(t, \bx)$ in the Eulerian description. Its Lagrangian counterpart is $\mathcal{V}(t,\bX) = v(t,\bpsi(t,\bX))$. Since we are concentrating on the Eulerian description, we will focus on $v(t, \bx)$.  Thus, in our model, the elastic energy of the solid will depend on the Finger deformation tensor $b$ and the infinitesimal pore volume $v$. Physically, this assumption is equivalent to stating that the internal volume variable $v$ will encompass all the effects of microscopic deformations on the elastic energy. 

Let us now consider the volume occupied by the fluid in a given spatial domain. We assume that the fluid fills the pores completely, so the volume occupied by the fluid in any given spatial domain is equal to the net volume of pores in that volume. Let us take the  infinitesimal Eulerian volume $\mbox{d}^3 \bx$ and define the pore volume fraction $g(t,\bx)$, so that the volume of fluid is given by $g(t,\bx)\mbox{d}^3 \bx$. Then,  one must take into account the available volume to the fluid, namely, the local concentration of pores $c(t,\bx)$ and the infinitesimal pore volume $v(t,\bx)$. 
The total volume of pores is written in the spatial description as: 
\begin{equation} 
g(t,\bx) = c(b(t,\bx)) v(t,\bx) \, . 
\label{g_c_v_constraint} 
\end{equation}

If, for example, the pores are ``frozen" in the material, they simply move as the material moves. Then, the change of the local concentrations of pores $c(t,\bx)$ due to deformations is given by
\begin{equation}\label{change_c}
c\big(t,\bpsi(t,\bX)\big) J_{\bpsi}(t,\bX) = c_0(\bX)\,, \quad J_{\bpsi}= |{\rm det} ( \nabla \bpsi)| =  | {\rm det}(\mathbb{F})|\,,
\end{equation}
where $c_0(\bX)$ is the initial concentration of pores in the Lagrangian point $\bX$. Using the definition \eqref{b_def} of the Finger tensor $b$ gives ${\rm det} \,b(t,\bx) = |{\rm det} \,\mathbb{F}(t,\bX_s)|^2$, hence we can rewrite the previous relation as
\[
c(t,\bx)\sqrt{{\rm det} \,b(t,\bx) } = c_0(\bX).
\]
In the case of an initially uniform porous media, \emph{i.e.}, $c_0=$ const, this formula shows that the concentration $c(t,\bx)$ is a function of the value $b(t,\bx)$ of the Finger deformation tensor
\begin{equation}
c(b)=\frac{c_0}{\sqrt{{\rm det}b}} \, .
\label{c_b_particular_neq} 
\end{equation}
Note that from \eqref{change_c}, the concentration of pores satisfies
\[
\partial_t c+ \operatorname{div}(c\bu_s)=0\,.
\]

\paragraph{Conservation law for the fluid.} 
In what follows, we will consider an incompressible fluid. The density of the fluid itself is denoted as $\bar\rho_f^0=$ const. We can thus discuss the conservation of the volume of fluid rather than the mass. Let us now look at the volume of fluid $g(t,\bx) \mbox{d}^3 \bx$ from a different point of view. The fluid must fill all the available volume completely, and it must have come from the initial point  $\bY=\bvarphi^{-1}(t,\bx)$. If the initial volume fraction at that point was $g_0(\bY) \mbox{d}^3 \bY$, then at a point $t$ in time we have 
\begin{equation} 
g(t,\bx) = g_0 \big(\bvarphi^{-1}(t,\bx)\big)J_{\bvarphi^{-1}}(t,\bx)  \, , \quad 
J_{\bvarphi^{-1}} := {\rm det} \big(\nabla \bvarphi^{-1}\big) \, . 
\label{cons_law_fluid} 
\end{equation} 
Differentiating \eqref{cons_law_fluid}, we obtain the conservation law for $g(t,\bx)$ written as 
\begin{equation} 
\partial_t g + \operatorname{div}  ( g\,\mathbf{u}_f ) =0 \, . 
\label{g_cons} 
\end{equation} 
The mass of the fluid in the given volume is $\bar\rho_f^0 g \mbox{d}^3 \bx$. 
Note that the incompressibility condition of the fluid \emph{does not} mean that $\operatorname{div}\bu_f=0$. That statement is only true for the case where no elastic matrix is present, \emph{i.e.}, for pure fluid. In the porous media case, a given spatial volume contains both fluid and elastic parts. The conservation of volume available to the fluid is thus given by \eqref{g_cons}.

\paragraph{Conservation law for the elastic body.} The mass density of the elastic body, denoted $\rho_s$, satisfies an equation analogous to \eqref{cons_law_fluid}, namely,
\begin{equation} 
\rho_s(t,\bx) = \rho_s^0\big(\bpsi^{-1}(t,\bx)\big)J_{\bpsi^{-1}}(t,\bx)  \, ,
\label{cons_law_elastic} 
\end{equation} 
where $\rho_s^0(\bX)$ is the mass density in the reference configuration. The corresponding differentiated form is
\begin{equation} 
\partial_t \rho_s + \operatorname{div}  ( \rho_s \mathbf{u}_s ) =0 \, . 
\label{rho_s_cons} 
\end{equation}

\subsection{The Lagrangian function and the variational principle in spatial variables}

\paragraph{Lagrangian.} For classical elastic bodies, the potential energy in the spatial description depends on the Finger deformation tensor $b$, \emph{i.e.}, $V=V(b)$. As we discussed above, in the porous media case, we consider the potential energy to depend on $b$ and $v$, and we write $V=V(b,v)$. Then, 
the Lagrangian of the porous medium is the sum of the kinetic energies of the fluid and elastic body minus the potential energy of the elastic deformations: 
\begin{equation} 
\ell(\bu_f,\bu_s, \rho_s,b,g,v)= \int_{\mathcal{B}}
\left[\frac{1}{2}  \bar\rho_f ^0 g |\bu_f|^2 + \frac{1}{2} \rho_s |\bu_s|^2 -V(b,v)\right]{\rm d}^3\bx \,. 
\label{Lagr_def} 
\end{equation} 
Note that the expression \eqref{Lagr_def} explicitly separates the contribution from the fluid and the elastic body in simple physically understandable terms. The interaction between the fluid and the media comes from the critical action principle involving the incompressibility of the fluid. We shall derive the equations of motion for an arbitrary  (sufficiently smooth) expression for $\ell(\bu_s,\bu_f, \rho_s,b,g,v)$, and will use 
 the physical Lagrangian \eqref{Lagr_def} for all computations in the paper.

\paragraph{Variational principle and incompressibility constraint.}
Condition \eqref{g_c_v_constraint} represents a scalar constraint for every point of an infinite-dimensional system. Formally, such constraint can be treated in terms of Lagrange multipliers. The application of the method of Lagrange multipliers for an infinite-dimensional system is quite challenging, see the recent review papers \cite{dell2018lagrange,bersani2019lagrange}. In terms of classical fluid flow, in the framework of Euler equations, the variational theory introducing incompressibility constraint has been developed by V. I. Arnold \cite{arnold1966geometrie} on diffeomorphism groups, with the Lagrange multiplier for incompressibility related to the pressure in the fluid. We will follow in the footsteps of Arnold's method and introduce a Lagrange multiplier for the incompressibility condition \eqref{g_c_v_constraint}. By analogy with Arnold, we will also treat this Lagrange multiplier as related to pressure, as it has the same dimensions, and denote it $p$.  Since \eqref{g_c_v_constraint} refers to the fluid content, the Lagrange multiplier $p$ relates to the fluid pressure. The equations of motion \eqref{eq_gen}, connecting pressure with the derivatives of the potential energy with respect to the pores' volume, will further justify this below. Note that $p$ may be different from the actual physical pressure in the fluid depending on the implementation of the model.  From the Lagrangian \eqref{Lagr_def} and the constraint  \eqref{g_c_v_constraint}, we define the action functional in the Eulerian description as
\begin{equation} 
S= \int_0^T\left[ \ell(\bu_f,\bu_s, \rho_s,b,g,v) -  \int_{{\cal B}} p\big( g- c(b) v\big) \mbox{d}^3 \bx\right] \mbox{d} t \,. 
\label{action_p}
\end{equation}

The equations of motion are obtained by computing the critical points of $S$ with respect to constrained variations of the Eulerian variables induced by free variations of the Lagrangian variables. 
Indeed, it is in the Lagrangian description that the variational principle is justified, as being given by the Hamilton principle with constraint. 
\revision{R3Q2}{This is one of the most important technical differences of our work with the previous work in the field, as all variations are obtained explicitly in the Eulerian frame, and all equations are therefore in Eulerian frame as well.}

The constrained variations of the Eulerian variables induced by the free variations $\delta\bpsi$, $\delta\bvarphi$ vanishing at $t=0,T$  are computed as 
\begin{equation} 
\begin{aligned}
\de \bu_f &= \partial_t \boldeta_{f} + \bu_f  \cdot \nabla\boldeta_f - 
\boldeta_f \cdot  \nabla \bu_f \\ 
\de \bu_s &= \partial_t \boldeta_{s} + \bu_s  \cdot \nabla\boldeta_s - 
\boldeta_s \cdot  \nabla\bu_s \\
\delta g&= -  \operatorname{div}(g\boldeta_f) \\ 
\delta\rho_s&= - \operatorname{div}(\rho_s\boldeta_s) \\ 
\delta b &= - \pounds_{\boldeta_s}b \, ,
\label{g_rhos_b_var}
\end{aligned} 
\end{equation} 
where  $\boldeta_{f}$ and $\boldeta_{s}$ are defined
\begin{equation} 
\boldeta_{f}=\de \bvarphi \circ \bvarphi^{-1} \, , \quad 
\boldeta_{s}=\de \bpsi \circ \bpsi^{-1} \, .
\label{eta_def} 
\end{equation} 
The variations $\delta v$ and $\delta p$ are arbitrary.

 In the case of the boundary conditions \eqref{free_slip} it follows from \eqref{eta_def} that $\boldeta_{f}$ and $\boldeta_{s}$ are  arbitrary time dependent vector fields vanishing at $t=0,T$ and tangent to the boundary $\partial\mathcal{B}$:
\begin{equation} 
\label{non-permeable} 
\boldeta_s\cdot\bn=0\, , \quad  \boldeta_f\cdot\bn=0 \, . 
\end{equation}
In the case of no-slip boundary conditions \eqref{no_slip}, we have
\begin{equation} 
\label{no_slip_var} 
\boldeta_f|_{ \partial \mathcal{B} }=0\, , \quad  \boldeta_s|_{ \partial \mathcal{B} }=0 \, , \quad \text{(or only $\boldeta_s|_{ \partial \mathcal{B} }=0$)}.
\end{equation}

\paragraph{Incorporation of external and friction forces.} Frictions forces, or any other forces, acting on the fluid $\bF_f$ and the media $\bF_s$ can be incorporated into the variational formulation by using the Lagrange-d'Alembert principle for external forces.  This principle reads
\begin{equation} 
\de S+ \int_0^T\!\!\int_{{\cal B}}\left(  \bF_f \cdot \boldeta_f + \bF_s \cdot \boldeta_s \right)  \mbox{d}^3\bx \, \mbox{d} t =0 \, , \quad  
\label{Crit_action} 
\end{equation} 
where $S$ is defined in \eqref{action_p} and the variations are given by \eqref{g_rhos_b_var}. 
Such friction forces are usually postulated from general physical considerations. If these forces are due exclusively to friction, the forces acting on the fluid and media at any given point must be equal and opposite, \emph{i.e.} $\bF_f=-\bF_s$, in the Eulerian treatment we consider here. For example, for porous media, it is common to posit the friction law
\begin{equation} 
\bF_f = - \bF_s= \mathbb{K} (\bu_s - \bu_f)\,,
\label{Darcy_law} 
\end{equation} 
with $\mathbb{K}$ being a positive definite matrix potentially dependent on material parameters and variables representing the media. In particular, the matrix $\mathbb{K}$ depends on the local porosity, composition of the porous media, deformation and other variables. The general functional form of dependence of $\mathbb{K}$ on the variables should be of the form $\mathbb{K}=\mathbb{K}(b,g)$. 
For example, when  deformations of porous media are neglected, \emph{i.e.}, assuming an isotropic and a non-moving porous matrix with $b={\rm Id}$, Kozeny-Carman equation is often used, which in our notation is  written in the form  $\mathbb{K} = \kappa g^3/(1-g)^2$, with $\kappa$ being a constant, see \cite{costa2006permeability} for discussion.
In general, the derivation of the dependence of tensor $\mathbb{K}$ on variables $g$ and $b$ from the first principles is difficult, and should presumably be obtained from experimental observations.

\subsection{Equation of motion}\label{equ_motion}

\paragraph{General form of the equations of motion.} In order to derive the equations of motion, we take the variations in the Lagrange-d'Alembert principle \eqref{Crit_action} as 
\begin{equation} 
\label{Crit_action_explicit} 
\begin{aligned} 
\de S+ &\int_0^T\!\!\int_{{\cal B}} \left( \bF_f \cdot \boldeta_f + \bF_s \cdot \boldeta_s \right)  \mbox{d}^3\bx\,  \mbox{d} t 
\\ 
=& \int_0^T\!\!\int_{{\cal B}} \left[ \dede{\ell}{\bu_f} \cdot \delta  \bu_f +  \dede{\ell}{\bu_s} \cdot \delta \bu_s + \dede{\ell}{\rho_s} \de \rho_s + \left( \dede{\ell}{b}+ pv \pp{c}{b} \right) :\de b   \right. \\ 
& \qquad\quad  + \left( \dede{\ell}{g} - p\right) \de g  +  \left( \dede{\ell}{v} + p c(b)\right) \de v+  \big(g-c(b) v\big)  \de p \\ 
&\qquad \left. \phantom{ \frac{\delta }{\delta }}  +\bF_f \cdot \boldeta_f + \bF_s \cdot \boldeta_s \right] \mbox{d}^3 \bx \,  \mbox{d} t=0 \,.
\end{aligned} 
\end{equation} 
The symbol $``:"$ denotes the contraction of tensors on both indices. Substituting the expressions for variations \eqref{g_rhos_b_var}, integrating by parts to isolate the quantities $\boldeta_f$ and $\boldeta_s$, and dropping the boundary terms leads to the expressions for the balance of the linear momentum for the fluid and porous medium, respectively, written in the Eulerian frame. This calculation is tedious yet straightforward for most terms and we omit it here.
For compactness of notation, we denote the contribution of the Finger tensor $b$ in the elastic  momentum equation with the \emph{diamond} operator 
\begin{equation}\label{diamond_coord}
(\Pi\diamond b)_k= - \Pi_{ij} \pp{b^{ij}}{x^k} - 2 \pp{}{x^i} \left( \Pi_{k j}b^{ij} \right)
\end{equation}
whose coordinate-free form reads 
\begin{equation}  
 \Pi \diamond b = - \Pi : \nabla b - 2 \operatorname{div} \left( \Pi \cdot b \right) \,.
\label{diamond_coord_free} 
\end{equation} 
\rem{ %%%BEGIN REM 
\begin{equation}\label{boundary_b} 
\int_ \mathcal{B} \Pi : \delta b= \int _ \mathcal{B} ( \Pi \diamond b) \cdot \boldsymbol{\eta} \,{\rm d} ^3 \bx + 2 \int_{ \partial \mathcal{B} } [ (\Pi \cdot  b ) \cdot \mathbf{n} ] \cdot \boldsymbol{\eta} \,  {\rm d} s.
\end{equation} 
} %%END REM 
The equations of motion also naturally involve the expression of the Lie derivative of a momentum density, whose global and local expressions are
\begin{equation}\label{Lie_der_momentum}
\begin{aligned} 
\pounds_{\bu}\bm &= \bu \cdot \nabla \bm + \nabla\bu ^\mathsf{T} \cdot \bm + \bm \operatorname{div}\bu\\
(\pounds_{\bu}\bm)_i&= \partial_j m_i u^j + m_j \partial_i u^j+ m_i \partial_j u^j\,.
\end{aligned} 
\end{equation}

With these notations, the Lagrange-d'Alembert principle \eqref{Crit_action_explicit} yields the system of equations
\begin{equation} 
\label{eq_gen} 
\left\{
\begin{array}{l}
\displaystyle\vspace{0.2cm}\partial_t\frac{\delta \ell}{\delta \bu_f}+ \pounds_{\bu_f} \frac{\delta \ell}{\delta \bu_f} = g \nabla \left( \frac{\delta  {\ell}}{\delta g}-  p\right)+\bF_f\\
\displaystyle\vspace{0.2cm}\partial_t\frac{\delta\ell}{\delta \bu_s}+ \pounds_{\bu_s} \frac{\delta\ell}{\delta \bu_s} =  \rho_s\nabla \frac{\delta\ell}{\delta \rho_s} + \left(\frac{\delta\ell}{\delta b}+ p v\frac{\partial c}{\partial b}\right)\diamond b+ \bF_s\\
\displaystyle\vspace{0.2cm} \frac{\delta\ell}{\delta v}= -  pc(b)\,,\qquad g= c(b)v\\
\vspace{0.2cm}\partial_tg + \operatorname{div}(g\bu_f)=0\,,\qquad\partial_t\rho_s+\operatorname{div}(\rho_s\bu_s)=0\,,\qquad \partial_tb+ \pounds_{\bu_s}b=0\,.
\end{array}\right.
\end{equation}
When the boundary conditions \eqref{no_slip} are used, no additional boundary condition arise from the variational principle. In the case of the free slip boundary condition \eqref{free_slip}, the variational principle yields the condition
\begin{equation}\label{BC_general} 
[\sigma _p \cdot \bn ] \cdot \boldsymbol{\eta} =0,\quad \text{for all $ \boldsymbol{\eta} $ parallel to $ \partial \mathcal{B} $},
\end{equation} 
where
\begin{equation}\label{def_sigma_p} 
\sigma _p := -2  \left( \frac{\delta\ell}{\delta b}+ p v\frac{\partial c}{\partial b} \right)  \cdot b.
\end{equation}
Physically, the condition \eqref{BC_general} states that the force  $ \mathbf{t}= \sigma _p \cdot \bn $ exerted at the boundary must be normal to the boundary (free slip). With \eqref{def_sigma_p} and \eqref{diamond_coord_free}, the solid momentum equation can  be written as
\begin{equation}\label{solid_momentum_rewritten} 
\partial_t\frac{\delta\ell}{\delta \bu_s}+ \pounds_{\bu_s} \frac{\delta\ell}{\delta \bu_s} =  \rho_s\nabla \frac{\delta\ell}{\delta \rho_s} - \left(\frac{\delta\ell}{\delta b}+ p v\frac{\partial c}{\partial b}\right): \nabla  b+ \operatorname{div} \sigma _p + \bF_s.
\end{equation}

The first equation in \eqref{eq_gen} arises from the term proportional to $\boldeta_f$  in the application of the Lagrange-d'Alembert principle. The second condition and the boundary condition \eqref{BC_general} arise from the term proportional to $\boldeta_s$. The third and fourth equations arise from the variations $\delta v$ and $\delta p$.
The last three equations follow from the definitions \eqref{cons_law_fluid}, \eqref{cons_law_elastic}, \eqref{intrinsic_def_b}, respectively.
In the derivation of \eqref{eq_gen}, we have used the fact that on the boundary $\partial\mathcal{B}$, $\boldeta_s$ and $\boldeta_f$ satisfy the boundary condition \eqref{non-permeable}.

\begin{remark}[Discussion of the form of the Lagrangian] 
{\rm 
Equations \eqref{eq_gen} allow for an arbitrary form of the dependence of the Lagrangian on the variables. The derivatives of the Lagrangian with respect to the variables entering \eqref{eq_gen}  should be considered to be variational derivatives. For example, if the integrand of the Lagrangian depends on both $\rho_s$ and its spatial derivatives $\nabla \rho_s$,  \emph{e.g.} 
\[ 
\ell = \int_{\mathcal{B}} \ell_0 ( \rho_s, \nabla \rho_s, \bu_s, \ldots) \mbox{d} ^3  \bx
\] 
then 
\[ 
\dede{\ell}{\rho_s}=\pp{\ell_0}{\rho_s}-\div \pp{\ell_0}{\nabla \rho_s} \, , 
\] 
and similarly with other variables such as $\bu_s$, $\brho_f$, $v$ \emph{etc}. 
Thus, equations \eqref{eq_gen} are capable of incorporating very general physical models of the porous media. However, it is important to note that in our model, we do not assume that the energy of the fluid depends on any kind of strain measure of the solid or the fluid. These energy considerations only refer to the fluid; the energy of the solid, of course, depends on $b$, the measure of strain of the solid.
The pressure $p$ in \eqref{eq_gen} is obtained purely from the  action principle with the action \eqref{action_p}. In that sense, our paper follows the framework of fluid description due to Arnold \cite{arnold1966geometrie}. }
\end{remark} 

\begin{remark}[On energy dissipation]
{\rm
In spatial coordinates, the forces applied to the fluid and solid must be equal and opposite, so $\bF_f = - \bF_s$. Under the appropriate boundary conditions, equations \eqref{eq_gen} dissipate energy. It can be shown that for arbitrary Lagrangian in \eqref{eq_gen}, we have 
\begin{equation}
E=\int_{{\cal B}} \left[ \bu_f \!\cdot\! \dede{\ell}{\bu_f} + 
\bu_s\! \cdot\! \dede{\ell}{\bu_s}
+ 
\dot v \dede{\ell}{\dot v} - \mathcal{L} \right] \mbox{d} ^3   \bx  \, 
\Rightarrow \, 
\dot E =  \int_{\cal B} \bF_s \cdot (\bu_s - \bu_f) \mbox{d} ^3   \bx ,
\label{diss_energy}
\end{equation}
so $\dot E \leq 0$ for arbitrary Darcy-like friction $\bF_s = \mathbb{K} (\bu_f-\bu_s)$, as long as the matrix $\mathbb{K}$ is positive definite. 
}
\end{remark}

\paragraph{Specific form of the equations.}

For the Lagrangian defined in  \eqref{Lagr_def}, using the third and fourth equations in \eqref{eq_gen}, the second term on the right hand side of \eqref{solid_momentum_rewritten} can be written as
\begin{align*} 
- \left(\frac{\delta\ell}{\delta b}+ p v\frac{\partial c}{\partial b}\right): \nabla  b&= \frac{\partial V}{\partial b}: \nabla b - pv \frac{\partial c}{\partial b}: \nabla b \\
&= \nabla \left( V - \frac{\partial V}{\partial v}v \right) + v \left( \nabla \frac{\partial V}{\partial v}- p \nabla c \right) \\
&  =g\nabla p + \nabla \left( V - \frac{\partial V}{\partial v}v\right).
\end{align*}
Then, the equations of motion \eqref{eq_gen} become 
\begin{equation}
\label{expressions_explicit} 
\hspace{-3mm} 
\left\{
\begin{array}{l}
\displaystyle\vspace{0.2cm}\bar\rho_f^0(\partial_t \bu_f+ \bu_f\cdot\nabla \bu_f )  = - \nabla  p + \frac{1}{g} \bF_f\\
\displaystyle\vspace{0.2cm}\rho_s (\partial_t \bu_s \!+\! \bu_s\cdot\nabla \bu_s) =g\nabla p \!+\! \nabla \left( V\! - \!\frac{\partial V}{\partial v}v\right) + \operatorname{div}  \sigma _p + \bF_s\\
\displaystyle\vspace{0.2cm} \frac{\partial V}{\partial v}=  pc(b),\qquad g= c(b)v\\
\vspace{0.2cm}\partial_tg + \operatorname{div}(g\bu_f)=0,\qquad \partial_t\rho_s+\operatorname{div}(\rho_s\bu_s)=0,\qquad \partial_tb+ \pounds_{\bu_s}b=0\,,
\end{array}\right.
\end{equation}
where the stress tensor $ \sigma _p $ defined in \eqref{def_sigma_p} becomes  
\begin{equation} 
\label{sigma_p} 
\sigma_p= -2 \left(  p v \frac{\partial c}{\partial b}\! -\! \frac{\partial V}{\partial b}\right)\cdot b\,, \qquad (\sigma_p)_k^i= -2 \left(  p v \frac{\partial c}{\partial b^{kj}}\! -\! \frac{\partial V}{\partial b^{kj}}\right)b^{ij}\,,
\end{equation}
and with the boundary condition \eqref{BC_general}.
 
With \eqref{sigma_p}, the divergence term in the media momentum equation (second equation above) is the analogue of the divergence of the stress tensor for an ordinary elastic media. This term, however, contains the contribution from both the potential energy and the fluid pressure. In what follows, we are going to extend these equations to the case of doubly incompressible media (fluid and solid), and we will also introduce the muscle stress. 

\revision{R1Q1}{
\paragraph{On the analysis of dimensionless parameters and relevant time scales.} 
It is interesting to estimate the relative importance of the various terms appearing in the equations of motion. The easiest way to proceed is through the analysis of time scales. \\
Suppose $\mu_f$ is the dynamic viscosity of the fluid, and $d$ is the typical size of the pores. Then, it is natural to form the pore-based Reynolds number comparing the order of magnitude of inertia to viscous terms. Unlike the single fluid case, where this number would be defined as $R = \rho_f u_f d/\mu_f$, this number should incorporate the relative motion of the fluid and solid media to accurately account for friction in the media. So, a natural way to define the porous media Reynolds number would be something like 
\begin{equation} 
R_{\rm pm} = \frac{\rho_f |\bu_f-\bu_s| d}{\mu_f} = \frac{|\bu_f-\bu_s| d}{\nu_f}\, , 
\label{PM_R}
\end{equation} 
where $\nu_f = \mu_f/\rho_f$ is the kinematic viscosity of the fluid. 
Then, $R_{pm} \gg 1$ will define the fully inertial regime whereas $ R_{pm} \ll 1$ will correspond to the fully viscous regime.  For a typical velocity difference $|\bu_f-\bu_s|\sim$1 mm/s, pore size $d \sim 1$mm, and the fluid being water, $R_{\rm pm} \sim 1$ so inertial and viscous terms are similar in order.
\\
Unfortunately, it is usually not easy to evaluate the typical scale of $|\bu_f-\bu_s|$, as they follow from the equations of motion. We shall thus develop an alternative approach by comparing the relevant time scales. 
\\
Suppose at $t=0$ there is an instantaneous velocity field created in one of the media, solid or fluid. A typical time scale for velocity field to converge to its equilibrium value $\bu_f \sim \bu_s$ due to friction can be estimated as 
\begin{equation} 
T_{\rm fr} \simeq \frac{\rho_f d^2}{\mu_f}= \frac{d^2}{\nu_f}.
\label{T_friction}
\end{equation} 
Taking $d=1$mm and the fluid being water with $\nu_f \sim 0.01$cm$^2$/s yields $T_{\rm fr} \sim$1 sec. 
\\
A second time scale can be derived from considerations of sound propagation in the media. This estimate is useful as the sound waves come from the balance of the elastic and inertial terms in the equations. As is described in \cite{FaFGBPu2020}, typical sound velocity in the fluid-filled porous media can be taken to be of the order $c \sim \sqrt{E/\rho_s}$, with $E$ being elastic modulus, with several corrections due to the nature of porous media and fluid interactions.  For cells-based materials, the typical value of elastic modulus is in the range of 100Pa to 100kPa \cite{demichelis2013study,vinckier1998measuring}. For biological tissues made mostly of water with $\rho_s \sim 1000$ kg/m$^3$, the typical sound velocity is then about $0.3 -30$m/s. For the purpose of estimate, if we take the sound speed as above, and the typical scale we are interested in is $L$, then the typical time scale is 
\begin{equation} 
T_{el}= \frac{L}{c} \sim L \sqrt{\frac{\rho_s}{E}}.
\label{T_sound}
\end{equation} 
For a typical scale of interest $L \sim 10$cm=$0.1$m, the typical time scale of sound wave propagation from one side of the system to another is going to be from $0.01$s for harder materials, to $1$s for very soft materials. The relevance of inertia to viscous terms can thus be described by the dimensionless number 
\begin{equation} 
{\cal P} = \frac{T_{fr}}{T_{el}},
\label{inertia_number} 
\end{equation} 
with ${\cal P} \gg 1$ corresponding to the case when elastic wave takes many oscillations before dissipating due to friction. For typical biological materials and fluid being water as described above ${\cal P} \sim 1 -100$ and the inertia terms are thus important for considerations of biological materials.  
}
\revision{EQ1\\R1Q4}{\subsection{Connection to previously derived models of porous media} 
\label{sec:previous_models} }

\subsubsection{Static media filled with ideal gas and connection to the porous medium equation (PME)}
\label{sec:PME} 
%https://www.imo.universite-paris-saclay.fr/~egert/Barenblatt.pdf
Let us consider a physical system that describes a polytropic flow of an ideal gas through a homogeneous static porous medium. For such a system, the kinetic and potential energies of the solid matrix are no longer relevant and we only have the terms describing the gas component. The state equation for pressure has the form
\begin{equation}
    p = p_0 \rho^\gamma,
\end{equation} where $\gamma \ge 1$ is the polytropic exponent. The choice of the exponent could describe among others isentropic (adiabatic and reversible, $\gamma = \gamma_0 > 0$) and isothermal (with $\gamma = 1$) flows. To apply the variational geometric formulation to such system, we need to compute the potential energy of the gas. In case $p_0(\bx) = \mathrm{const}$ we simply have
\begin{equation}
V(\rho, T) = V_0 \rho^{\gamma}.
\end{equation}

The corresponding Lagrangian takes the form  

\begin{equation}
\ell(\bu, \rho)= \int_{{\mathcal B}}  \left[\frac{1}{2}\rho|\bu|^2 -V_0 \rho^{\gamma}\right]{\rm d}^3\bx.
\label{lagr_simple_fluid}
\end{equation}

With these notations, the variations in the Lagrange-d'Alembert principle \eqref{Crit_action} yields the system of equations
\begin{equation} 
\label{eq_pme_dynamic} 
\left\{
\begin{array}{l}
\displaystyle\vspace{0.2cm}\partial_t\frac{\delta\ell}{\delta \bu}+ \pounds_{\bu} \frac{\delta\ell}{\delta \bu} =  \rho\nabla \frac{\delta\ell}{\delta \rho} + \bF\\
\vspace{0.2cm}\partial_t \rho + \operatorname{div}(\rho \bu)=0\,,
\end{array}\right.
\end{equation}
where for the friction force we posit $\bF := -\mu \rho \bu,$ according to the linear Darcy's law.

After computing the explicit form of functional derivatives and collecting the terms, the equations of motion \eqref{eq_pme_dynamic} become 
\begin{equation}
\label{pme_dynamic_explicit} 
\hspace{-3mm} 
\left\{
\begin{array}{l}
\displaystyle\vspace{0.2cm}\rho(\partial_t \bu+ \bu\cdot\nabla \bu )  = - \rho \nabla  \pp{V}{\rho} - \mu \rho \bu\\
\vspace{0.2cm}\partial_t \rho + \operatorname{div}(\rho \bu)=0\,.
\end{array}\right.
\end{equation}
Let us consider the special case of a degenerate Lagrangian without kinetic energy term
\begin{equation}
\ell(\bu, \rho)= -\int_{{\mathcal B}}   V_0 \rho^{\gamma}{\rm d}^3\bx.
\label{potential_only_lagrangian}
\end{equation}
The application of the Lagrange-d'Alembert principle to \eqref{potential_only_lagrangian} will formally yield a system with no dynamic terms in the left hand side of the first equation of \eqref{eq_pme_dynamic}, namely
\begin{equation}
\hspace{-3mm} 
\left\{
\begin{array}{l}
\displaystyle\vspace{0.2cm}\mu \bu  = - \nabla  \pp{V}{\rho} = - \gamma V_0 \rho^{\gamma - 1} \nabla \rho = -\frac{V_0}{p_0} \nabla p\\
\vspace{0.2cm}\partial_t \rho + \operatorname{div}(\rho \bu)=0\,.
\end{array}\right.
\label{PME_system}
\end{equation}

We substitute the velocity from the first equation into the continuity equation in the system \eqref{PME_system} to get a closed form equation for the density
\begin{equation}
\label{pme_explicit} 
\partial_t \rho 
=  \frac{V_0}{\mu} \operatorname{div}(\rho \nabla \rho^\gamma).
\end{equation}

The equation \eqref{pme_explicit} is known as the porous medium equation (PME).

We should notice, that while this result was obtained simply by dropping the inertial term $\frac{1}{2}\rho|\bu|^2$ from the Lagrangian, the correctness of this approach goes beyond the scope of this paper, as it employs a degenerate Lagrangian. The use of a small parameter $\epsilon \rightarrow 0$ as a coefficient for the inertial term would lead to the singular perturbations and will not yield the same result immediately. The solutions of systems with singular perturbations may employ asymptotic methods, see for example \cite{verhulst2005methods}.

The porous medium equation \eqref{pme_explicit} could be nondimensionalized by scaling out the constant and rewritten in the form 
\begin{equation}
 \partial_t u = \Delta (u^m), \quad m := 1 + \gamma > 1,
 \label{PME}
\end{equation} which is a nonlinear heat equation, formally of parabolic type \cite{vazquez2007porous}. The interesting property of this porous medium equation, that does not hold for the linear heat equation $\partial_t u = \Delta u$ is that \eqref{PME} has finite speed for the propagation of disturbances from the level $u = 0.$

The PME \eqref{PME} has a special solution representing mass or heat release from a point source \cite{vazquez2007porous}, that was obtained by Zeldovich, Kompaneets and Barenblatt, and the terms \textit{source solution}, \textit{Barenblatt-Pattle solution}, \textit{Barenblatt solution} or \textit{ZKB solution} are often used for this solution.  The ZKB solution has the self-similar form:  
$u=t^{-\alpha} f(\xi)$ with $\xi = \|\bx\| t^{-\beta}$, see \cite{barenblatt_1996}. For equation \eqref{PME}, this self-similar solution can be written explicitly as 
\begin{equation}
    u(\bx, t) = t^{-\alpha} \left( C - k \|\bx\|^2 t^{-2\beta}\right)^{\frac{1}{m-1}}_+,
\label{ZKB}
\end{equation}
where $(f)_+ := \max\{f, 0\},$
\begin{equation}
    \alpha = \frac{d}{d(m-1) + 2}, \quad \beta = \frac{\alpha}{d}, \quad k = \frac{\alpha(m-1)}{2md},
\end{equation}
$C > 0$ is an arbitrary constant and $d$ is the dimension of the space. The constant $C$ is chosen in such a way that the mass of the solution $M$ is equal to a given value. 
The initial data in \eqref{ZKB} as $t \rightarrow 0$ is a Dirac mass $u(\bx, t) \rightarrow M \delta(\bx),$ where $M = M(C, m, d).$ The source solution has compact support in space for every fixed time, in other words, the solution is non-zero only for $\| \bx\|\leq R(t)$, with the radius of the support $R(t) = t^\beta \sqrt{C/k},$ and the mass of the solution conserved for all times, $\int_{\mathbb{R}^d} u(\bx, t) \mathrm{d}\bx = M = \mathrm{const}.$

\revision{R2Q1}{
\subsubsection{Connections with Biot's equations for wave propagation}
It is worth noting that the equations \eqref{expressions_explicit} reduce to the Biot's model for propagation of waves in porous media \cite{biot1956theory1}, see also \cite{Fellah2004ultrasonic} for current application to biomedical field. This result was explained in details in \cite{FaFGBPu2020}, and we refer the reader to that paper, as the derivation of linearization is quite technical and cumbersome. To achieve that correspondence, one has to linearize equations \eqref{expressions_explicit} about a uniform homogeneous state $g=g_0$, yielding: 
\begin{equation}\label{PDE_dispersion} 
\left\{ 
\begin{array}{l}
\displaystyle\vspace{0.2cm}\rho_f g_0 \frac{\partial^2}{\partial t^2} \bu_f +\beta \frac{\partial}{\partial t}(\bu_f - \bu_s) -g_0 \zeta \nabla {\rm div} \bu_f + g_0 (\zeta+ \xi)   \nabla {\rm div} \bu_f=\mathbf{0}\\
\displaystyle\vspace{0.2cm}\rho_s \frac{\partial^2}{\partial t^2}\bu_s- \beta \frac{\partial}{\partial t}(\bu_f - \bu_s) + g_0 {(\zeta + \xi)} \nabla {\rm div} \bu_f\\
\displaystyle \qquad   - \left( g_0 (\zeta +2 \xi) + \Lambda+ G \right) \nabla {\rm div} \bu_f -G  \Delta \bu_s = \mathbf{0} \,, 
\end{array} 
\right. 
\end{equation} 
where $\Lambda$ and $G$ are the Lam\'e coefficients for the dry porous media, and  $\zeta$ and $\xi$ come from the microscopic energy of the pores. These variables have the  dimension of elastic modulus (\emph{i.e.}, pressure).  The coefficient $\xi$ is related to the dry media bulk modulus and is coming from the second derivative $\partial^2_{gg}V$ at the equilibrium. The cross-derivative  $\partial^2_{g b}V$ at the equilibrium  is a tensor assumed to be proportional to the identity matrix for a uniform isotropic media, with a coefficient of proportionality being equal to $\xi$ with an appropriate coefficient making $\xi$ having the dimension of elastic modulus. 
\\
 The corresponding Biot's system for acoustic waves in porous media is given by
\begin{equation} 
\label{linear_biot_system_0}
\hspace{-0.33cm}\left\{
\begin{array}{l}
\displaystyle\vspace{0.2cm}\frac{\partial^2}{\partial t^2}(\rho^{(f)}_{22} \bu+ \rho_{12} \bv) + \beta \frac{\partial}{\partial t}(\bu_f - \bu_s) - \nabla\div (R\bu_f+ Q\bu_s )  =  \mathbf{0}, \\
\displaystyle\vspace{0.2cm}\frac{\partial^2}{\partial t^2}(\rho^{(s)}_{11} \bu_s + \rho_{12} \bu_f) - \beta \frac{\partial}{\partial t}(\bu_f - \bu_s)\\
\displaystyle \qquad  - \nabla\div (Q\bu_f + P\bu_s) + N  \nabla \times \nabla \times  \bu_s = \mathbf{0},
\end{array} 
\right. 
\end{equation}
with $N$ being shear modulus of the elastic body. 
We shall note that Biot's equations is not directly applicable to an incompressible fluid, since the expressions for the variables $P,Q$ and $R$ in \eqref{linear_biot_system_0} involve explicitly the bulk modulus of the fluid.  However, if we proceed formally and use the equations from the literature and put $K_f= \infty$ for an incompressible fluid, the expressions for $P,Q$ and $R$ in terms of the bulk moduli of the porous skeleton $K_b$ and the elastic body itself $K_s$, see \textit{e.g.}, \cite{Fellah2004ultrasonic} are given by 
\begin{equation} 
\label{PQR_def} 
P=(1-g_0) K_s + \frac{4}{3} N \,, \quad Q= g_0 K_s \, , \quad R= \frac{g_0^2 K_s}{1- g_0 - K_b/K_s } \, . 
\end{equation} 
Let us now compare this system with the linearization of our equations  \eqref{PDE_dispersion}, where we have set $\rho_{12}=\rho_{21}=0$. The case of $\rho_{12} \neq 0$ and $\rho_{21} \neq 0$ can be easily incorporated by considering a more general inertia matrix in the Lagrangian.  There is also an exact correspondence between the friction terms. Thus, we need to compare the coefficients of the spatial derivative terms. A direct comparison between Biot's linearized system \eqref{linear_biot_system_0} and \eqref{PDE_dispersion} gives 
$R= g_0 \zeta$ by observing the coefficients of the terms proportional to $\nabla {\rm div} \bu_f$ from the equations \eqref{linear_biot_system_0}. From the term proportional to  $\nabla {\rm div} \bv$ in the first equation of \eqref{linear_biot_system_0}, we obtain $Q=-g_0 (\xi + \zeta)$. Finally by using $\nabla \times \nabla \times  \bv = \nabla \div \bv - \Delta \bv$ we obtain the expressions of $N$ and $P$. To summarize, the Biot's coefficients $(P,Q,R,N)$ are given by 
\begin{equation} 
\begin{aligned}
\label{Biot_correspondence}
R &= g_0\zeta,\quad 
%\\
Q & = -g_0(\xi+\zeta),\quad 
%\\
N & = G,  % + g_0\mu,\   
\quad 
%\\
P&=\Lambda +2 G  + g_0 ( \zeta + 2 \xi) \,.
\end{aligned}
\end{equation}
The expression $\Lambda+2 G$ is also known as the $P-$wave modulus. In our case, this $P$-wave modulus is modified by  additional terms $\xi$ and $\zeta$  coming from the microscopic elasticity properties of the porous matrix. We refer the reader to \cite{FaFGBPu2020} for more details. 
}
\revision{R2Q2}{\subsubsection{Connections with Biot's quasi-static equations of porous media}
Other important equations related to porous media are the so-called poroelasticity equations \cite{bociu2016analysis, bociu2020nonlinear}, which are sometimes also called the equations for the porous media, or the (nonlinear) Biot model for quasi-static porous media. See also \cite{showalter2000diffusion} for literature review and mathematical analysis of solutions for related models. These equations describe slow, inertia-less deformation of the porous media filled with an incompressible fluid. If we neglect the kinetic energy terms in the Lagrangian and consider the Lagrange-d'Alembert principle
\[
\delta \int_0^T\!\!\!\int_ \mathcal{B} \left[ -e_s(b, g ) + p\left(g - g_0 \circ \bvarphi^{-1} J_{\bvarphi^{-1}} \right)   \right] {\rm d} ^3 \mathbf{x} \,{\rm d} t + \int_0^T\!\!\!\int_ \mathcal{B} \big( \bF _f \cdot \boldsymbol{\eta} _f + \bF _s \cdot \boldsymbol{\eta} _s \big) {\rm d} ^3 \mathbf{x}\, {\rm d} t=0
\]
for variations $ \delta b = - \pounds _ { \boldsymbol{\eta} _s} b$ and free variations $ \delta g$, $ \delta p$, $ \delta \bvarphi= \boldsymbol{\eta} _f \circ \bvarphi$, we get
%%%%%%%
\begin{equation} 
\left\{ 
\begin{aligned} 
0= & - g \nabla p + \mathbb{K} (\bu_s-\bu_f)\\ 
0 =&  - (1-g) \nabla p + \operatorname{div}  \left[ \big(e_s +  p (1- g)\big){\rm Id} +  \sigma_e\right]  + \mathbb{K} (\bu_f-\bu_s) 
\\ 
0 = & -\pp{e_s}{g}(b,g) +p\,,
\end{aligned} 
\right. 
\label{no_inertia} 
\end{equation} 
where $(\sigma_e)^i_k=  2 \frac{\partial e_s}{\partial b^{kj}}b^{ij}$.
If we linearize the elastic stress tensor $\sigma_e$ in the above equations as
\begin{equation} 
\sigma_e \simeq \Lambda \epsilon +G \mbox{div} \epsilon - (1-\phi) p  {\rm Id}, \quad \epsilon \simeq \frac{1}{2} \left(b - {\rm Id} \right) 
\end{equation} 
and add the first two equations demonstrating the balance of momenta for the fluid and the solid, we obtain the equations for the quasi-static porous media equations that are quite close to (but not exactly the same as) those presented in \cite{bociu2016analysis} without the viscoelastic terms and  \cite{bociu2020nonlinear}.
%\rem{%%BEGIN REM 
%\begin{equation} 
%\label{eq_quasi-static} 
%g \nabla p = \mathbb{K} (\bu_f-\bu_s) \, ,\quad 
%\mbox{div} \sigma_{\rm tot} + 
%\mathbb{K} (\bu_f-\bu_s) =0 \,, \quad 
%\sigma_{\rm tot}:= \sigma_e+(e_s-p g) {\rm I} 
%\end{equation} 
%} %%%END REM 
The last equation of our reduced equations \eqref{no_inertia}, obtained with respect to variations $ \delta g$, connects the pressure, which is the Lagrange multiplier for incompressibility, with the fluid content and deformations. 
In contrast, the nonlinear Biot model uses an alternative \emph{Bio-Willis} relationship for the quantity (fluid content) $\zeta(t, \bx) = g(t,\bx)-g_0(\bx)$, where $g_0(\bx)$ is a baseline local value of the porosity \cite{bociu2020nonlinear}. In our notation, assuming $\bpsi(0,\bx)=\bx$, the Biot-Willis relationship for the fluid content $\zeta$ reads 
\begin{equation}
\zeta=c_0 p + \alpha  {\rm div} \left(\bpsi^{-1}(t,\bx)-\bx \right) \,.
\label{Biot_Willis}
\end{equation} 
\revision{R1Q2}{
For an incompressible fluid and solid, one takes $c_0=0$ and $\alpha=1$, attributing all the change in available volume to the dilation of the media. In our opinion, \eqref{Biot_Willis} is somewhat difficult to justify mathematically for large deformations. For example, if the system consisting of incompressible fluid and solid is undergoing a shift as a rigid body with the velocity $\mathbf{U}$, then $\bpsi^{-1}(t,\bx) = \bU t + \bx$. Thus, the equation \eqref{Biot_Willis}  states that $g(t, \bx)=g_0(\bx)$. However, the true statement is that $g$ simply moves with the media, so $g(t,\bx) = g_0(\bx - \mathbf{U} t)$, where we have identified, for simplicity, the coordinates $\bX$ and $\bx$ at $t=0$. Thus, if $g_0(\bx)$ is non-uniform in space,  it is difficult to interpret the equation \eqref{Biot_Willis}. It is especially true for the biological materials which have inherently non-uniform porosity, such as 'sneezing' sponges  \emph{Ephydatia muelleri} described in \cite{ludeman2014evolutionary}, which served as an inspiration for this paper. In our description, we do not need an extra physical assumption such as \eqref{Biot_Willis}. The full condition for porosity for any motion for incompressible fluid and solid reads 
\begin{equation} 
\left\{
\begin{aligned} 
&g(t,\bx) = g(0,\bvarphi^{-1} (t,\bx)) J_{\bvarphi^{-1} } (t,\bx)
\\
&(1-g(t,\bx)) = (1-g(0,\bpsi^{-1} (t,\bx)))J_{\bpsi^{-1} } (t,\bx)\,.
\end{aligned} 
\right. 
\label{doubly_incompr} 
\end{equation}
Thus,  the true condition for doubly incompressible flow involves both $g(t,\bX)$, $\psi(t,\bX)$, and $\bvarphi(t,\bX)$. A differential form of these incompressibility conditions is given by equation \eqref{g_uf_us_constr} below. This additional incompressibility condition closes the system and no additional physical assumptions such as \eqref{Biot_Willis} are necessary. We shall postpone the discussion of the doubly incompressible case until Sec.~\ref{sec:doubly_incompr}. Here, we just notice that in our opinion, a direct application of Biot-Willis' porosity conditions \eqref{Biot_Willis} to realistic biological systems is difficult, especially when additional muscle stress is introduced for large deformations and in the presence of non-uniform porosity --
a fact of which the sponge \emph{Ephydatia muelleri} is happily unaware.
}
\\
In contrast, in our model we do not need the additional constraint \eqref{Biot_Willis} as the relationship connecting $\zeta$, $p$ and $\phi$ automatically follows from the last equation of \eqref{no_inertia}. \revision{R2Q3}{Moreover, we have an automatic dissipation of the energy due to the existence of variational principle. In contrast, finding the energy-like quantity for the nonlinear porous media (Biot) model is highly nontrivial, as the works  \cite{bociu2016analysis, bociu2020nonlinear} illustrate. 
We thus believe that in spite of higher apparent complexity compared to the simplified models used in the literature in the field, our equations are actually mathematically simpler to analyze from the point of view of functional analysis, and hope that experts in analysis of PDEs will have an opportunity to perform rigorous analysis of existence and uniqueness of solutions to our equations. 
}
}
\revision{R2Q3}{\subsubsection{Perforated domains and the nature of friction terms} }
It is also interesting to connect our works to the literature on homogenizations of perforated domains and subsequent discussions of the nature of friction terms in porous media. Perforated domains are defined to be a two-phased media, which contains of a large number of small holes made out of solid with the remainder of the domain filled with fluid \cite{allaire1991homogenization1,allaire1991homogenization2}. The concept of perforated domains was originally introduced as a means to rigorously derive the origin of friction terms in porous media, such as Darcy's law containing fluid velocity only and corresponding Darcy-Brinkman models, involving velocity and Navier-Stokes like friction terms involving gradients of velocity \cite{brinkman1949calculation,brinkman1952viscosity}, see also \cite{srinivasan2014thermodynamic}. Since then, perforated domains have enjoyed active studies especially regarding rigorous analysis of systems that are close to the Navier-Stokes equations, in the context of vanishing viscosity and very small hole size \cite{lacave2016vanishing}. The vanishing velocity leads to boundary layers and thus very interesting mathematical analysis of the problem. One can view our system as a possible homogenization of the system where the holes in space are not static, but are connected to each other by springs. The discussion of exact nature of the friction terms is beyond the scope of this paper. Two of the authors of this paper have recently outlined general possibilities for the friction terms in \cite{FGBPu2021}, and we refer the reader to that paper for details.  Since our equations allow for arbitrary nonlinearities and  accurately take into account inertia, we believe that the connection between homogenization of equations in perforated domains is of interest and should be pursued, since the friction terms in our system are posited and not derived. 

%%%END REM 

\color{black} 

\section{The case of living organisms and active porous media}

\subsection{Introduction of elastic muscle stress}\label{Introduction_muscle}

Let us now consider a porous media that can generate its own stress in the solid in addition to the elastic stress experienced by the solid from the deformation. The physical context of this work are the sponges generating their own internal stress to contract and expunge water from themselves. The sponges generate their own internal muscle stress which we call $\bar{s}$, when we compute that stress in the Eulerian frame of reference.
%\todoFGB{Not sure if $\bar{S}$ is in the Lagrangian frame. You mean $\bar{S}(t,\bX)^{ij}$, right? with Eulerian indices $ij$ at the point $\bx=\bpsi(t,\bX)$ and expressed at the Lagrangian point $\bX$, so a mixture of Eulerian-Lagrangian frame because it is "along $\bpsi$". A more geometric material object is a stress $\bar P^{Ai}$ which appears in the unreduced Lagrange-d'Alembert principle simply as $\bar P:  \nabla \delta \bpsi$. Then $\int_ { \mathcal{B} _s} \bar P( \bX):  \nabla \delta \bpsi (\bX){\rm d}^3\bX= \int_{\mathcal{B}} \bar s(\bx): \nabla \boldsymbol{\eta} _s{\rm d}^3\bx$, where $\bar P$ and $\bar s$ are related via the Piola transform. Not sure what we want to say, but maybe it is confusing to say that $\bar S$ is in the Lagrangian frame.
%\\
%VP: Yes, that is a good point. Indeed, $\bar S$ has coordinates that are computed in Eulerian frame. We actually never use that Lagrangian frame muscle stress, so I reformulated it to streamline the discussion. So I removed the text below: 
%\\
%\texttt{Because the muscle action is controlled by the living organism, we can assume that $\bar{S}$ is a given function of the Lagrangian label $\bX$ (material particle of the sponge) and time: $\bar{S}=\bar{S}(t, \bX)$. The corresponding stress recalculated in the spatial frame will be denoted by $\bar{s}(t,\bx)=\bar{S}(t, \bX(t,\bx))$.}
%\\
%and just refererred to the Eulerian stress. 
%}
The forces $\operatorname{div} \bar{s}$ generated by this stress in the spatial frame have to be considered as external in Hamilton's principle \eqref{Crit_action}, acting only on the solid part of the system.

In this work, we shall only consider the mechanical effect of the muscle stress, without getting into the details of the actual mechanism of generation of stress itself. The microscopic dynamics of the muscle itself is highly complex \cite{geeves1999structural}, and is not essential at this point. It may, however, become important later when thermodynamics effects are considered.  

There are two ways to introduce the muscle force by modifying \eqref{Crit_action}. One way is to consider the muscle stress through the typical way of adding external forces via the Lagrange-d'Alembert principle. In that modification, the muscle forces ${\rm div}\overline{s}$ are added to the force acting on the solid $\bF_s$, as: 
\begin{equation} 
\de S+\int_0^T\!\!\int_{{\cal B}} \left(   \bF_f \cdot \boldeta_f + \left( \bF_s + \operatorname{div} \bar{s} \right) \cdot \boldeta_s \right)  \mbox{d}^3\bx \, \mbox{d} t =0 \, .\quad  
\label{Crit_action_muscle} 
\end{equation} 
Another way is to consider the muscle action as an \emph{internal} force 
through an alternative formulation of the critical action principle as 
\begin{equation} 
\label{Crit_action_muscle_2} 
\de S + \int_0^T\!\!\int_{{\cal B}} \left( \bF_f \cdot \boldeta_f+ \bF_s \cdot  \boldeta_s - \bar{s} : \nabla  \boldeta_s \right) \mbox{d}  ^3  \bx \,\mbox{d} t =0 \,.
\end{equation} 
At the first sight, both formulations are equivalent since application of integration over the volume to the last term brings \eqref{Crit_action_muscle_2} to \eqref{Crit_action_muscle}, \emph{except for the boundary terms}. The action \eqref{Crit_action_muscle_2} generates an additional force on the boundary given by $-\overline{s}\cdot \bn$, with $ \bn ( \mathbf{x} )$ the normal to the boundary at $\bx \in \partial {\mathcal B}$.

\revision{R3Q1}{This variational principle illustrates the first key difference with \cite{FaFGBPu2020}, which did not comprise the theory of internal muscle stress. As far as we are aware, this is a novel variational principle for the inclusion of internal muscle stress.  In this section, we apply this variational principle to the case of compressible solid and incompressible fluid. 
} 

Let us now consider the following thought experiment: a volume of solid is acted upon with the uniform muscle stress $\overline{s}= \text{const}$. Then, $\operatorname{div}\overline{s}=0$ inside the volume
${\mathcal B}$. However, uniform stress will affect the boundary in the formulation \eqref{Crit_action_muscle_2} and not in the  formulation \eqref{Crit_action_muscle}. Thus, in our opinion, it is more natural to consider the approach \eqref{Crit_action_muscle_2} and not \eqref{Crit_action_muscle}.

This variational principle leads to the modified version of \eqref{eq_gen} for the active porous media: 
\begin{equation} 
\label{eq_gen_muscle} 
\left\{
\begin{array}{l}
\displaystyle\vspace{0.2cm}\partial_t\frac{\delta \ell}{\delta \bu_f}+ \pounds_{\bu_f} \frac{\delta \ell}{\delta \bu_f} = g \nabla \left( \frac{\delta  {\ell}}{\delta g}-  p\right)+\bF_f\\
\displaystyle\vspace{0.2cm}\partial_t\frac{\delta\ell}{\delta \bu_s}+ \pounds_{\bu_s} \frac{\delta\ell}{\delta \bu_s} =  \rho_s\nabla \frac{\delta\ell}{\delta \rho_s} + \left(\frac{\delta\ell}{\delta b}+ p v\frac{\partial c}{\partial b}\right)\diamond b+ \operatorname{div}  \bar{s} +\bF_s\\
\displaystyle\vspace{0.2cm} \frac{\delta\ell}{\delta v}= -  pc(b)\,,\qquad g= c(b)v\\
\vspace{0.2cm}\partial_tg + \operatorname{div}(g\bu_f)=0\,,\qquad\partial_t\rho_s+\operatorname{div}(\rho_s\bu_s)=0\,,\qquad \partial_tb+ \pounds_{\bu_s}b=0\,.
\end{array}\right.
\end{equation}
For the particular case of the Lagrangian given by \eqref{Lagr_def} the equations become 
\begin{equation}
\label{expressions_explicit_muscle} 
\hspace{-3mm} 
\left\{
\begin{array}{l}
\displaystyle\vspace{0.2cm}\bar\rho_f^0(\partial_t \bu_f+ \bu_f\cdot\nabla \bu_f )  = - \nabla  p + \frac{1}{g} \bF_f\\
\displaystyle\vspace{0.2cm}\rho_s (\partial_t \bu_s \!+\! \bu_s\cdot\nabla \bu_s) =g\nabla p + \nabla \left( V - \frac{\partial V}{\partial v}v\right) + \operatorname{div} \left( \sigma_p + \bar{s}  \right) + \bF_s\\
\displaystyle\vspace{0.2cm} \frac{\partial V}{\partial v}=  pc(b),\qquad g= c(b)v\\
\vspace{0.2cm}\partial_tg + \operatorname{div}(g\bu_f)=0,\qquad \partial_t\rho_s+\operatorname{div}(\rho_s\bu_s)=0,\qquad \partial_tb+ \pounds_{\bu_s}b=0\,.
\end{array}\right.
\end{equation}
In the case of the free slip boundary condition \eqref{free_slip}, the variational principle yields boundary conditions similar to \eqref{BC_general}
\begin{equation}\label{BC_general_2} 
\left[(\sigma _p + \overline{s}) \cdot \bn \right] \cdot \boldsymbol{\eta} =0\,,\quad \text{for all $ \boldsymbol{\eta} $ parallel to $ \partial \mathcal{B} $}\,,
\end{equation}
and the same definition of the stress tensor $ \sigma _p $ in \eqref{def_sigma_p}. 
When the boundary conditions \eqref{no_slip} are used, no additional boundary condition arise from the variational principle.
Thus, physically, the muscle stress $\bar{s}$ is simply added to the effective stress $\sigma_p$ in equations \eqref{expressions_explicit_muscle} and the boundary conditions \eqref{BC_general_2}.

\begin{remark}[On the free flow of fluid through the boundary]
\label{remark:free_bnd}
{\rm 
Suppose the elastic solid is fixed in space in a domain $ \mathcal{D}_s \subset \mathbb{R} ^3  $. If the fluid can leave or enter the domain, it is natural to assume that all fluid particles are included in a domain $\mathcal{D}_f$ that always contains the solid, $\mathcal{D}_s \subset \mathcal{D}_f$, for example, $ \mathcal{D}_f = \mathbb{R}^3$, \emph{i.e.} the fluid occupies the whole space. Then, the back-to-labels maps in that case are given by 
\begin{equation} 
\bX(t, \_\, ) : \mathcal{D}_s \rightarrow \mathcal{B} _s \quad\text{and}\quad \bY(t,\_\,): \mathcal{D}_s \rightarrow \mathbb{R} ^3\,,
\label{back_to_labels_map_free_flow} 
\end{equation} 
where $\bX(t, x) \in \mathcal{B}_s$ is a diffeomorphism with $\mathcal{B}_s$ the set of all solid labels as before, and $\bY(t, x)$ is an embedding with $\mathcal{B} (t):=\bY(t, \mathcal{D}_s )$ the set of all fluid labels corresponding to fluid particles that are in the elastic solid at time $t$. Consequently, fluid labels in $ \mathbb{R} ^3-\mathcal{B} (t) $ correspond to fluid particles that are not in the elastic solid at time $t$. Such particle may, \emph{e.g.}, have already left the solid at time $t$ or will penetrate into the solid at a later time. We shall also note that the accurate representation of the moving material boundary needs to be treated carefully, see \cite{dell2009boundary}, as the boundary of the elastic matrix material  presents a singularity. We shall postpone the quite complex and technical discussion of the free fluid outflow of the boundary to a follow up work.}
\end{remark} 

%\todo{VP: It is quite technical, I think - I would rather not get into these technical details. Moreover, accurate representations of boundary conditions in the physical case can depend on how exactly the boundary was obtained. For example, one can imagine that at the boundary, all the pores are pointing normal to the boundary, vs having random direction, or are restricted somehow compared to the pores inside the volume \emph{etc.}. One can imagine these things happen, for example, if a plastic matrix was cut with a hot blade, which would certainly change the structures of the pores at the boundary.  }

Under the assumptions leading to the concentration dependence $c=c_0/\sqrt{\operatorname{\det} b}$, equations \eqref{expressions_explicit_muscle} simplify further to 
\begin{equation}
\label{expressions_explicit_simple} 
\hspace{-3mm} 
\left\{
\begin{array}{l}
\displaystyle\vspace{0.2cm}\bar\rho_f^0(\partial_t \bu_f+ \bu_f\cdot\nabla \bu_f )  = - \nabla  p + \frac{1}{g} \bF_f\\
\displaystyle\vspace{0.2cm}\rho_s (\partial_t \bu_s \!+\! \bu_s\cdot\nabla \bu_s) =g\nabla p   
  + \operatorname{div} \left( \sigma_e  + \bar{s} \right) \!+\! \bF_s\\
\displaystyle\vspace{0.2cm} \frac{\partial V}{\partial v}=  pc(b),\qquad g= c(b)v\\
\vspace{0.2cm}\partial_tg + \operatorname{div}(g\bu_f)=0,\qquad \partial_t\rho_s+\operatorname{div}(\rho_s\bu_s)=0,\qquad \partial_tb+ \pounds_{\bu_s}b=0\,,
\end{array}\right.
\end{equation}
where $ \sigma _e$ is the elastic stress
\begin{equation}\label{sigma_e} 
\sigma _e = 2 \frac{\partial V}{\partial b} \cdot b + V{\rm Id}\,.
\end{equation} 
This formula follows from the equality $\nabla \left( V\! - \!\frac{\partial V}{\partial v}v\right) + \operatorname{div} ( \sigma_p)= \operatorname{div} \sigma _e$ which holds when $c=c_0/\sqrt{\operatorname{\det} b}$ since in this case $ \frac{\partial c}{\partial b} \cdot b = - \frac{1}{2} c{\rm Id}$.
 We can further put $\bF_f = - \bF_s = \mathbb{K} (\bu_s-\bu_f)$ to introduce the friction according to the Darcy law. 

For the developments below, we also need to introduce the equation for the back-to-labels maps. Recall that the Lagrangian label of the solid is denoted $\bX$ and we use the same notation for the back-to-labels map defined by $ \bX(t,\bx) = \bPsi ^{-1} (t,\bx)$ where $\bPsi(t, \bX)$ denotes the configuration map for the solid. Recall also that the Eulerian velocity of the solid is given by
\begin{equation} 
\bu_s   =  \partial _t {\bpsi} \circ  \bpsi^{-1} \quad\text{i.e.}\quad \bu_s (t,\bx) =  \partial _t {\bpsi}\big(t, \bpsi^{-1}(t,\bx)\big)\,.
\label{solid_vel} 
\end{equation} 
Differentiating the identity $\bpsi \circ \bX = {\rm Id}$, or, in coordinates, $\bpsi(t, \bX(t, \bx)) = \bx$, with respect to time, we obtain: 
\begin{equation} 
\nabla \bpsi\circ \bX \cdot \partial _t \bX+ \bu_s = \mathbf{0} 
\quad 
\Leftrightarrow 
\quad 
\partial _t \bX   = - \left(  \nabla \bpsi \circ \bX \right)^{-1} \cdot \bu_s 
\label{evolution_X_0} 
\end{equation} 
which can be written as 
\begin{equation} 
\label{evolution_X} 
\partial _t \bX   = - \nabla \bX \cdot  \bu_s \quad \Leftrightarrow \quad 
\bu_s = -  \left( \nabla \bX \right)^{-1} \cdot \partial _t \bX\, . 
\end{equation} 
The strain tensor $b$ and, correspondingly, stress tensor $\sigma_e$ depend on the spatial gradients of that map, \emph{i.e.},   $\nabla \bX(t,\bx)$, as well as possibly $\bx$ and $v$, since $V$ depends on both $b$ and $v$. Thus, the second equation of \eqref{expressions_explicit_simple} becomes a set of coupled elliptic equations for $\nabla \bX$ and $v$. On the other hand, the muscle action depends only on $\bX$ and $t$. 

\subsection{Simplification: one-dimensional dynamics} 
Let us now compute the equations \eqref{expressions_explicit_simple} for one dimensional reduction. The physical model is as follows: suppose there is an elastic porous media filled with an incompressible fluid, positioned inside a perfectly slippery one-dimensional pipe with a circular cross-section.  The elastic media has a muscle inside which can contract and expand. 
Since the space occupied by the muscle is fixed in space, its back-to-labels map is a diffeomorphism of the form $X(t,\_\,): [-L,L] \rightarrow [-L,L]$, $X=X(t,x)$. The fluid can leave or enter the muscle hence its back-to-labels map is an embedding $Y(t, \_\,):[-L,L] \rightarrow \mathbb{R} $, $Y=Y(t,x)$. 

Equation \eqref{evolution_X} for solid and fluid become simply 
\begin{equation}
X_t(t,x) = - X_x(t,x) u_s (t,x) \, , \quad
Y_t(t,x) = - Y_x(t,x) u_f (t,x).
\label{evolution_XY_1D}
\end{equation}
Using the assumption of the Darcy law of friction in one dimension, we derive the one-dimensional version of \eqref{expressions_explicit_simple}  as 
\begin{equation}
\label{expressions_explicit_1D_simplified} 
\left\{
\begin{array}{l}
\displaystyle\vspace{0.2cm}\bar\rho_f^0(\partial_t u_f+ u_f \partial_x u_f )  = -  \partial_x p + \frac{1}{g}  K(u_s - u_f)\\
\displaystyle\vspace{0.2cm}\rho_s (\partial_t u_s + u_s \partial_x u_s) =g\partial_x p   
  + \partial_x \left( \sigma_e  + \bar{s} \right) \!+ K(u_f-u_s) \\
\displaystyle\vspace{0.2cm} \frac{\partial V}{\partial v}=  pc(b),\qquad g= c(b)v\\
\vspace{0.2cm}\partial_tg + \partial_x (g u_f)=0,\qquad \partial_t\rho_s+\partial_x (\rho_s u_s)=0,\qquad \partial_tb+ \pounds_{u_s}b=0\,.
\end{array}\right.
\end{equation}
In one dimension, the Finger tensor becomes simply
\begin{equation}\label{Finger_1_D} 
b(t,x)=(X_x(t,x))^{-2}.
\end{equation}
The densities $g$ and $ \rho  _s$ with reference values $ g^0(X)$ and $ \rho  _s^0(X)$ are expressed as
\begin{equation}\label{rho_g_1_D} 
\rho  _s(t,x) = \rho  _s^0( X(t,x)) X_x(t,x) \quad\text{and}\quad g(t,x) =g^0( X(t,x)) X_x(t,x)\,.
\end{equation} 
The third equation in \eqref{expressions_explicit_1D_simplified} expresses $p$ as an explicit function of $v$ and $b$ hence from  \eqref{Finger_1_D} the pressure becomes a function $p=p(X_x,v)$.
Since $v=g/c(b)$ from the fourth equation in \eqref{expressions_explicit_1D_simplified}, we obtain explicitly the pressure as a function $p= p(X_x,g)$. In one dimension, the elastic stress tensor \eqref{sigma_e} is
\begin{equation}\label{sigma_e_1_D} 
\sigma _ e = 2 \frac{\partial V}{\partial b}(b,v) b + V(b,v).
\end{equation} 
From  \eqref{Finger_1_D} and  $v=g/c(b)$, we get an explicit expression
\[
\sigma _e= \sigma _e( X_x, g).
\]

\subsection{A reduction of the equation of motion for Lagrangian variables}

We now show how to reduce the system \eqref{expressions_explicit_1D_simplified} to two coupled PDEs for the two variables describing the Lagrangian coordinates of the solid and fluid. \revision{R3Q1}{This exact reduction of the system to one dimension is novel to this paper and was not undertaken in our previous paper \cite{FaFGBPu2020}.} Suppose that at $t=0$, the state is non-deformed so $X(0,x)=x$ and $\rho_s(0,x)=\rho_s^0$ for all $x$, where $\rho_s^0$ is the reference density of the porous media assumed to be a constant in space. Thus from the first equation in \eqref{rho_g_1_D}, we have 
\begin{equation} 
\rho_s(t,x)=\rho_s^0 X_x(t,x)\,. 
\label{rho_s_int} 
\end{equation}

Similarly, we also assume that at $t=0$, the fluid is undisturbed, $Y(0,x)=x$ and $g(0,x)=g_0$, for some constant $g_0$.
From the second equation in \eqref{rho_g_1_D}, we have 
\begin{equation} 
g(t,x)=g_0 Y_x(t,x)\,. 
\label{g_eq}
\end{equation} 
On the other hand, the velocities can also be expressed from \eqref{evolution_XY_1D} as 
\begin{equation} 
u_s = - \frac{X_t}{X_x} \, , \qquad u_f = - \frac{Y_t}{Y_x} \, . 
\label{uf_us_XY}
\end{equation} 
We also note that since the pressure depends on $X_x$ and $g$, from \eqref{g_eq} we have 
\begin{equation} 
p=p(X_x,g) = P(X_x,Y_x) 
\label{P_sol}
\end{equation}
a given function of $(X_x,Y_x)$.

Substitution of \eqref{uf_us_XY} together with \eqref{rho_s_int} and \eqref{g_eq} into the first two equations of 
\eqref{expressions_explicit_1D_simplified} gives a closed system of two coupled PDEs for the back-to-labels maps $X$ and $Y$: 
\begin{equation} 
\left\{ 
\begin{aligned} 
&\bar\rho_f ^0g_0   \left( -Y_{tt} + 2 \frac{Y_t Y_{tx}}{Y_x} - \frac{Y_t^2 Y_{xx}}{Y_x^2} \right) \\
& \hspace{1cm}= - g_0 Y_x \partial_x \big(P(X_x,Y_x)  
\big)+K \left( \frac{Y_t}{Y_x}
- \frac{X_t}{X_x} \right) 
\\
& \rho_s^0 \left( -X_{tt} + 2 \frac{X_t X_{tx}}{X_x}  - \frac{X_t^2 X_{xx}}{X_x^2} \right) \\ 
& \hspace{1cm}= g_0 Y_x \partial_x \big(P(X_x,Y_x)\big) +
\partial_x \big( \sigma_e (X_x, Y_x) + \overline{s} \big) +K \left( \frac{X_t}{X_x}
- \frac{Y_t}{Y_x} \right) \,.
\end{aligned} 
\right. 
\label{PDE_XY}
\end{equation} 
In addition to lowering the number of equations, in our opinion equation \eqref{PDE_XY} is easier to implement than its Eulerian version since in nature, one would expect the muscle stress to be a function of the Lagrangian variable of the solid $X$, \emph{i.e.}, a given muscle fiber, rather than the spatial coordinate $x$, so $\overline{s}=\overline{s}(X,t)$. It can be written as an explicit function of $x$ only for given solution $X(t,x)$, which is awkward from the point of view of numerical solution, as the last equation of \eqref{expressions_explicit_1D_simplified} describing evolution of $X$ has to be explicitly computed at every time step. In the formulation \eqref{PDE_XY}, where the equation is formulated in terms of back-to-labels maps $(X,Y)$, specifying $\overline{s}=\overline{s}(X,t)$ to be a given function of $X$ does not present any fundamental problem for implementing the numerical solution. 

\begin{remark}[Conservation of total momentum]
\label{rem:cons_mom}
{\rm 
The total momentum of the fluid is $M_f=\int _{-L}^L\bar\rho_f^0 g u_f \mbox{d} x= -\int_{-L}^L \bar\rho_f^0 g_0 Y_t \mbox{d} x$, and the total momentum of the solid is $M_s=\int _{-L}^L \rho_s  u_s \mbox{d} x= - \int _{-L}^L \rho_s X_t \mbox{d} x$. Then, the total fluid+solid momentum
\begin{equation} 
M=\int _{-L}^L \left(  \bar\rho_f^0 g u_f +\rho_s  u_s \right)  \mbox{d} x = - \int _{-L}^L \left(  \bar\rho_f^0 g_0  Y_t +\rho_s^0  X_t \right)  \mbox{d} x
\label{fluid_solid_momentum} 
\end{equation} 
is conserved for periodic boundary conditions. Indeed, 
\begin{equation} 
\begin{aligned} 
\dot M & =\int _{-L}^L \left(  - 2 \rho  _s^0 \frac{X_t X_{tx}}{X_x} +  \rho  _s^0\frac{X_t^2X_{xx}}{X_x^2} 
\right. 
\\ 
& \left. \qquad - 2 \bar\rho_f^0  g _0 \frac{Y_t Y_{tx}}{Y_x} + \bar\rho_f^0 g _0\frac{Y_t^2Y_{xx}}{Y_x^2}+ 
\partial_x \left( \sigma_e + \overline{s} \right)  \right) \mbox{d} x 
\\ 
& = \int _{-L}^L \partial_x \left( - \rho  _s^0 \frac{X_t^2}{X_x}  -   \bar\rho_f^0 g _0\frac{Y_t^2}{Y_x} +
 \sigma_e + \overline{s} \right) \mbox{d} x 
 \\
 & = \int _{-L}^L \partial_x \left( - \rho  _s u_s ^2  - \bar\rho_f^0 gu_f ^2  +
 \sigma_e + \overline{s} \right) \mbox{d} x\\
&=  \big[ - \rho  _s u_s ^2  -  \bar\rho_f^0gu_f ^2  +
 \sigma_e + \overline{s} \big] _{-L}^L 
\, . 
\end{aligned} 
\label{cons_momentum} 
\end{equation} 
The boundary terms vanish for periodic boundary conditions. For other cases, for example, fixed boundary conditions for the solid, $u_s(\pm L,t)=0$, the conservation of momentum depends on the cancellation of the fluid momentum through the boundary, elastic stress and muscle stress contributions in the boundary term of \eqref{cons_momentum}.
}
\end{remark}

\subsection{Numerical solution of equations (\ref{PDE_XY})}\label{sec_Numerics}

\paragraph{Choice of potential.} For the numerical solution, we postulate the following choice of potential energy:
\begin{equation} 
V (X_x,g)=\frac12 \alpha \left(X_x - 1\right)^{2} + \frac12 \beta \big(g + \left(1 - g_{0}\right) X_x - 1\big)^{2}\,. \label{_sp_potential_0}
\end{equation} 
The terms have the following physical sense: the expression proportional to $\alpha$ corresponds to the linear elasticity term (Hooke's law), the second term, proportional to $\beta$, is the difference of $g$ and the ``totally-incompressible porosity", a quantity, that would be equal to the porosity in the case if the solid was totally incompressible. The potential energy ``penalizes" changes in microscopic volume of the solid. 

\begin{remark}
{\rm Note that \eqref{_sp_potential_0} is the physical description of the potential as a function of $X_x$ and $g$, consistent with our description $V=V(b,v)$ in general case. For initially uniform system, using $g=g_0 Y_x$ we can express $V$ as
\begin{equation}\label{V_Y_x} 
V(X_x,Y_x)=\frac12 \alpha \left(X_x - 1\right)^{2} + \frac12 \beta \big(g_{0} Y_x + \left(1 - g_{0}\right) X_x - 1\big)^{2}
\end{equation} 
in terms of $X_x$ and $Y_x$, the latter related to the state of fluid. However, we do not assume that the potential energy of the solid is dependent on the state of the fluid here, it is an inference based on the properties of the solution. Thus, the potential energy defined by \eqref{_sp_potential_0} depends only on the properties of the solid, consistent with our description. This is in contrast, for example, with the works \cite{sciarra2008variational,madeo2008variational} where the energy of the porous media is dependent on the states of fluid and solid. The expression \eqref{V_Y_x} provides a formal connection between two approaches, as the solid energy \emph{formally} depends on the state of the fluid after the substitution $g=g_0 Y_x$, even though the physics of two approaches is quite different. }
\end{remark}

In order to compute the associated stress $ \sigma _0$ and pressure $p$, we need to express $V$ as a function $V(b,v)$.
Recalling that in one dimension, $b=X_x^{-2}$, and using
\[
g= c(b)v= \frac{c_0}{\sqrt{b}}v 
\]
we can rewrite \eqref{_sp_potential_0} as  
\begin{equation}  V(b,v) = \frac{\alpha}{2} \left( \frac{1}{\sqrt{b}} - 1\right)^{2} + \frac{\beta}{2} \left(\frac{c_{0} v}{\sqrt{b}} + \frac{1 - g_{0}}{\sqrt{b}} -1\right)^{2}.
\label{_sp_potential_1}
\end{equation} 
Now we are able to compute the stress terms from \eqref{sigma_e_1_D} as follows 
\begin{equation}\label{_sp_sigma_e_0}
\begin{aligned}\sigma_e& = 2 b \frac{\partial}{\partial b} V{\left(b, v \right)} + V{\left(b, v \right)}\\
& = \left(1 - \frac{1}{\sqrt{b}}\right) - \frac{\beta }{\sqrt{b}}   \left({c_{0} v} + 1 - g_{0}\right) \left(-1 + \frac{c_{0} v}{\sqrt{b}} + \frac{1 - g_{0}}{\sqrt{b}}\right) +V(b,v)\\
&=\frac{\alpha}{2} \left(1 - \frac{1}{b}\right) - \frac{\beta}{2b} \left(- b + (c_0v + (1 - g_0))^2\right)
\end{aligned}
\end{equation} 
and the pressure is found as 
\begin{equation}\label{_sp_pressure_0}
pc(b) = \pp{V}{v} = \frac{\beta}{\sqrt{b}} c_{0} \left(-1 + \frac{c_{0} v}{\sqrt{b}} + \frac{1 - g_{0}}{\sqrt{b}}\right).
\end{equation}
Substituting $b = X_x ^{-2}$ and $ c_0v = g\sqrt{b} =  g_0 Y_x X_x ^{-1} $ in \eqref{_sp_sigma_e_0} and \eqref{_sp_pressure_0} yields  $ \sigma _e$ and $p$ as functions of $X_x$ and $Y_x$ as
\begin{align}
\sigma_e(X_x, Y_x)& = - \frac{\alpha}{2} (X_x^2-1) - \frac{\beta}{2} \left( ( X_x (1-g_0) + Y_x g_0)^2-1  \right) \label{_sp_sigma_e_2}\\
p(X_x, Y_x) &= \beta g_{0} \left(g_{0} Y_x + \left(1 - g_{0}\right) X_x- 1\right). \label{_sp_pressure_1} 
\end{align}
We will use also the formula
\[
\partial_{x} \sigma_e =  - \alpha X_x X_{xx} 
 - \beta \left(g_{0} Y_x + (1-g_{0})X_x \right) \left(g_{0} Y_{xx} + (1-g_{0}) X_{xx}\right).
\]

\medskip

\noindent 
For simulations, it is useful to define $ \xi(t,x)$ and $ \phi (t,x)$ such that
\begin{equation} 
X(t, x) = x + \xi(t, x), \qquad Y(t, x) = x+ \phi(t, x)\, . 
\label{xi_phi_def}
\end{equation} 
Then $V$ is a quadratic, positive definite function of $(\xi_x,\phi_x)$, as expected: 
\begin{equation} 
V (\xi_x,\phi_x)= \frac12 \alpha \xi_x^{2} + \frac12 \beta \big(g_{0} \phi_x + \left(1 - g_{0}\right) \xi_x \big)^{2}\, .
\label{_sp_potential_phi_xi}
\end{equation}

\medskip 
\noindent 
For the temporally and spatially bound muscle stress, we take 
\begin{equation}\label{s_eq_sims} 
\overline{s}(t, X) = S_0 e^{-t/T-X^2/W^2},
\end{equation} 
where $S_0$, $T$ and $W$ are the given parameters of amplitude, time scale and width of applied muscle stress.

\revision{R1Q3}{\paragraph{Order of variables and non-dimensionalization.} 
The consideration above is valid for the case when the quantities of interest are either dimensional or non-dimensionalized. In what follows, we consider all units to be dimensionless. The Lagrangian function has the dimension of energy per volume, \emph{i.e.}, pressure. For simplicity, we rescale the Lagrangian function by a typical value of the elastic modulus. Then, $\alpha$ and $\beta$ in \eqref{_sp_potential_phi_xi} are of order $1$. The typical velocities can be scaled, for example, by the typical value of the speed of sound  in the dry media $c \sim \sqrt{E/\rho_s}$. The time is then scaled by $T = L/c$, where $L$ is the typical value of the length of interest, for example, typical size of the muscle. We can also assume that $\rho_f \sim \rho_s$ since the cells in biological materials mostly consist of water and are also filled with water-based fluid.  The value of the total momentum $M$ defined by \eqref{fluid_solid_momentum} is then expressed in terms of units $\rho_s c L$. For example, for water-based sponges $\rho_f \simeq \rho_s \simeq 1000$kg/m$^3$ with $E \sim 1$~kPa, $c \sim 1$ m/s. If $L \sim 0.1 $m then the one-dimensional total momentum is expressed in units of 100 kg$\cdot$m$^2$/s. Additional unit of length comes from the fact that the momentum is integrated over the length of the material. 
}
\color{black}
\paragraph{Discretization.} Let us consider the system \eqref{PDE_XY} on a finite space interval $x \in [-L,L]$.  
We discretize the system by considering the $N+2$ discrete data points $\left\{ x_0, \ldots x_{N+1} \right\} $, with $x_0=-L$ and $x_{N+1}=L$. We assume, for simplicity, a uniform discretization step $h=x_{i+1}-x_i$.

There are several types of boundary conditions: fixed, free or periodic. Let us for simplicity assume that the total extent of the $x$-domain occupied by the system is fixed, due to implemented boundary conditions holding the elastic material in a fixed place in space and not preventing the escape of fluid. Hence regarding the discretization of $X(t, x)$ we assume $X(t, -L)=X_1(t)=-L$ and $X(t,L)=X_{N+1}(t)=L$. Since $X_0$ and $X_{N+1} $ don't have any dynamics, we only consider the dynamics of $X_1, X_2,\ldots X_{N}$. Then, if the outflow of fluid from the boundaries is blocked, then we will have $Y(t, -L)=-L$ and $Y(t, L)=L$. If there is a fluid outflow from the boundaries, the nature of the boundary conditions will depend on many factors, \emph{i.e.} the need to overcome external pressure of the outside fluid, and the exact nature of the outflow. Thus, even for the fixed boundary conditions for the solid, the setting of boundary conditions for the fluid is non-trivial. The simplest ones are periodic boundary conditions, where $\xi=X-x$ and $\phi=Y-x$ are periodic with period $2 L$. We shall thus use periodic boundary conditions in our simulations.

For periodic boundary conditions, $(X_x,X_t,Y_x,Y_t)$ and their spatial derivatives are periodic function in $x$ with the period $2 L$. The forward $\Delta_i^f$ and backward $\Delta_i^b$ derivatives of any periodic function $F$ for the periodic boundary conditions is 
\begin{equation} 
\!\!\!\left\{ 
\begin{aligned} 
\Delta_i^f F & = \frac{F_{i+1}-F_i }{h}  , \;\; 1 \leq i<N, \;\;  \Delta^f_N F =\Delta^f_1 F
\\
\Delta_i^b F & = \frac{F_{i}-F_{i-1}}{h}  ,  \;\;  1 <i\leq N,  \;\; \Delta^b_1 F = \Delta^b_N F \, . 
\end{aligned} 
\right. 
\label{derivs_XY}
\end{equation} 
Then, we can approximate the first derivative by $\Delta_i^0=(\Delta_i^f+\Delta_i^b)/2$ and second derivative by $\Delta_i^2=\Delta_i^f *\Delta_i^b$.

\rem{ %%%BEGIN REM 
In terms of $\xi(t, x)=X(t, x)-x$ and $\phi(t,x)=Y(t, x)-x$, see \eqref{xi_phi_def}, the boundary conditions  \eqref{derivs_XY}  are homogeneous. Alternatively, we can use periodic boundary conditions for $\xi(t,x) $ and $\phi(t,x)$ on the interval $-L<x<L$. 

and given by: 
\begin{equation} 
\left\{ 
\begin{aligned} 
\Delta_i^f \xi & = \frac{1}{h} \left(\xi_{i+1}-\xi \right)\, , \quad ( 1 \leq i<N), \quad  \Delta^f_N \xi =- \frac{1}{h}  \xi   \quad \mbox{since $\xi_{N+1}=0$} 
\\
\Delta_i^b \xi & = \frac{1}{h} \left( \xi_{i}-\xi_{i-1} \right)\, ,  \quad ( 1 <i\leq N),  \quad \Delta^b_0 \xi = - \frac{1}{h}  \xi  \quad \mbox{since $\xi_0=0$} 
\\
\Delta_i^f \phi & = \frac{1}{h} \left( \phi_{i+1}-\phi \right)\, , \quad ( 1 \leq i<N), \quad  \Delta^f_N \phi = 0 \quad \mbox{since $\partial_x \phi (x=x_N) =0$} 
\\
\Delta_i^b \phi & = \frac{1}{h} \left( \phi_{i}-\phi_{i-1} \right)\, , \quad ( 1 <i\leq N), \quad  \Delta^b_0 \phi = 0 \quad \mbox{since $\partial_x \phi (x=x_1) =0$ } \,.
\end{aligned} 
\right. 
\label{derivs_XY_homogeneous}
\end{equation} 
} %%%END REM 

\begin{figure}[htbp]
\centering 
\captionsetup{width=.8\linewidth}
\includegraphics[width=0.95 \textwidth]{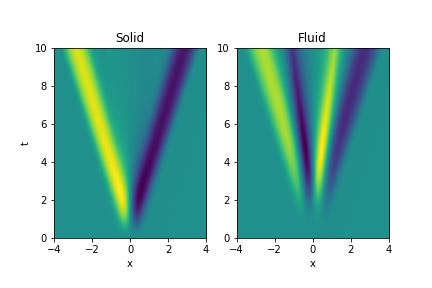}
\caption{
\label{fig:solution} 
\small Example of numerical solution of \eqref{PDE_XY}. Left panel:  solution for $\xi (t,x)=X(t,x)-x$ (solid), right panel: $ \eta (t,x) = Y(t,x)-x$ (fluid). For simulations presented here, the parameters for the stress in  \eqref{s_eq_sims} are $S_0=0.1$, $W=1$, $T=1$. The material parameters are $g_0=0.5$, $K=1$ and $\alpha=1$, $\beta=1$ in \eqref{_sp_potential_0}. 
}
\end{figure}

The results for numerical solution for $(\xi,\phi)$ with periodic boundary conditions  are presented on Fig.~\ref{fig:solution}. An initial disturbance caused by the muscle action on the matrix in the center of the elastic body is propagating along the matrix, both for fluid and for elastic material, although the shape of wave propagation is different. \revision{R3Q1}{Note that the numerical solutions were also not presented in \cite{FaFGBPu2020}, as that paper was focused on the propagation of sound waves in porous media as a particular application. }

\subsection{Self-propulsion by periodic motion of the stress}\label{sec:self_propulsion}

\begin{figure}[htbp]
\centering 
\captionsetup{width=.8\linewidth}
\includegraphics[width=0.95 \textwidth]{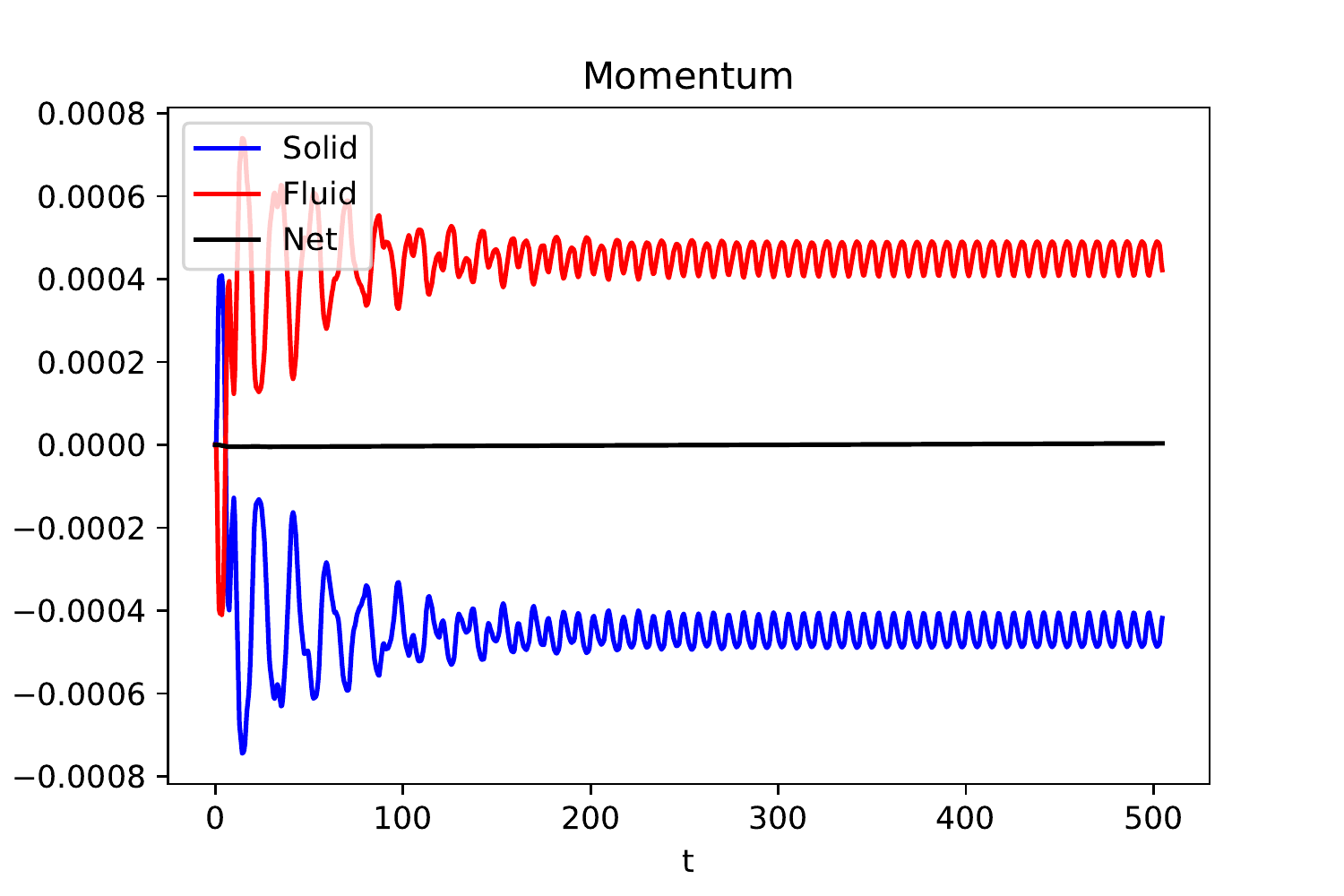}
\caption{
\label{fig:momenta} 
\small Momenta for solid $M_s = - \rho_s^0   \int _{-L}^LX_t \mbox{d} x $ (blue line) and fluid
$M_f = -  \bar\rho_f^0 g_0 \int _{-L}^L Y_t \mbox{d} x $ (red line) and total momentum $M=M_f+M_s$ (black line) for a given numerical solution with zero initial conditions and prescribed traveling muscle force given by \eqref{prescribed_S} with parameters $S_0=1$, $W=1$ and $U=1$. The total momentum $M_s+M_f$ is close to $0$ with expected accuracy throughout the simulation, starting at $10^{-6}$ and increasing to a fraction of $10^{-5}$ during the computation presented. 
}
\end{figure}

Let us now consider periodic boundary conditions, $\xi(t, x+2L)=\xi(t,x)$ and $\phi(t,x+2 L) = \phi(t,x)$, with $\xi(t,x)=X(t,x)-x$ and $\phi(t,x)=Y(t,x)-x$, and also consider the case when there is a periodic motion of the muscle's stress along the porous media. More precisely, let us consider the prescribed motion of the muscle stress in the form 
\begin{equation} 
\begin{aligned} 
&\overline{s}(t,X) = S_0 e^{ - (X-U t)_{\rm per}^2/W^2 } \, , 
\\
& (X-U t)_{\rm per}:=\left( X-U t \, {\rm mod} \, 2 L \right)-L\,.
\end{aligned} 
\label{prescribed_S}
\end{equation}
One can also express \eqref{prescribed_S} by saying $(X-Ut)_{\rm per}$ is a periodic function of $X$ with the same period as $X$, with the values contained in the interval between $-L$ and $L$. It is interesting to see if such a motion of the stress along the porous body can create self-propulsion of the solid. Of course, due to the conservation of momentum given in Remark~\ref{rem:cons_mom}, equation \eqref{cons_momentum}, it is not possible to accelerate both the fluid and solid in the same direction. However, it is possible to have opposite, and non-zero net momenta of solid and fluid, as shown on Fig.~\ref{fig:momenta}. One can see that the amplitude of the net momentum is quite small. 

\revision{R1Q5}{
It is important to note that the persistent oscillations in each of the momentum observed for large times on Fig.~\ref{fig:momenta} (as well as Fig.~\ref{fig:momenta_incompressible} below) are not numerical artifacts, but are due to the periodic motion of the muscle action along the media as described in \eqref{prescribed_S}. However, the equations did appear somewhat stiff, likely because of the several time scales present in the media. We used the package \emph{LSODA} from Pythons' \texttt{scipy.integrate} package, implementing Adams/backward differentiation formula (BDF) method with automatic stiffness detection and switching.
}
\color{black}

One could conjecture that the system is converging to a traveling wave solution, with small persistent oscillations about a steady state as illustrated on Fig.~\ref{fig:momenta}. Presumably, for biological applications, organisms would optimize the efficiency of motion and not smoothness. Thus, while considerations of traveling wave solutions are certainly possible, we will skip them here as they have, in our opinion, limited value for applications.

\section{Equations for the case when both the fluid and the solid are incompressible}
\label{sec:doubly_incompr}
\subsection{Physical justification and derivation of equations}\label{Variational_1}

From the physical point of view we may notice that for many biological materials the bulk modulus $K$ has the same order of magnitude or sometimes higher than the bulk modulus of water (2.2 GPa). The physics of this effective incompressibility can be understood from the fact that the elastic matrix consists of cells which are en large composed of incompressible water. Thus, the porous matrix can be effectively treated as incompressible elastic material. Physically,  if we select an arbitrary region in porous media filled with fluid and 'lock' the fluid inside the porous matrix and won't let it escape, the volume of such region shall not change under the assumption of total incompressibility. We can express the incompressibility of the solid as follows
\begin{equation}
\label{total_incompress}
1 - g(t, \bx) = (1 - g_0)\big(\bPsi^{-1}(t,\bx)\big)   J_{\bPsi^{-1}}(t,\bx)\,.
\end{equation}
Notice the similarity with the incompressibility of fluid given by \eqref{cons_law_fluid}. One way to include the incompressibility of the solid given in \eqref{total_incompress} is by adding an extra term in the action enforcing this condition with a Lagrange multiplier. There is although a simpler way to reach the answer. We differentiate \eqref{total_incompress} with respect to time to get
\begin{equation}
\label{total_g_cons}
    \partial_t (1-g) + \div\big((1-g)\bu_s\big) = 0\,.
\end{equation}
 (compare with \eqref{g_cons} and \eqref{rho_s_cons}) which can be written as 
\begin{equation} 
\operatorname{div} ( g \bu_f + (1-g) \bu_s ) =0 \, . 
\label{g_uf_us_constr} 
\end{equation} 
Since the constraint on the velocities is holonomic, we can also infer the following relationship between the variations $\boldeta_s$ and $\boldeta_f$ 
\begin{equation} 
\operatorname{div} \big( g \boldeta_f + (1-g) \boldeta_s \big) =0 \, . 
\label{eta_uf_us_constr} 
\end{equation} 
We introduce a Lagrange multiplier $\mu$ for \eqref{eta_uf_us_constr} and add it to the action $S$ in \eqref{action_p} as 
\begin{equation}\label{action_add}  
\begin{aligned} 
S_I &= S - \int_0^T\!\!\int_ \mathcal{B}   \mu \operatorname{div} ( g \boldeta_f + (1-g) \boldeta_s ) \mbox{d}^3 \bx\, \mbox{d}t \\
&= S+  \int_0^T\!\!\int_ \mathcal{B} \left( g \nabla \mu \cdot \boldeta_f + (1-g) \nabla \mu \cdot \boldeta_s\right)  \mbox{d}^3 \bx\,\mbox{d}t \, . 
\end{aligned} 
\end{equation} 
The Lagrange-d'Alembert principle  with friction forces applied to $S_I$ reads, similarly to \eqref{Crit_action_muscle_2},
\begin{equation} 
\de S_I+  \int_0^T\!\!\int_ \mathcal{B} \left(   \bF_f \cdot \boldeta_f +  \bF_s - \bar{s} : \nabla  \boldeta_s \cdot \boldeta_s \right)  \mbox{d}^3\bx \, \mbox{d} t =0 
\label{Crit_action_muscle_incompressible} 
\end{equation} 
and yields the system
\rem{ 
\todo{TF: We should also notice, that upon imposing \eqref{total_incompress}, we no longer have pore volume as an independent variable. So do we need to exclude it from the equations? At least dependence $V=V(b,v)$ makes no sense in such configuration and probably should be replaced with $V = V(b).$ Perhaps we need to use fluid incompressibility constraint in form \eqref{cons_law_fluid} and not introduce $c(b)$ and $v$ into the Lagrangian? Then all terms involving $v$ will be excluded from final equations, as well as the extra equation for the pore volume. \\ 
VP: You can either exclude pore volume explicitly, in which case there is no need for another Lagrange multiplier. Or, alternatively, we can keep it and have another Lagrange multiplier. Because the constraint of incompressibility is holonomic, these approaches are equivalent. They would not be if the constraint were non-holonomic. }
After collecting terms with corresponding variations, we end up with the following equations of motion. Notice the extra constraint, in comparison with \eqref{eq_gen}. The form of constraint below is the sum of the fluid and the solid incompressibility constraints. 
}
\begin{equation} 
\label{eq_gen_incomress} 
\!\!\!\!\!\!\left\{
\begin{array}{l}
\displaystyle\vspace{0.2cm}\partial_t\frac{\delta \ell}{\delta \bu_f}+ \pounds_{\bu_f} \frac{\delta \ell}{\delta \bu_f} = g \nabla \left( \frac{\delta  {\ell}}{\delta g}-  (p  + \mu)\right)+\bF_f\\
\displaystyle\vspace{0.2cm}\partial_t\frac{\delta\ell}{\delta \bu_s}+ \pounds_{\bu_s} \frac{\delta\ell}{\delta \bu_s} = -(1-g)\nabla \mu + \rho_s\nabla \frac{\delta\ell}{\delta \rho_s}   + \left(\frac{\displaystyle \delta\ell}{\displaystyle \delta b}+ p v\frac{\displaystyle \partial c}{\displaystyle \partial b}\right)\diamond b+
\operatorname{div}  \overline{s} +\bF_s\\
\displaystyle\vspace{0.2cm} \frac{\delta\ell}{\delta v}= -  pc(b)\,,\qquad g= c(b)v\\
\vspace{0.2cm}\partial_tg + \operatorname{div}(g\bu_f)=0\,,\qquad\partial_t\rho_s+\operatorname{div}(\rho_s\bu_s)=0\,,\qquad \partial_tb+ \pounds_{\bu_s}b=0\\
\partial_t(1-g)+\operatorname{div}((1-g)\bu_s)=0\,\Rightarrow\quad\operatorname{div}(g\bu_f+(1-g)\bu_s)=0\,.
\end{array}\right.
\end{equation}
\revision{R1Q4}{In the equation \eqref{eq_gen_incomress}, the incompressibility conditions for fluid and solid are enforced by the Lagrange multipliers $p$ and $\mu$ respectively having the physical meaning of pressures in the fluid and solid. In addition, the last equation of that system describes an additional conservation law due to double incompressibility.   } 
Using the physical Lagrangian \eqref{Lagr_def}, the expanded form of totally incompressible equations of motion becomes
\begin{equation}
\label{expressions_explicit_doubly_incompress} 
\!\!\!\!\!\!\left\{
\begin{array}{l}
\displaystyle\vspace{0.2cm} \bar\rho_f^0(\partial_t \bu_f+ \bu_f\cdot\nabla \bu_f )  = - \nabla (p + \mu) + \frac{1}{g} \bF_f\\
\displaystyle\vspace{0.2cm}\rho_s (\partial_t \bu_s \!+\! \bu_s\cdot\nabla \bu_s) =g\nabla p - (1-g)\nabla\mu +  \nabla \left( V\! - \!\frac{  \partial V}{ \partial v}v\right) +   \operatorname{div}  (\sigma _p + \overline{s}) + \bF_s\\
\displaystyle\vspace{0.2cm} \frac{\displaystyle \partial V}{ \partial v}=  pc(b),\qquad g= c(b)v\\
\vspace{0.2cm}\partial_tg + \operatorname{div}(g\bu_f)=0,\qquad \partial_t\rho_s+\operatorname{div}(\rho_s\bu_s)=0,\qquad \partial_tb+ \pounds_{\bu_s}b=0\\
\vspace{0.2cm}\operatorname{div}(g\bu_f+(1-g)\bu_s)=0\,.
\end{array}\right.
\end{equation} 
As before, under the assumptions leading to the concentration dependence $c=c_0/\sqrt{\operatorname{\det} b}$ given by \eqref{c_b_particular_neq}, the solid momentum equation in \eqref{expressions_explicit_doubly_incompress} simplifies further to 
\begin{equation}\label{momentum_solid_ii} 
\rho_s (\partial_t \bu_s \!+\! \bu_s\cdot\nabla \bu_s) =g\nabla p   - (1-g)\nabla\mu+ \operatorname{div} \left( \sigma_e  + \bar{s} \right)+ \bF_s
\end{equation} 
with $ \sigma _e$ the elastic stress given in \eqref{sigma_e}.

Note that in the above equations, the pressure, while being a Lagrange multiplier for the fluid incompressibility condition, is actually specified due to the $\delta v$ condition, \emph{i.e.}, the third equation of \eqref{expressions_explicit_doubly_incompress}, exactly as in \eqref{expressions_explicit} and \eqref{expressions_explicit_muscle}. In contrast, $\mu$, the Lagrange multiplier for the incompressibility of elastic matrix, does not have an explicit expression and must be found so the solution satisfies the last equation of \eqref{expressions_explicit_doubly_incompress}. 
The equation for $\mu$ can be derived explicitly by taking a time derivative of the last equation of \eqref{expressions_explicit_doubly_incompress}. For simplicity, let us rewrite the equations for velocities $\mathbf{u}_f$ and $\mathbf{u}_s$ as 
\begin{equation} 
\partial_t \mathbf{u}_f = \mathbf{R}_f - \frac{1}{\rho_f} \nabla \mu \, , \quad 
\partial_t \mathbf{u}_s = \mathbf{R}_s - \frac{1-g}{\rho_s} \nabla \mu \, , 
\label{uf_us}
\end{equation} 
where $\mathbf{R}_{f,s}$ are the right-hand sides of the fluid and solid equations excluding the $\mu$-term, which depend on the variables $(\bu_f,\bu_s,b,g,\rho_s)$ but not on their time derivatives.   Differentiating the last equation of \eqref{expressions_explicit_doubly_incompress} with respect to time, we obtain 
\begin{equation}  
\!\!\!\!\!\operatorname{div} \left[ \left(  \frac{g}{\bar\rho_f^0}+ \frac{(1-g)^2}{\rho_s} \right) \nabla \mu \right]
=\operatorname{div} \big(-  \operatorname{div} ( g \mathbf{u}_f ) (\mathbf{u}_f-\mathbf{u}_s ) 
+g \mathbf{R}_f +(1-g) \mathbf{R}_s \big)  
\label{eq_incompress_constraint} 
\end{equation} 
which is an elliptic equation for $\mu$, reminding of the regular pressure equation in a fluid. 
It is also useful to interpret the equation for the fluid pressure $p$ in the doubly incompressible media. When interpreting the physical nature of the potential energy $V=V(b,v)$, one notices that if the solid is incompressible as well, then $v$ is no longer a free variable, but has a dynamics slaved to that of $b$ which, in the simplest case of uniform initial conditions, is written as $v=v(b)$, so $V(b,v) = V(b,v(b))=W(b)$. Thus, the equation for the fluid pressure seemingly would give $p=0$ since $W(b)$ does not depend on $v$. That conclusion, however, would be incorrect. One has to \emph{first} write the expression for the potential energy in terms of the microscopic volume $v$, and \emph{only then} connect $v$ to the Finger tensor $b$ after taking the derivative in the pressure equation. Thus, in general, the pressure in the fluid is not going to vanish. While this approach requires careful consideration, in our opinion, it does have merit since it is easier to compute $V(b,v)$ from general principles and then substitute $v=v(b)$. If one insists on using the expression for potential energy $W=W(b)=V(b,v(b))$ then one needs to accurately compute the derivatives of $W(b)$ as a complex function of $b$, leading to the same terms as in \eqref{expressions_explicit_doubly_incompress}. 

\revision{R3Q1}{The incompressibility of both the fluid and the solid is another novel development as compared to \cite{FaFGBPu2020}. As in the previous section, we will continue the investigation incorporating the internal muscle stress using the modified Lagrange-d'Alembert principle.  }

We shall further note that equations \eqref{expressions_explicit_doubly_incompress}, while correct, are somewhat difficult to interpret physically because of the presence of two pressures, $p$ and $\mu$ being the Lagrange multipliers for incompressibility of fluid and solid, respectively. With these two pressures, the interpretation of Terzaghi's principle of equating pressures within the matrix and the fluid becomes non-apparent. In Appendix~\ref{app:approach2}, we derive an alternative formulation of the equations of motion, based on thermodynamics, elucidating the nature of the two pressures and connection between them, in compressible and incompressible cases for both fluid and solid.

\revision{R1Q5}{
\begin{remark}[On incompressibility and stiffness of equations]
{\rm It is worth noting that taking the approach of strictly incompressible solid is advantageous when taking the limit of the solid to be progressively closer to being incompressible, \emph{i.e.} the bulk modulus of the material going to infinity. This could be understood on a simple example of a pendulum with a rod that is either fully rigid or very close to being rigid. A pendulum with a rigid constraint leads to a familiar equation for the angle -- the pendulum equation -- that is certainly not stiff. However, taking the limit of a softer pendulum going to an infinitely rigid limit will indeed lead to stiff equations, as we would be trying to describe the vanishingly small and increasingly fast motions of an almost rigid rod. 
}
\end{remark}
}
\color{black} 
\subsection{Reduction for 1D motion} 

In what follows, we shall proceed with further simplification of equation \eqref{expressions_explicit_doubly_incompress}, see also \eqref{eqs_incompr_incompr_final}, to one dimension and its subsequent numerical analysis.
We follow the derivation of \eqref{PDE_XY} applied now to the doubly incompressible system \eqref{expressions_explicit_doubly_incompress}, with the solid momentum equation given in \eqref{momentum_solid_ii}, which leads to 
\begin{equation} 
\left\{ 
\begin{aligned} 
& \bar\rho ^0_f g_0   \left( -Y_{tt} + 2 \frac{Y_t Y_{tx}}{Y_x} - \frac{Y_t^2 Y_{xx}}{Y_x^2} \right)\\
& \hspace{2.5cm} = - g_0 Y_x  \partial_x (p+\mu)
+K \left( \frac{Y_t}{Y_x}
- \frac{X_t}{X_x} \right) 
\\
& \rho_s^0 \left( -X_{tt} + 2 \frac{X_t X_{tx}}{X_x}  - \frac{X_t^2 X_{xx}}{X_x^2} \right) 
=  g_0 Y_x \partial_x p - (1-g_0 Y_x) \partial_x \mu \\ 
& \hspace{2.5cm} 
+\partial_x \left( \sigma_e (X_x,Y_x) + \overline{s} \right) +K \left( \frac{X_t}{X_x}
- \frac{Y_t}{Y_x} \right) 
\\
& \partial _x\left( g_0 Y_t + (1-g_0 Y_x) \frac{X_t}{X_x} \right) =0, \quad  p = \frac{1}{c(b)}\pp{V}{v} \, , \quad b=X_x^{-2} 
\end{aligned} 
\right. 
\label{PDE_XY_2_incompressible}
\end{equation}
where, as before, $\sigma_e:=2   \pp{V}b b+ V $ is the elastic stress tensor reduced to the 1-dimensional case.

%%%END REM 

The last equation of \eqref{PDE_XY_2_incompressible} follows from the last equation of \eqref{expressions_explicit_doubly_incompress}  since $g=g_0 Y_x$ by the incompressibility of fluid \eqref{cons_law_fluid} in one dimension. We can further use the reduction of solid incompressibility \eqref{total_incompress} to one dimension to get 
\begin{equation} 
1-g = 1-g_0 Y_x= (1-g_0) X_x \, , 
\label{eq_X_x} 
\end{equation} 
so the last equation of \eqref{incompressibility_cond_XY} reduces to 
\begin{equation} 
\partial _x \left(  (1-g_0) X_t + g_0 Y_t  \right) \quad \Rightarrow \quad (1-g_0) X_t + g_0 Y_t = C(t) ,
\label{incompr_cond_simplified} 
\end{equation} 
where the integration 'constant' in the right hand side can depend on time.

One can see that the net momentum defined as
\[
M:=\int_{-L}^L  \left( \bar\rho_f^0 g u_f +\rho_s  u_s\right)  \mbox{d} x = - \int_{-L} ^L \left( \bar\rho^0_f g_0  Y_t +\rho_s^0  X_t \right) \mbox{d} x
\] is still conserved: 
\begin{equation} 
\begin{aligned} 
\dot M&= -\int_{-L}^L \left( \rho_f g_0 Y_{tt} + \rho_s^0 X_{tt} \right) \mbox{d} x\\
&= \int_{-L}^L \partial _x  \left( - \rho  _s^0 \frac{X_t^2}{X_x}  -   \bar\rho_f^0 g _0\frac{Y_t^2}{Y_x} - p +
 \sigma_e + \overline{s} \right) \mbox{d}x
 \\
 & = 
 \left. \left( 
 - \rho  _s^0 \frac{X_t^2}{X_x}  -   \bar\rho_f^0 g _0\frac{Y_t^2}{Y_x} - p +
 \sigma_e + \overline{s}  \right) \, 
 \right|_{-L}^L \, , 
 \end{aligned}
\label{cons_tot_mom}
\end{equation} 
provided the boundary conditions are periodic, or chosen in such a way that the boundary terms in \eqref{cons_tot_mom} vanish.

Using the initial conditions for the Lagrangian variables $X(t=0,x)=x$ and $Y(t=0,x)=x$, we obtain a connection between the Lagrangian variables \emph{for all} $x$ and time 
\begin{equation} 
(1-g_0) X(t,x)+ g_0 Y(t,x) = D(t) + x \, , \quad D(t) = \int_0^t C(s) \mbox{d} s \, . 
\label{Lagr_var_connect} 
\end{equation} 
This connection between the Lagrangian variables in porous media is, in our opinion, quite unexpected. 

To proceed, we also need to compute the pressure $p$ which can be done from the double incompressibility condition, \emph{i.e.}, the last equation of 
\eqref{PDE_XY_2_incompressible}. We rewrite the first two equations of that system as 
\begin{equation} 
X_{tt}=R_X + \frac{1-g_0Y_x}{\rho_s^0} \partial_x \mu \, , \quad 
Y_{tt}=R_Y +  \frac{Y_x}{\bar\rho_f^0} \partial_x \mu\, . 
\label{eq_XY_short}
\end{equation} 
where $R_X$ and $R_Y$ are the right-hand sides of the corresponding equations \eqref{PDE_XY_2_incompressible}
without the $p$ terms. 
Differentiating equation  \eqref{incompr_cond_simplified} with respect to time gives 
\begin{equation} 
 \left[ \frac{Y_x g_0}{\rho_f} + (1-g_0) \frac{(1-g_0 Y_x) }{\rho_s^0}\right] \mu_x 
= - \left( g_0 R_Y + (1-g_0) R_X \right) +  C'(t) \, ,
\label{incompressibility_cond_XY}
\end{equation} 
from which we express $\mu_x$ as
\begin{equation}
\begin{aligned}
&  \mu_x(X_t, X_x, X_{xt}, Y_x, Y_{xt}, R_X, R_Y, C'(t); g_0, \rho_s^0, \rho_f) = 
-{\mathcal A} + C'(t) { \mathcal B} \, \quad \mbox{with} 
\\ & {\mathcal A} :=
\frac{ g_0 R_Y + (1-g_0) R_X  }{ \frac{Y_x g_0}{\rho_f} + (1-g_0) \frac{(1-g_0 Y_x) }{\rho_s^0} }, 
\quad  { \mathcal B}:=\frac{ 1  }{ \frac{Y_x g_0}{\rho_f} + (1-g_0) \frac{(1-g_0 Y_x) }{\rho_s^0} }  \, . 
%\qquad + C'(t) X_{x} \left( \frac{\left(- X_{x} Y_{x} g_{0} \rho^{0}_{s} + \rho_{f} \left(- Y_{x}^{2} g_{0}^{2} + 2 Y_{x} g_{0} - 1\right)\right)}{\rho_{f} \rho^{0}_{s}} \right)^{-1} 
\end{aligned}
\label{incompressibility_mu}
\end{equation} 
Note that \eqref{incompressibility_mu} is the one-dimensional analogue of \eqref{eq_incompress_constraint} which also uses \eqref{incompr_cond_simplified} for the definition of $C(t)$.

The solution for $\mu_x$ computed from the condition \eqref{incompressibility_cond_XY} can be put back into the first two equations of \eqref{PDE_XY_2_incompressible} to form a closed system in terms of $(X,Y,X_t,Y_t)$ and its spatial derivatives. The value of $C'(t)$ is computed in such a way that the mean value of $\mu_x$ is zero for periodic boundary conditions, which gives: 
\begin{equation} 
C'(t)= \frac{\int_{-L}^L{\mathcal A}  \mbox{d} x}{\int_{-L}^L{\mathcal B}  \mbox{d} x} \,.
\label{mu_periodic} 
\end{equation} 
This adjustment is necessary since we have implicitly assumed that all functions, including the Lagrange multipliers, are periodic and thus all their derivatives have zero mean. The modification \eqref{mu_periodic} is not necessary for simulations on the line.

\begin{figure}[htbp]
\centering 
\captionsetup{width=.8\linewidth}
\includegraphics[width=0.7 \textwidth]{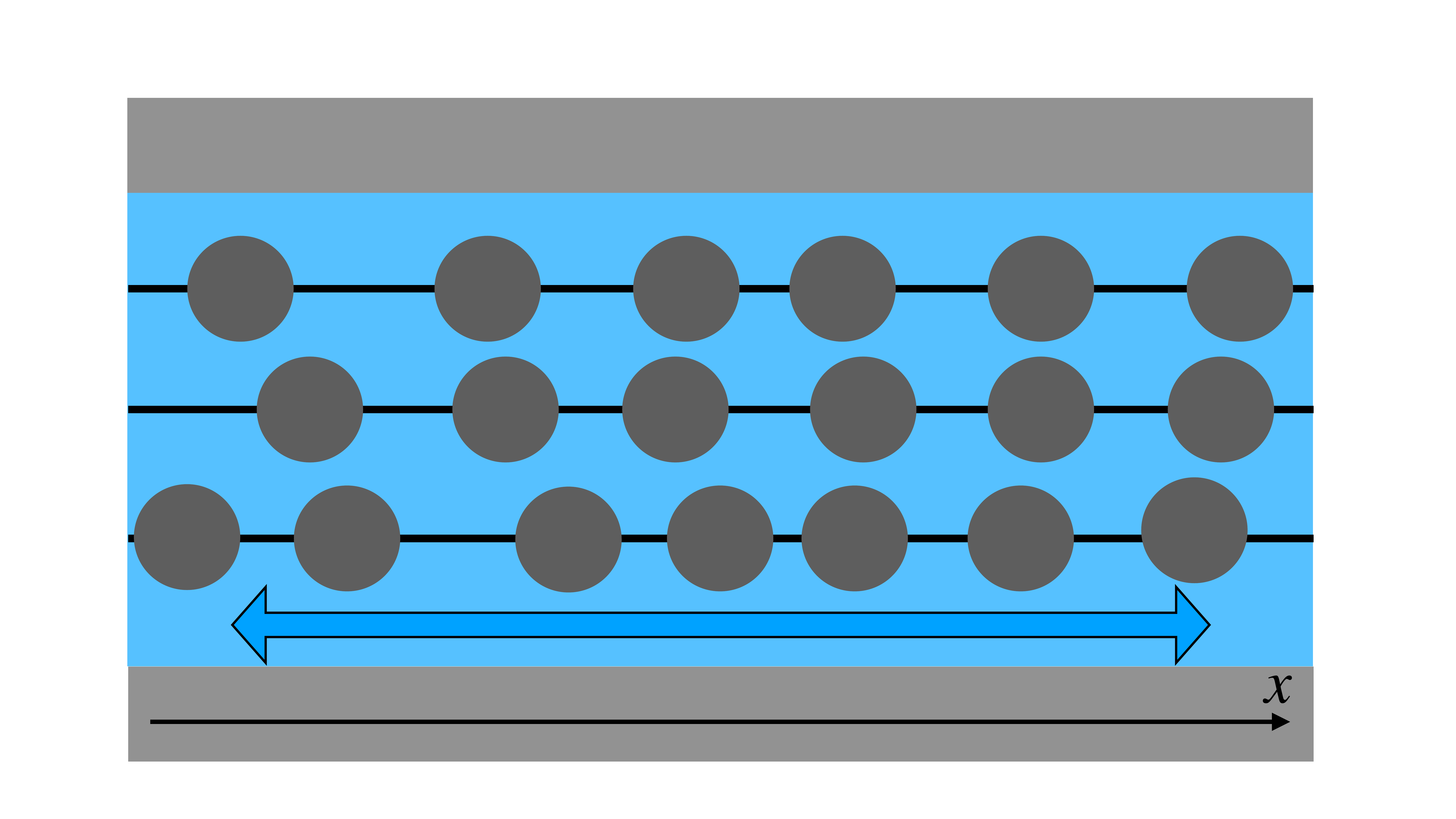}
\caption{
\label{fig:2_incompr_1d} 
\small A sketch of the physical realization of a one dimensional, doubly incompressible case, justifying the potential \eqref{pot_incompressible}. A channel is filled with incompressible fluid and holds solid particles on elastic strings running parallel to the axis of the channel (the $x$-axis). The space available to the fluid is dependent on the density of solid particles in a given interval $[x, x+ \mbox{d} x]$. The potential $V$ depends on the local deformations of the strings caused by the motion of the particles,  the elastic properties of the strings and the number of the strings running through the channel. }
\end{figure}

To derive the potential, we consider the following physical realization of one dimensional, doubly incompressible porous media. Consider a tube filled with an incompressible fluid, and suppose there are elastic muscle threads of negligible volume that are running along the axis of the tube. On each thread, there are rigid (and hence incompressible) beads attached to a given point on a particular thread, as illustrated on Fig.~\ref{fig:2_incompr_1d}. When the threads are stretched, the beads move inside the fluid and change the local volume of the fluid $g$ at the given Eulerian point. Then, the elastic energy is proportional to the deformation energy of the thread times the number of threads per given interval $x.$

%%%END REM 

\revision{R1Q4}{Based on the considerations above, we suggest to use the following potential: 
\begin{equation} 
V(b,v)  = \frac{\alpha}{2} (X_x-1)^2 = \frac{\alpha}{2} \left( \frac{1}{\sqrt{b}} -1 \right)^2 \, ,
\label{pot_incompressible} 
\end{equation} 
where in the physical realization presented in Fig.~\ref{fig:2_incompr_1d}, the constant $\alpha$ is dependent on the number of the springs for a cross-section of the tube and a typical elasticity of each spring. The potential can be viewed as the lowest power expansion in terms of the extension of each spring $X_x-1$, as the potential must be convex and smooth about the equilibrium $X_x=1$. Physically, this potential is simply the combined potential of multiple springs illustrated on Fig.~\ref{fig:2_incompr_1d}, expressed in terms of $X_x$ in the first equal sign, and then expressed in terms of $b$ in the second part of the equation \eqref{pot_incompressible}. 
Alternatively, one can view the potential \eqref{pot_incompressible} as a particular case of \eqref{_sp_potential_0} with $\beta=0$, since the second term in \eqref{_sp_potential_0} proportional to $\beta$ describes the elastic energy due to the deformation of the pores. } With the potential \eqref{pot_incompressible}, we obtain 
\begin{equation} 
 \sigma_e = 2 \pp{V}{b}b+V= -\frac{\alpha}{2}(X_x^2-1)\, , \quad p=\frac{1}{c}\pp{V}{v}=0\, . 
\label{p_sigma_incompressible} 
\end{equation} 
We now present the results of numerical solutions obtained for the potential \eqref{pot_incompressible}.

First, we present a doubly incompressible computation equivalent to the case of the compressible solid presented in Fig.~\ref{fig:solution}, for the same values of parameters except setting $\beta=0$ in the the potential given by \eqref{_sp_potential_0}, \emph{i.e.}, using the potential \eqref{pot_incompressible}. Notice that the motion of the solid and the fluid acquires  'jerkiness' in the double incompressible case, since the motion is less smooth than that illustrated on the Fig.~\ref{fig:solution}.
\begin{figure}[htbp]
\centering 
\captionsetup{width=.8\linewidth}
\includegraphics[width=0.95 \textwidth]{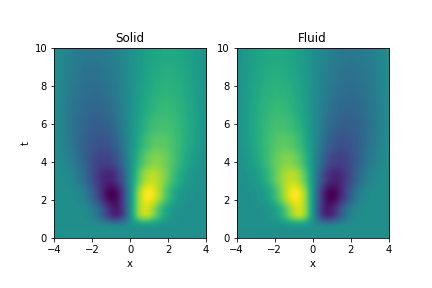}
\caption{
\label{fig:solution_incompressible} 
\small Example of numerical solution of \eqref{PDE_XY_2_incompressible}. Left panel:  solution for $X(t,x)-x$ (solid), right panel: $Y(t,x)-x$ (fluid). All parameters are as for solution presented in Fig.~\eqref{fig:solution} except for the extra incompressibility condition for the solid in \eqref{PDE_XY_2_incompressible} and change $\beta=0$ in potential \eqref{_sp_potential_0}, \emph{i.e.}, taking the the potential \eqref{pot_incompressible} with elastic stress $\sigma_e$ given by \eqref{p_sigma_incompressible}. 
}
\end{figure} 

\begin{figure}[htbp]
\centering 
\captionsetup{width=.8\linewidth}
\includegraphics[width=0.95 \textwidth]{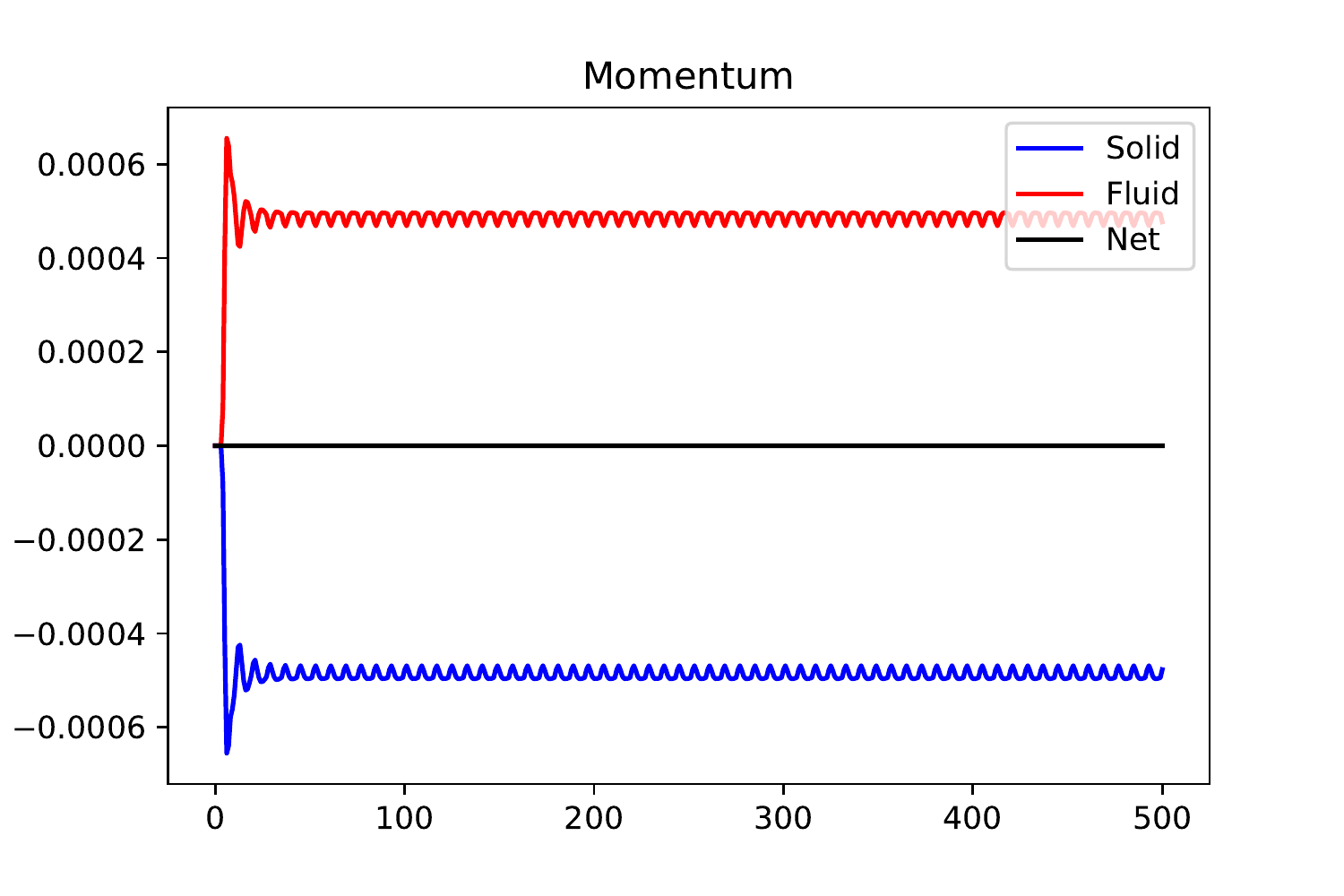}
\caption{
\label{fig:momenta_incompressible} 
\small Momenta for solid $M_s = - \rho_s^0   \int_{-L}^L X_t \mbox{d} x $ (blue line), 
$M_f = - \rho_f g_0 \int_{-L}^L Y_t \mbox{d} x $ (red line) and the total $M_f+M_s$ (black line) for a given numerical solution with zero initial conditions and prescribed traveling muscle force given by \eqref{prescribed_S} with parameters $S_0=0.1$, $W=1$ and $U=1$. As in Fig.~\ref{fig:momenta_incompressible}, the net momentum $M_s+M_f$ is close to $0$ with expected accuracy throughout the simulation. All other parameters for simulations are taken exactly the same as in Fig.~\ref{fig:momenta}, \emph{i.e.} $g_0=0.5$, $K=1$, with the same total time of simulation $t=500$. We note that compared to \eqref{fig:momenta}, the stabilization of motion occurs on a much faster time scale compared with the compressible case. 
}
\end{figure}

\begin{figure}[htbp]
\centering 
\captionsetup{width=.8\linewidth}
\includegraphics[width=0.95 \textwidth]{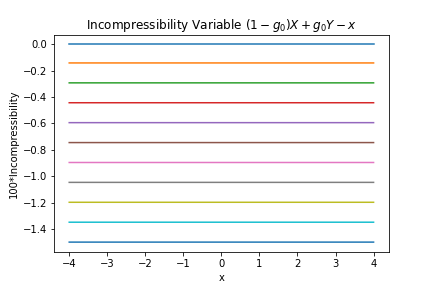}
\caption{
\label{fig:incompressibility} 
\small The value of $(1-g_0) X + g_0 Y-x$ computed for a set of values of $t$ in simulation output ($t=0,50,100,\ldots,500$), presented as a function of $x$. According to the  equation \eqref{incompr_cond_simplified}, that expression should be independent of $x$, although can vary with $t$, which is consistent with results presented here. All solutions $(1-g_0) X + g_0 Y-x$ presented as a function of $x$  are close to horizontal lines. The distance of these horizontal lines from the $x$-axis is dependent on time as is permitted by \eqref{incompr_cond_simplified}. 
}
\end{figure}

\begin{figure}[htbp]
\centering 
\captionsetup{width=.8\linewidth}
\includegraphics[width=0.95 \textwidth]{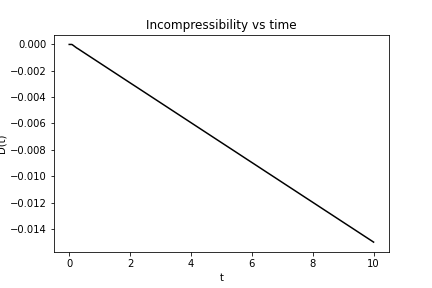}
\caption{
\label{fig:incompressibility_vs_time} 
\small The value of $D(t)$, computed as the mean value of $(1-g_0) X + g_0 Y -x $ (mean computed with respect to the $x$-variable), as a function of time $t$, for the results presented in 
Fig.~\ref{fig:incompressibility}. 
}
\end{figure}

Next, we show the self-propulsion due to the generation of momentum due to the traveling wave motion of the muscle stress as was discussed in Section~\ref{sec:self_propulsion} and further illustrated in \eqref{fig:momenta}. Fig.~\ref{fig:momenta_incompressible} shows the possibility of generating self-propulsion of the solid from rest due to non-zero momenta of the solid and fluid, while the net momentum of both solid and fluid is conserved and equal to $0$. 
Furthermore, on Fig.~\ref{fig:incompressibility}, we illustrate how well the incompressibility condition is satisfied. In other words, we plot the term $(1-g_0) X + g_0 Y -x$ as a function of $x$ for all available values of $t$, which, according to the last equation in \eqref{PDE_XY_2_incompressible}, must be equal to $D(t)$, which is a constant as a function of $x$, but can vary in time. 
Finally, on Fig.~\ref{fig:incompressibility_vs_time}, we present the variable $D(t)=(1-g_0) X + g_0 Y-x$, expected to be independent of $x$, as defined by the last equation of \eqref{PDE_XY_2_incompressible}.  Fig.~\ref{fig:incompressibility} confirms that this variable, taken as a function of $x$ for a fixed $t$, is indeed almost a constant within numerical accuracy. Thus, we compute $D(t)$ as the mean value of $(1-g_0) X + g_0 Y-x$.

%%%END REM 

\revision{EQ1\\R2Q4\\R3Q2}{\section{Conclusions}}

\noindent In this paper, we have demonstrated the following new results:
\medskip 

 \begin{enumerate} 
 \item Incompressibility of both the fluid and the matrix is introduced using a variational principle with Lagrange multiplier. The advantage of this method is the consistent approach to both phases (fluid and solid) and automatic coupling between microscopic and macroscopic variables. In addition, the balance of pressures on the fluid-solid interface is incorporated automatically by the variation with respect to the fluid volume fraction, or an equivalent variable. 
 \item Our theory is derived in spatial (Eulerian) coordinates which makes it more appropriate for such applications as wave propagation in porous media, in contrast to the theories based on Lagrangian coordinates. 
 \item Our theory is valid for arbitrary Lagrangians and arbitrary deformations of the media, and can incorporate the case of incompressible fluid and either  compressible or incompressible solid. 
 \item An additional advantage of the variational method developed here is the automatic derivation of boundary conditions for the media in the no-stress and fixed cases. 
 \item We state a novel (as far as we are aware) analogue of the Lagrange-d'Alembert method for external forces to incorporate the internal stress caused by the muscle, for the case of biological materials. 
 \item We demonstrate that the equations of motion can be reduced exactly to one dimension and perform analytical and numerical studies of the resulting model, in the case of incompressible fluid and when the solid is either compressible or incompressible.   
 \item Using numerical simulations, we show the possibility of self-propulsion of the porous matrix (solid) while preserving the net-zero momentum of the fluid and solid.
\end{enumerate}

\revision{EQ1//R2Q3//R2Q4}{There are several avenues for further work based on the ideas developed in our paper.
\\
\paragraph{Biological applications.} In the future, it will be interesting to combine this work to include additional biologically relevant problems. For example, one could combine the variational methods introduced here with the previous work by the authors on the geometric variational approach to elastic tubes conveying fluid \cite{gay2014exact,gay2015flexible,gay2018stability}. Making the tube's wall porous will be relevant to other engineering \cite{heshmati2019vibration} and biological applications like arterial flow \cite{bukac2015effects}. 
\\
\paragraph{Analysis of PDEs.} Due to the variational approach used here, our equations  dissipate energy in the absence of muscle work. The dissipation of energy naturally gives the bounds on the solution in the phase space. While the focus of this paper is not on the rigorous mathematical analysis, it seems that the existence of solutions in the appropriate functional space can be proven quite easily. Physically, we expect the solution to be unique if the Lagrangian has appropriate convexity properties. This is just a conjecture, however, which will need to be investigated more closely. 
\\
\paragraph{Boundary conditions.} For a successful implementation of variational methods for fluid-structure interaction problems, and, in particular, for understanding the outflow from the porous media, the boundary conditions for moving boundaries need to be considered in more detail, as outlined in Remark~\ref{remark:free_bnd}. Boundary conditions for porous media using variational methods have been considered before \cite{dell2009boundary,serpieri2016general}, and using geometric variational methods in the context of purely elastic media in \cite{GBMaRa12}. Thus, further investigations of boundary conditions for incompressible fluid and solid using variational methods presented here are certainly of interest. 
\\
\paragraph{Variational approach to thermodynamics of porous media.} It is possible to incorporate irreversible processes and thermodynamics in variational methods  \cite{gay2017variational,gay2017lagrangian,gay2017lagrangian2,GBYo2018,GBYo2019}. We have recently submitted a paper \cite{FGBPu2021} outlining the thermodynamics of porous media with heat exchange between fluid and solid. Based on the second law of thermodynamics, we have derived the equations of motion incorporating heat exchange and also derived the general form of the friction terms based on the second law of thermodynamics. Further work in the variational analysis  of porous media with thermodynamics also appears to be a promising direction of future research. 
\\
\paragraph{Discrete variational methods for accurate long-term computations of porous media.} It is worth noting that our numerical scheme does not take into account the variational structure of equations. For the time scales we have computed the solutions for, $t \sim 10^3$, no apparent loss of accuracy was observed in simulations. However, the computation of much longer times may need the use of variational integrators. Variational integrators conserve momenta-related quantities to machine precision and allow for accurate introduction of the friction to equations. In our case, since the friction is strictly internal to the system, variational methods can  be beneficial for ensuring the long-term preservation of momenta of combined fluid-solid system. It is especially important for computations of the motion in two or three dimensions, and in the case where the fluid viscosity, and, correspondingly, the friction forces, is relatively small. We refer the reader to \cite{WeMa1997,MaWe2001} for the general theory of variational integrators and to \cite{gay2016variational,FGBPu2020} for the derivations of variational integrators for fluid-structure interactions with incompressibility constraint, in particular, for the application to the motion of an elastic tube filled with an incompressible fluid. }

\color{black} 
%\todo{VP: Actually, we should be able to do it - porous garden hose! We need to assume something like the fact that the fluid leaking through the walls is disappearing, e.g. evaporating or is merged with the outside fluid if the whole thing is submerged in water. Any other ideas? }

\appendix

\section{Alternative derivation emphasizing the physical nature of pressure}\label{app:approach2}

In this Appendix, we give an alternative derivation which will allow elucidating the physical meaning behind the pressure-like terms coming from the Lagrange multipliers. We show that these multipliers are related to the physical pressure in the fluid. While the resulting equations are the same as derived above, we believe that this derivation is useful since it appeals to the physical meaning behind the quantities, and not just their formal mathematical origin. Moreover, the derivation presented in this Section will be useful for further studies of thermodynamic effects in porous media which will be undertaken in follow-up work. 
Let us start by introducing the actual densities of the material 
\begin{equation} 
\overline{\rho}_f:=\frac{\rho_f}{g} \, , \qquad 
\overline{\rho}_s:=\frac{\rho_s}{1-g} \, . 
\label{rho_bar_def}
\end{equation} 
These densities correspond to the physical density of the fluid or the matrix, taken by themselves, for example, water only, or the elastic matrix with all voids filled with the same material as the matrix itself. Next, the fluid and solid specific internal energy should depend only on the combinations $\overline{\rho}_f$ and $\overline{\rho}_s$, respectively, and not, for example, on the densities and $g$. In addition, the internal energy of the solid should also depend on the Finger deformation tensor $b.$ Therefore, we put the specific internal energies of the fluid and solid to be $E_f=E_f(\overline{\rho}_f)$ and $E_s=E_s(\overline{\rho}_s,b)$ and take the Lagrangian
\begin{equation} 
\begin{aligned}
\!\!\!\ell( \mathbf{u} _f , \mathbf{u} _s , \rho  _f , \rho  _s ,b, g) & =  \int_ \mathcal{B} \Big[ \frac{1}{2} \rho  _f | \mathbf{u} _f | ^2 +  \frac{1}{2} \rho  _s | \mathbf{u} _s | ^2  - \rho  _f E_f (\bar \rho_f)  -   \rho  _s E_s (\bar \rho_s,b) \Big] \mbox{d} ^3 \bx.
\end{aligned} 
\label{lagr_energy}
\end{equation} 
Recall that the specific internal energy is the energy density per mass. Since the mass of the solid is proportional to the local density, we have $\mbox{d} m_s = \rho_s \mbox{d} \bx$, so the energy density per volume is then $V_s=\rho_s E_s(\bar \rho_s,b)$, and not, for example, $V \neq \bar \rho_s E_s$. Similar consideration also applies to the fluid, with the energy density per volume is $V_f = \rho_f E_f(\bar \rho_f)$.

\paragraph{The case of fluid and matrix both being compressible. } Neglecting the thermal effects, this is the simplest case to consider. The action is defined as
\begin{equation} 
S= \int_0^T \ell( \mathbf{u} _f , \mathbf{u} _s , \rho  _f , \rho  _s ,b, g){\rm d}^3 \mathbf{x}\, {\rm d} t 
\label{action_compressible} 
\end{equation} 
and we use the Lagrange-d'Alembert principle 
\eqref{Crit_action_muscle_2}
\begin{equation} 
\de S+ \int_0^T\!\!\int_ \mathcal{B} \left(   \bF_f \cdot \boldeta_f +  \bF_s  \cdot \boldeta_s - \bar{s} : \nabla  \boldeta_s \right) \mbox{d}^3\bx \, \mbox{d} t =0 \, .\quad  
\label{Crit_action_compressible} 
\end{equation} 
Here the variation $\de g$ is arbitrary and the variations of the other variables $(\de \bu_f, \de \bu_s, \de \rho_s, \de \rho_f) $ are given by 
\begin{equation} 
\begin{aligned}
\de \bu_f &= \partial_t \boldeta_{f} + \bu_f  \cdot \nabla\boldeta_f - 
\boldeta_f \cdot  \nabla \bu_f \\ 
\de \bu_s &= \partial_t \boldeta_{s} + \bu_s  \cdot \nabla\boldeta_s - 
\boldeta_s \cdot  \nabla\bu_s \\
\delta\rho_s&= - \operatorname{div}(\rho_s\boldeta_s) \\ 
\delta\rho_f&= - \operatorname{div}(\rho_f\boldeta_f) \\ 
\delta b &= - \pounds_{\boldeta_s}b \,.
\label{g_rhos_b_var_2}
\end{aligned} 
\end{equation} 
The variation with respect to $g$ is particularly interesting and gives 
\begin{equation} 
\overline{\rho}_f^2 \pp{E_f}{\rho_f}- \overline{\rho}_s^2 \pp{E_s}{\rho_s}=0  \quad 
\Rightarrow \quad 
\overline{\rho}_f^2 \pp{E_f}{\rho_f}= \overline{\rho}_s^2 \pp{E_s}{\rho_s}=:P \, .
\label{var_g} 
\end{equation} 
Thus, from the variation with respect to $g$ we obtain that the thermodynamic pressure in the fluid and solid are equal. The equations of motion are thus:
\begin{equation} 
\!\!\!\!\!\!\!\!\left\{\!\!
\begin{array}{l}
\vspace{0.2cm}\displaystyle \rho_f (\partial_t \bu_f+ \bu_f\cdot \nabla \bu_f) = - g \nabla \left(\bar\rho_f^2 \frac{\partial E_f}{\partial\bar\rho_f}\right) + \bF_f 
\\
\vspace{0.2cm}\displaystyle\rho_s( \partial_t \bu_s+ \bu_s\cdot \nabla \bu_s) = - (1-g) \nabla \left(\bar\rho_s^2 \frac{\partial E_s}{\partial\bar\rho_s}\right) 
+  2\operatorname{div} \left( \rho_s\frac{\displaystyle \partial E_s}{\displaystyle \partial  b}\cdot b  \right) 
+ \operatorname{div}  \bar{s} + \bF_s \\
\vspace{0.2cm}\displaystyle\partial_t \rho_f+ \operatorname{div}(\rho_f \bu_f)=0,\qquad \partial_t \rho_s+ \operatorname{div}(\rho_s \bu_s)=0,\qquad\partial_t b+ \pounds_{\bu_s}b=0\\
\displaystyle  \bar\rho_f^2\frac{\partial E_f}{\partial\bar\rho_f} =  \bar\rho_s^2\frac{\partial E_s}{\partial\bar\rho_s}=:P\,.
\end{array}
\right.
\label{compressible_eqs} 
\end{equation} 
 The variable $g$ is determined in terms of the other variables from the last equality, which can be written as a condition $g=g( \rho  _f, \rho  _s,b )$.

\paragraph{The case of a compressible solid and incompressible fluid.} In this case, we need to augment the action principle by adding the fluid incompressibility condition with a Lagrange multiplier $p_f$. The Lagrange-d'Alembert principle now reads
\begin{equation} 
\begin{aligned} 
& \de  \int_0^T \left[ \ell( \mathbf{u} _f , \mathbf{u} _s , \rho  _f , \rho  _s , b,g) +\int_ \mathcal{B}  p_f \left( g - (g^0 \circ \varphi_f ^{-1})J_{\varphi_f^{-1}}\right)  {\rm d} ^3  \mathbf{x}\right] {\rm d} t\\
& \hspace{3cm} + \int_0^T\!\!\int_ \mathcal{B} \left(   \bF_f \cdot \boldeta_f +  \bF_s \cdot \boldeta_s  - \bar{s} : \nabla  \boldeta_s  \right) \mbox{d}^3\bx \, \mbox{d} t =0 \, ,
\end{aligned} 
\label{Crit_action_incompr_compr} 
\end{equation} 
with arbitrary variations $ \delta g$ and $\delta p_f$, the same Lagrangian \eqref{lagr_energy} as before and variations of other variables given by \eqref{g_rhos_b_var_2}.  This yields the following system: 
\begin{equation} 
\!\!\!\!\!\!\!\!\!\left\{\!\!
\begin{array}{l}
\vspace{0.2cm}\displaystyle \rho_f (\partial_t \bu_f+ \bu_f\cdot \nabla \bu_f) = - g \nabla \left(\bar\rho_f^2 \frac{\partial E_f}{\partial\bar\rho_f}\right) - g \nabla p _f +\bF_f\\
\vspace{0.2cm}\displaystyle\rho_s( \partial_t \bu_s+ \bu_s\cdot \nabla \bu_s) = - (1-g) \nabla \left(\bar\rho_s^2 \frac{\partial E_s}{\partial\bar\rho_s}\right) +  2\operatorname{div} \left(  \rho_s \frac{\displaystyle \partial E_s}{\displaystyle \partial  b}\cdot b \right) 
+ \operatorname{div}  \bar{s} + \bF_s \\
\vspace{0.2cm}\displaystyle\partial_t \rho_f+ \operatorname{div}(\rho_f \bu_f)=0,\qquad \partial_t \rho_s+ \operatorname{div}(\rho_s \bu_s)=0,\qquad\partial_t b+ \pounds_{\bu_s}b=0\\
\vspace{0.2cm}\displaystyle\partial_t g+ \operatorname{div}(g \bu_f)=0\\
\displaystyle  \bar\rho_f^2\frac{\partial E_f}{\partial\bar\rho_f} + p_f =  \bar\rho_s^2\frac{\partial E_s}{\partial\bar\rho_s}=:P\,.
\end{array}
\right.
\label{incompr_compr_eqs} 
\end{equation} 
Note the difference between the last equality in \eqref{incompr_compr_eqs} and that of system \eqref{compressible_eqs}. Note also that with the last equality, the first two equations of motion for fluid can be written as 
\begin{equation} 
\hspace{-2mm} 
\left\{
\begin{array}{l}
\vspace{0.2cm}\displaystyle \rho_f (\partial_t \bu_f+ \bu_f\cdot \nabla \bu_f) = - g \nabla P + \bF_f \\
\displaystyle\rho_s( \partial_t \bu_s+ \bu_s\cdot \nabla \bu_s) = - (1-g) \nabla P+  2\operatorname{div} \left(  \rho_s \frac{\partial E_s}{\partial  b}\cdot b \right) + \operatorname{div}  \bar{s} + \bF_s \,,
\end{array}
\right.
\label{fluid_eqs_motion} 
\end{equation} 
with $P= \bar\rho_s^2\frac{\partial E_s}{\partial\bar\rho_s}$.  Hence the first six equations in \eqref{incompr_compr_eqs}, in which the first two equations are rewritten as \eqref{fluid_eqs_motion} with $P= \bar\rho_s^2\frac{\partial E_s}{\partial\bar\rho_s}$,  can be solved for the six variables $ \bu _f$, $\bu _s$, $\rho  _f$, $\rho  _s$, $b$, $g$. Then $p_f$ is found from the last equality $p_f=   \bar\rho_s^2\frac{\partial E_s}{\partial\bar\rho_s}-\bar\rho_f^2\frac{\partial E_f}{\partial\bar\rho_f}$.

This shows that the internal energy of the fluid $ \rho  _f E_f(\bar \rho  _f)$ can be neglected in the Lagrangian without affecting the dynamics.
Furthermore, notice that from the $ \rho  _f $ and $g$ equations, we get $ \partial _t \bar  \rho  _f  + \bu_f \cdot \nabla  \bar\rho  _f =0$, so $\bar \rho  _f $ is a constant, if it is a constant $\bar\rho_f^0$ at the initial time $t=0$. This corresponds to the case of a homogeneous incompressible fluid.

As we will show below, the link with our variable $v$ describing internal volume, used in \eqref{Lagr_def}, is 
\begin{equation} 
\label{v_rho_link} 
 \kappa v= v_s - \bar v _s =\frac{1}{ \rho  _s } - \frac{1}{\bar \rho  _s } \, ,
\end{equation} 
where the constant $ \kappa $ is given by $ \kappa = c_0/ \rho  _s^0$.
Thus, the volume $v$ introduced in  Section~\ref{sec:3D_eqs} is proportional to the effective specific volume of solid $v_s$ minus the microscopic specific volume of solid $\bar v _s $. Note that $v_s$ and $\bar v_s$ defined in \eqref{v_rho_link} have the dimensions of inverse density, with the physical meaning of available free volume per mass.
%Technically speaking, that variable $v$ is not equal to the variable $v$ introduced in Section~\ref{sec:3D_eqs}, and are proportional with a coefficient $\alpha$ having dimensionality of inverse mass, $v_{\rm new} = \kappa v_{\rm old}$.  That coefficient $\kappa$ may be taken as $c_0/\rho^*$ where $c_0$ is the typical initial concentration of the pores, and $\rho^*$ is the typical value of the density (either solid or fluid).   
As we shall also explain below, in this description $ \rho  _s(b) $ takes place of the concentrations of pores $c(b)$ with the relation $c(b)=\kappa\rho_s(b)$.

\rem{ %%%BEGIN REM 
\todo{VP: I was getting a bit confused with formulas here, so I tried to straighten it out. Concentration $c(b)$ has dimensions $1/m^3$, and density is $kg/m^3$. In our original definition of $v$, it had dimensions of volume, so $c v$ is dimensionless, whereas in the new definition \eqref{V_E_relationship} it has dimensions of inverse density, so $\rho_s(b) v$ is dimensionless, but $c(b) v$ is not. I was thinking of changing the definition of $v$ in the discussion above with the dimensional coefficients, but that was pretty ugly. So I put the text above and hope there will be no confusion. I think both definitions of $v$ are important for different physical cases.  } 
} %%%END REM 

\paragraph{The case when both fluid and solid are incompressible.} In this case, we take the Lagrange-d'Alembert action principle to be enforcing both the incompressibility of the fluid and the solid using the Lagrange multipliers $p_f$ and $p_s$ as  
\begin{equation} 
\begin{aligned} 
& \de  \int_0^T \bigg[ \ell( \mathbf{u} _f , \mathbf{u} _s , \rho  _f , \rho  _s , b,g)  + \int_ \mathcal{B}  p_f \left( g - (g^0 \circ \varphi_f ^{-1})J_{\varphi_f^{-1}}\right) \mbox{d}^3 \bx\\
& \hspace{4cm} + \int_ \mathcal{B} p_s \left( (1-g) - ((1-g^0) \circ \varphi_s ^{-1})J_{\varphi_s^{-1}}\right)  {\rm d} ^3\mathbf{x} \\
& \hspace{4cm} + \int_ \mathcal{B} \left(   \bF_f \cdot \boldeta_f +  \bF_s \cdot   \boldeta_s- \bar{s} : \nabla  \boldeta_s  \right) \mbox{d}^3\bx \bigg]\mbox{d} t =0 \,,  
\end{aligned} 
\label{Crit_action_incompr_incompr} 
\end{equation} 
with arbitrary variations $ \delta g$,  $\delta p_f$, and $\delta p_s$, and the same Lagrangian \eqref{lagr_energy} as before. We get the following system of equations: 
\begin{equation}
\left\{
\begin{array}{l}
\vspace{0.2cm}\displaystyle \rho_f (\partial_t \bu_f+ \bu_f\cdot \nabla \bu_f) = - g \nabla \left(\bar\rho_f^2 \frac{\partial E_f}{\partial\bar\rho_f}\right) - g \nabla p _f + \bF_f \\
\vspace{0.2cm}\displaystyle\rho_s( \partial_t \bu_s+ \bu_s\cdot \nabla \bu_s) = - (1-g) \nabla \left(\bar\rho_s^2 \frac{\partial E_s}{\partial\bar\rho_s}\right)  - (1- g )\nabla p _s
\\
\vspace{0.2cm} \hspace{4cm} \displaystyle +  2\operatorname{div} \left(  \rho_s \frac{\displaystyle \partial E_s}{\displaystyle \partial  b}\cdot b \right) 
+ \operatorname{div}  \bar{s} + \bF_s 
\\
\vspace{0.2cm}\displaystyle\partial_t \rho_f+ \operatorname{div}(\rho_f \bu_f)=0,\qquad \partial_t \rho_s+ \operatorname{div}(\rho_s \bu_s)=0,\qquad\partial_t b+ \pounds_{\bu_s}b=0\\
\vspace{0.2cm}\displaystyle\partial_t g+ \operatorname{div}(g \bu_f)=0, \qquad \partial_t (1-g)+ \operatorname{div}((1-g) \bu_s)=0\\
\displaystyle  \bar\rho_f^2\frac{\partial E_f}{\partial\bar\rho_f} + p_f =  \bar\rho_s^2\frac{\partial E_s}{\partial\bar\rho_s}+p_s=:P\,.
\end{array}
\right.
\label{eq_incompr_incompr} 
\end{equation} 
The last equation, obtained from the variations $\delta g$,  defines the effective pressure $P$ expressed in terms of two Lagrange multipliers $(p_f,p_s)$ enforcing the incompressibility of fluid and solid, respectively.  We note that we can rewrite the system in an equivalent way as
\begin{equation}
\left\{
\begin{array}{l}
\vspace{0.2cm}\displaystyle \rho_f (\partial_t \bu_f+ \bu_f\cdot \nabla \bu_f) = - g \nabla P + \bF_f \\
\vspace{0.2cm}\displaystyle\rho_s( \partial_t \bu_s+ \bu_s\cdot \nabla \bu_s) = - (1-g) \nabla P\\ 
 \displaystyle \hspace{3.8cm} +  2\operatorname{div} \left( \rho_s \frac{\partial E_s}{\partial  b}\cdot b \right) + \operatorname{div}  \bar s+ \bF_s  
\\
\vspace{0.2cm}\displaystyle\partial_t \rho_f+ \operatorname{div}(\rho_f \bu_f)=0,\quad \partial_t \rho_s+ \operatorname{div}(\rho_s \bu_s)=0,\quad\partial_t b+ \pounds_{\bu_s}b=0\\
\vspace{0.2cm}\displaystyle\partial_t g+ \operatorname{div}(g \bu_f)=0\\
\vspace{0.2cm}\displaystyle \operatorname{div}( g \bu_f + (1-g) \bu_s)=0\\ 
\displaystyle  \bar\rho_f^2\frac{\partial E_f}{\partial\bar\rho_f} + p_f =  \bar\rho_s^2\frac{\partial E_s}{\partial\bar\rho_s}+p_s=P\,.
\end{array}
\right.
\label{eqs_incompr_incompr_final} 
\end{equation} 
The pressure $P$ appearing in the first two equations is computed from the elliptic equation deduced from taking the time derivative of the next to last equation of \eqref{eqs_incompr_incompr_final} which is the incompressibility constraint. The Lagrange multipliers $p_f$ and $p_s$ are then computed from the last equation. As earlier, the internal energy of the fluid can be neglected without changing the dynamics. From the $ \rho  _f $, $ \rho  _s$, $g$ and $g-1$ equations, we get $ \partial _t \bar  \rho  _f  + \bu_f \cdot \nabla  \bar\rho  _f =0$ and $ \partial _t \bar  \rho  _s  + \bu_s \cdot \nabla  \bar\rho  _s =0$, so $\bar \rho  _f $, $\bar \rho  _s$ are constant, if they are constant at the initial time $t=0$:
\begin{equation} 
\bar \rho_f = \bar\rho_f^0 = {\rm const}, \quad \bar \rho_s =\bar\rho_s^0 = {\rm const} \, . 
\label{bar_rho_const} 
\end{equation}
This corresponds to the case when both the incompressible fluid and solid are homogeneous.

%%%
\rem{
It is a very natural physical conclusion since these quantities describe the densities of 'pure' materials (fluid and solid) and are thus expected to be constant. Therefore, the derivatives with respect to the pressure\textcolor{magenta}{/densities?} in the last equation of \eqref{eq_incompr_incompr} can be neglected giving the balance of pressures: 
\begin{equation} 
p_f=p_s=p \, .
\label{pressure_equality} 
\end{equation} 
Thus, the equations of motion become: 
\begin{equation} 
\left\{
\begin{array}{l}
\vspace{0.2cm}\displaystyle \rho_f g (\partial_t \bu_f+ \bu_f\cdot \nabla \bu_f) = - g \nabla p + \bF_f \\
\vspace{0.2cm}\displaystyle\rho_s( \partial_t \bu_s+ \bu_s\cdot \nabla \bu_s) = - (1-g) \nabla p+  2\operatorname{div} \left(\rho_s \frac{\partial E_s}{\partial  b}\cdot b \right) + \operatorname{div}  \bar s+ \bF_s 
\\ 
\vspace{0.2cm}\displaystyle \partial_t b+ \pounds_{\bu_s}b=0
\\
\vspace{0.2cm}\displaystyle \operatorname{div}  ( g \bu_f + (1-g) \bu_s ) =0 
\end{array}
\right.
\label{eqs_incompr_incompr_final} 
\end{equation} 
}

\begin{remark}[On the mixture of $N$ incompressible materials]
{\rm The variational approach presented above generalizes to a mixture of $N$ incompressible or compressible fluids and solids, \emph{e.g.} if a single incompressible matrix is filled with $N-1$ incompressible fluids like oil and water, or incompressible water and compressible air. In that case, we  just consider $g_1,..., g_N$ with $g_1+ ... + g_N=1$ and as many constraints $p_k(...)$ as there are incompressible components.
}
\end{remark}

\paragraph{Relation with the previous variational derivation.} We shall now relate the variational treatment carried out in this paragraph with that of \S\ref{Introduction_muscle} and \S\ref{Variational_1} which uses $c(b)$ and $v$. We focus on the concentration dependence given in \eqref{c_b_particular_neq}. We start with the case of a compressible solid filled with an incompressible fluid and show that equations \eqref{expressions_explicit_simple} reduce to \eqref{incompr_compr_eqs} when $V$ and $E_s$ are related as  
\begin{equation} 
\label{V_E_relationship}
\begin{aligned} 
V(b,v) &= \rho  _s(b) E_s( \bar \rho  _s(b,v), b)\\
\bar \rho _s (b,v)&= \frac{ \rho  _s(b)}{1- \rho  _s(b) \kappa v} \, ,\quad \mbox{or} \quad \kappa v = \frac{1}{\rho_s}-\frac{1}{\bar \rho_s} \,,
\end{aligned} 
\end{equation} 
with $ \kappa = c_0/ \rho  _s^0$. Then $c(b)$ and $\rho_s(b)$ take the form 
\begin{equation} 
\label{c_rho_s_connection} 
c(b)=  \frac{ c^0}{\sqrt{ \operatorname{det}b}}, \qquad  
\rho  _s(b) =\frac{ \rho_s^0}{\sqrt{ \operatorname{det}b}}
\end{equation} 
hence they are related as
\begin{equation}\label{relation_c_rhos} 
c(b)= \kappa \rho  _s(b).
\end{equation} 
Additionally,  $ \rho  _f$ in \eqref{incompr_compr_eqs} defined by $ \rho  _f:= \bar \rho  _f^0 g$. 
As we have seen, we can assume $E_f=0$ without loss of generality.
Using the formulas
\[
\frac{\partial \rho  _s}{\partial b} \cdot b= - \frac{1}{2} \rho  _s   \quad\text{and}\quad  \frac{\partial \bar\rho  _s}{\partial b} \cdot b = - \frac{1}{2} \frac{\bar \rho  _s ^2 }{ \rho  _s } \, , 
\]
we have the relations
\begin{equation}\label{important_relations} 
\sigma _e= 2 \frac{\partial V}{\partial b} \cdot b + V {\rm Id} = - \bar \rho  _s ^2 \frac{\partial E_s}{\partial \bar \rho  _s }  {\rm Id} + 2 \rho  _s \frac{\partial E_s}{\partial b} \cdot b \quad\text{and}\quad \frac{\partial V}{\partial v} = \kappa \rho  _s \bar \rho  _s ^2 \frac{\partial E_s}{\partial \bar \rho  _s }\,.
\end{equation} 
From the second relation in \eqref{important_relations}, the third equation in \eqref{expressions_explicit_muscle} and \eqref{relation_c_rhos} yield $p= \rho  _s ^2 \frac{\partial E_s}{\partial \bar \rho  _s }$ and then using the first relation in \eqref{important_relations} and $ \rho  _f:= \bar \rho  _f^0 g$, we obtain that the two momentum equations in \eqref{expressions_explicit_simple} reduce to those of \eqref{incompr_compr_eqs}. The equations for $ \rho  _s $, $g$, and $b$ coincide while the equation for $ \rho  _f$ in \eqref{incompr_compr_eqs} follows from  the definition $ \rho  _f:= \bar \rho  _f^0 g$ and the equation for $g$. In particular, we have the relations $p=\rho  _s ^2 \frac{\partial E_s}{\partial \bar \rho  _s }=p_f=P$ between the pressures appearing in \eqref{expressions_explicit_simple} and \eqref{incompr_compr_eqs} (recall that $E_f=0$ here).  

The case when both the solid and fluid are incompressible, \emph{i.e.} the relation between equations \eqref{expressions_explicit_doubly_incompress} and  \eqref{eqs_incompr_incompr_final} is shown in a similar way. In this case the pressures are related as $p_f= P= p+ \mu $ and $ p_s= \mu $  (recall again that $E_f=0$ here).

\begin{remark}[On neglecting internal energies in the incompressible case] 
{\rm 
We have seen in the above calculation that in the incompressible case, the internal energy of the fluid can be neglected, \emph{i.e.} we can set $E_f=0$. One could therefore be tempted to conclude that in the case of an incompressible solid and fluid, the corresponding spatial derivatives $\pp{E_s}{\bar \rho_s}$ can be neglected as well. Setting 
\begin{equation} 
\pp{E_f(\bar \rho_f)}{\bar \rho_f}= 0 \, , \quad \pp{E_s(\bar \rho_s,b)}{\bar \rho_s} 
\stackrel{\displaystyle\text{?}}{=}  0 
\label{vanish_rho_deriv} 
\end{equation} 
in the last equation of \eqref{eqs_incompr_incompr_final} yields simply $p_f=p_s$ which gives, in the notation above, $\mu=P$
and $p=0$. That assumption, however, is  incorrect due to \eqref{important_relations}.  This is true even when the solid is incompressible, and $\bar \rho_s$ is an advected quantity, and thus can be a constant in space. However, setting the partial derivative of $E_s(\bar{\rho_s},b)$ with respect to $\bar \rho_s$ to zero in \eqref{vanish_rho_deriv} is incorrect, even in that case, since $E_s$ depends additionally on $b$. The derivatives $\pp{E_f}{\bar \rho_f}$ or $\pp{E_s}{\bar\rho_s}$ can be dropped only if there is no dependence of $E_f$ or $E_s$ on other quantities, like coordinates or, in our case, Finger's deformation tensor $b$. In our case, the derivative of $E_s$ with respect to $\bar \rho_s$ has to be taken first and then evaluated at the physically relevant value $\bar \rho_s$, leading, in general, to $p_f \neq p_s$ in the last equation of \eqref{eqs_incompr_incompr_final}. In general, the internal energies of either fluid $E_f$ or solid $E_s$ cannot be dropped if there is a dependence of the variables, such as $b$ in our case, or the coordinates $\bx$. An explicit dependence on coordinates for fluid $E_f$ and solid $E_s$ energies appears, for example, from the introduction of the effects of gravity for fluid and solid, leading to terms proportional to the vertical coordinate $x_3$ for both fluid and solid. In that case, neither $\pp{E_f}{\bar \rho_f}$ nor 
$\pp{E_s}{\bar \rho_s}$ can be neglected in  \eqref{eqs_incompr_incompr_final}.}
\end{remark}

\section*{Acknowledgments}
We are thankful for fruitful and productive discussions with Profs D. D. Holm, A. Ibraguimov, T. S. Ratiu and D. V. Zenkov. The idea to investigate the active porous media as an interesting application of variational methods was suggested to us by Prof. M. Lewis during the presentation of this work at the University of Alberta's math biology seminar. 
FGB is partially supported by the ANR project GEOMFLUID 14-CE23-0002-01.  TF and VP were partially supported by NSERC Discovery Grant and the University of Alberta. TF was also partially supported by MITACS program, project number IT15958.

\bibliographystyle{unsrt}

\begin{thebibliography}{10}

\bibitem{ludeman2014evolutionary}
Danielle~A. Ludeman, Nathan Farrar, Ana Riesgo, Jordi Paps, and Sally~P. Leys.
\newblock Evolutionary origins of sensation in metazoans: functional evidence
  for a new sensory organ in sponges.
\newblock {\em BMC evolutionary biology}, 14(1):3, 2014.

\bibitem{meyers2008biological}
Marc~Andr{\'e} Meyers, Po-Yu Chen, Albert Yu-Min Lin, and Yasuaki Seki.
\newblock Biological materials: structure and mechanical properties.
\newblock {\em Progress in Materials Science}, 53(1):1--206, 2008.

\bibitem{naleway2015structural}
Steven~E. Naleway, Michael~M. Porter, Joanna McKittrick, and Marc Andr{\'e} Meyers.
\newblock Structural design elements in biological materials: application to
  bioinspiration.
\newblock {\em Advanced materials}, 27(37):5455--5476, 2015.

\bibitem{khaled2003role}
Abdul-Rahim A. Khaled, and Kambiz Vafai.
\newblock The role of porous media in modeling flow and heat transfer in
  biological tissues.
\newblock {\em International Journal of Heat and Mass Transfer},
  46(26):4989--5003, 2003.

\bibitem{Te1943}
Karl Terzaghi.
\newblock {\em Theoretical Soil Mechanics}.
\newblock Wiley, New York, 1943.

\bibitem{biot1941general}
Maurice~A. Biot.
\newblock General theory of three-dimensional consolidation.
\newblock {\em Journal of applied physics}, 12(2):155--164, 1941.

\bibitem{biot1955theory}
Maurice~A. Biot.
\newblock Theory of elasticity and consolidation for a porous anisotropic
  solid.
\newblock {\em Journal of applied physics}, 26(2):182--185, 1955.

\bibitem{biot1957elastic}
Maurice~A. Biot and David~G. Willis.
\newblock The elastic coefficients of the theory of consolidation.
\newblock {\em J. appl. Mech}, 24:594--601, 1957.

\bibitem{biot1962mechanics}
Maurice~A. Biot.
\newblock Mechanics of deformation and acoustic propagation in porous media.
\newblock {\em Journal of applied physics}, 33(4):1482--1498, 1962.

\bibitem{biot1962generalized}
Maurice~A. Biot.
\newblock Generalized theory of acoustic propagation in porous dissipative
  media.
\newblock {\em The Journal of the Acoustical Society of America},
  34(9A):1254--1264, 1962.

\bibitem{biot1963theory}
Maurice~A. Biot.
\newblock Theory of stability and consolidation of a porous medium under
  initial stress.
\newblock {\em Journal of Mathematics and Mechanics}, pages 521--541, 1963.

\bibitem{biot1972theory}
Maurice~A. Biot and G.~Temple.
\newblock Theory of finite deformations of porous solids.
\newblock {\em Indiana University Mathematics Journal}, 21(7):597--620, 1972.

\bibitem{joseph1982nonlinear}
Daniel~D. Joseph, Donald~A. Nield, and George Papanicolaou.
\newblock Nonlinear equation governing flow in a saturated porous medium.
\newblock {\em Water Resources Research}, 18(4):1049--1052, 1982.

\bibitem{detournay1993fundamentals}
Emmanuel Detournay and Alexander H.-D. Cheng.
\newblock Fundamentals of poroelasticity.
\newblock In {\em Analysis and design methods}, pages 113--171. Elsevier, 1993.

\bibitem{dell1998micro}
Francesco dell'Isola, Luigi Rosa, and Cz~Wo{\'z}niak.
\newblock A micro-structured continuum modelling compacting fluid-saturated
  grounds: The effects of pore-size scale parameter.
\newblock {\em Acta mechanica}, 127(1-4):165--182, 1998.

\bibitem{brovko2007continuum}
George~L. Brovko, A. G.~Grishayev, and Olga~A. Ivanova.
\newblock Continuum models of discrete heterogeneous structures and saturated
  porous media: constitutive relations and invariance of internal interactions.
\newblock In {\em Journal of Physics: Conference Series}, volume~62, page~1.
  IOP Publishing, 2007.

\bibitem{carcione2010computational}
Jos{\'e}~M. Carcione, Christina Morency, and Juan~E Santos.
\newblock Computational poroelasticity: review.
\newblock {\em Geophysics}, 75(5):75A229--75A243, 2010.

\bibitem{grillo2014darcy}
Alfio Grillo, Melania Carfagna, and Salvatore Federico.
\newblock The {D}arcy-{F}sorchheimer law for modeling fluid flow in biological
  tissues.
\newblock {\em Theoretical \& Applied Mechanics}, 41(4), 2014.

\bibitem{showalter2000diffusion}
Ralph~E. Showalter.
\newblock Diffusion in poro-elastic media.
\newblock {\em Journal of mathematical analysis and applications},
  251(1):310--340, 2000.

\bibitem{bociu2016analysis}
Lorena Bociu, Giovanna Guidoboni, Riccardo Sacco, and Justin~T. Webster.
\newblock Analysis of nonlinear poro-elastic and poro-visco-elastic models.
\newblock {\em Archive for Rational Mechanics and Analysis}, 222(3):1445--1519,
  2016.

\bibitem{bastide2018penalization}
Alain Bastide, Pierre-Henri Cocquet, and Delphine Ramalingom.
\newblock Penalization model for {N}avier--{S}tokes--{D}arcy equations with
  application to porosity-oriented topology optimization.
\newblock {\em Mathematical Models and Methods in Applied Sciences},
  28(08):1481--1512, 2018.

\bibitem{rajagopal2007hierarchy}
Kumbakonam~R. Rajagopal.
\newblock On a hierarchy of approximate models for flows of incompressible
  fluids through porous solids.
\newblock {\em Mathematical Models and Methods in Applied Sciences},
  17(02):215--252, 2007.

\bibitem{wilmanski2006few}
Krzysztof Wilmanski.
\newblock A few remarks on {B}iot's model and linear acoustics of poroelastic
  saturated materials.
\newblock {\em Soil Dynamics and Earthquake Engineering}, 26(6-7):509--536,
  2006.

\bibitem{ChMo2010}
Dominique Chapelle and Philippe Moireau.
\newblock General coupling of porous flows and hyperelastic formulations --
  from thermodynamics principles to energy balance.
\newblock {\em Proceedings of INRIA}, 7395:1--31, 2010.

\bibitem{ChMo2014}
Dominique Chapelle and Philippe Moireau.
\newblock General coupling of porous flows and hyperelastic formulations --
  from thermodynamics principles to energy balance and compatible time schemes.
\newblock {\em European Journal of Mechanics - B/Fluids}, 46:82--96, 2014.

\bibitem{coussy1995mechanics}
Olivier Coussy.
\newblock {\em Mechanics of porous continua}.
\newblock Wiley, 1995.

\bibitem{seguin2019multi}
Brian Seguin and Noel~J. Walkington.
\newblock Multi-component multiphase flow through a poroelastic medium.
\newblock {\em Journal of Elasticity}, 135(1-2):485--507, 2019.

\bibitem{both2019gradient}
Jakub~Wiktor Both, Kundan Kumar, Jan~Martin Nordbotten, and Florin~Adrian Radu.
\newblock The gradient flow structures of thermo-poro-visco-elastic processes
  in porous media.
\newblock {\em arXiv preprint arXiv:1907.03134}, 2019.

\bibitem{bedford1979variational}
Anthony Bedford and Douglas~S. Drumheller.
\newblock A variational theory of porous media.
\newblock {\em International Journal of Solids and Structures},
  15(12):967--980, 1979.

\bibitem{aulisa2007variational}
Eugenio Aulisa, Adem Cakmak, Akif Ibragimov, and Alexander Solynin.
\newblock Variational principle and steady state invariants for non-linear
  hydrodynamic interactions in porous media.
\newblock {\em Dynamics of Continuous, Discrete and Impulsive Systems (Series
  A)}, 2007.

\bibitem{aulisa2010geometric}
Eugenio Aulisa, Akif Ibragimov, and Magdalena Toda.
\newblock Geometric framework for modeling nonlinear flows in porous media, and
  its applications in engineering.
\newblock {\em Nonlinear Analysis: Real World Applications}, 11(3):1734--1751,
  2010.

\bibitem{lopatnikov2004macroscopic}
Sergey~L. Lopatnikov and Alexander H.-D. Cheng.
\newblock Macroscopic {L}agrangian formulation of poroelasticity with porosity
  dynamics.
\newblock {\em Journal of the Mechanics and Physics of Solids},
  52(12):2801--2839, 2004.

\bibitem{lopatnikov2010poroelasticity}
Sergey~L. Lopatnikov and John~W. Gillespie.
\newblock Poroelasticity-{I}: governing equations of the mechanics of
  fluid-saturated porous materials.
\newblock {\em Transport in porous media}, 84(2):471--492, 2010.

\bibitem{dell2000variational}
Francesco dell'Isola, Massimo Guarascio, and Kolumban Hutter.
\newblock A variational approach for the deformation of a saturated porous
  solid. a second-gradient theory extending {T}erzaghi's effective stress
  principle.
\newblock {\em Archive of Applied Mechanics}, 70(5):323--337, 2000.

\bibitem{sciarra2008variational}
Giulio Sciarra, Francesco dell'Isola, Nicoletta Ianiro, and Angela Madeo.
\newblock A variational deduction of second gradient poroelasticity {I}:
  general theory.
\newblock {\em Journal of Mechanics of Materials and Structures},
  3(3):507--526, 2008.

\bibitem{madeo2008variational}
Giulio Sciarra, Francesco dell'Isola, Nicoletta Ianiro, and Giulio Sciarra.
\newblock A variational deduction of second gradient poroelasticity {II}: An
  application to the consolidation problem.
\newblock {\em Journal of Mechanics of Materials and Structures},
  3(4):607--625, 2008.

\bibitem{dell2009boundary}
Francesco dell'Isola, Angela Madeo, and Pierre Seppecher.
\newblock Boundary conditions at fluid-permeable interfaces in porous media: A
  variational approach.
\newblock {\em International Journal of Solids and Structures},
  46(17):3150--3164, 2009.

\bibitem{serpieri2011formulation}
Roberto Serpieri and Luciano Rosati.
\newblock Formulation of a finite deformation model for the dynamic response of
  open cell biphasic media.
\newblock {\em Journal of the Mechanics and Physics of Solids}, 59(4):841--862,
  2011.

\bibitem{serpieri2015variationally}
Roberto Serpieri, Francesco Travascio, Shihab Asfour, and Luciano Rosati.
\newblock Variationally consistent derivation of the stress partitioning law in
  saturated porous media.
\newblock {\em International Journal of Solids and Structures}, 56:235--247,
  2015.

\bibitem{serpieri2016general}
Roberto Serpieri and Francesco Travascio.
\newblock General quantitative analysis of stress partitioning and boundary
  conditions in undrained biphasic porous media via a purely macroscopic and
  purely variational approach.
\newblock {\em Continuum Mechanics and Thermodynamics}, 28(1-2):235--261, 2016.

\bibitem{auffray2015analytical}
Nicolas Auffray, Francesco dell'Isola, Victor~A. Eremeyev, Angela Madeo, and
  Giuseppe Rosi.
\newblock Analytical continuum mechanics {\`a} la {H}amilton--{P}iola least
  action principle for second gradient continua and capillary fluids.
\newblock {\em Mathematics and Mechanics of Solids}, 20(4):375--417, 2015.

\bibitem{serpieri2016variational}
Roberto Serpieri, Alessandro Della~Corte, Francesco Travascio, and Luciano
  Rosati.
\newblock Variational theories of two-phase continuum poroelastic mixtures: a
  short survey.
\newblock In {\em Generalized Continua as Models for Classical and Advanced
  Materials}, pages 377--394. Springer, 2016.

\bibitem{travascio2017analysis}
Francesco Travascio, Shihab Asfour, Roberto Serpieri, and Luciano Rosati.
\newblock Analysis of the consolidation problem of compressible porous media by
  a macroscopic variational continuum approach.
\newblock {\em Mathematics and Mechanics of Solids}, 22(5):952--968, 2017.

\bibitem{serpieri2017variational}
Roberto Serpieri and Francesco Travascio.
\newblock {\em Variational Continuum Multiphase Poroelasticity}.
\newblock Springer, 2017.

\bibitem{placidi2008variational}
Luca Placidi, Francesco dell'Isola, Nicoletta Ianiro, and Giulio Sciarra.
\newblock Variational formulation of pre-stressed solid--fluid mixture theory,
  with an application to wave phenomena.
\newblock {\em European Journal of Mechanics-A/Solids}, 27(4):582--606, 2008.

\bibitem{FaFGBPu2020}
Tagir Farkhutdinov, Fran\c{c}ois Gay-Balmaz, and Vakhtang Putkaradze.
\newblock Geometric variational approach to the dynamics of porous media filled
  with incompressible fluid.
\newblock {\em Acta Mechanica}, 231:3897--3924, 2020.

\bibitem{arnold1966geometrie}
Vladimir Arnold.
\newblock Sur la g{\'e}om{\'e}trie diff{\'e}rentielle des groupes de lie de
  dimension infinie et ses applications {\`a} l'hydrodynamique des fluides
  parfaits.
\newblock In {\em Annales de l'institut Fourier}, volume~16, pages 319--361,
  1966.

\bibitem{GBMaRa12}
Fran\c{c}ois Gay-Balmaz, Jerrold~E. Marsden, and Tudor~S. Ratiu.
\newblock Reduced variational formulations in free boundary continuum
  mechanics.
\newblock {\em J. Nonlin. Sci.}, 22(4):463--497, 2012.

\bibitem{marsden1994mathematical}
Jerrold~E. Marsden and Thomas~J.~R. Hughes.
\newblock {\em Mathematical foundations of elasticity}.
\newblock Courier Corporation, 1994.

\bibitem{dell2018lagrange}
Francesco dell'Isola and Fabio Di~Cosmo.
\newblock Lagrange multipliers in infinite-dimensional systems, methods of.
\newblock {\em Encyclopedia of Continuum Mechanics}, 2018.

\bibitem{bersani2019lagrange}
Alberto Bersani, Francesco dell'Isola, and Pierre Seppecher.
\newblock Lagrange multipliers in infinite dimensional spaces, examples of
  application.
\newblock {\em Encyclopedia of Continuum Mechanics}, 2019.

\bibitem{costa2006permeability}
Antonio Costa.
\newblock Permeability-porosity relationship: A reexamination of the
  kozeny-carman equation based on a fractal pore-space geometry assumption.
\newblock {\em Geophysical research letters}, 33(2), 2006.

\bibitem{demichelis2013study}
Alessia Demichelis, Stefano Pavarelli, Leonardo Mortati, Giudo Sassi, and Mariapaola Sassi.
\newblock Study on the AFM force spectroscopy method for elastic modulus
  measurement of living cells.
\newblock In {\em Journal of Physics: Conference Series}, volume 459, page
  012050. IOP Publishing, 2013.

\bibitem{vinckier1998measuring}
Anja Vinckier and Giorgio Semenza.
\newblock Measuring elasticity of biological materials by atomic force
  microscopy.
\newblock {\em FEBS letters}, 430(1-2):12--16, 1998.

\bibitem{verhulst2005methods}
Ferdinand Verhulst.
\newblock {\em Methods and applications of singular perturbations: boundary
  layers and multiple timescale dynamics}, volume~50.
\newblock Springer Science \& Business Media, 2005.

\bibitem{vazquez2007porous}
Juan~Luis V{\'a}zquez.
\newblock {\em The porous medium equation: mathematical theory}.
\newblock Oxford University Press on Demand, 2007.

\bibitem{barenblatt_1996}
Grigory~Isaakovich Barenblatt.
\newblock {\em Scaling, Self-similarity, and Intermediate Asymptotics:
  Dimensional Analysis and Intermediate Asymptotics}.
\newblock Cambridge Texts in Applied Mathematics. Cambridge University Press,
  1996.

\bibitem{biot1956theory1}
Maurice~A. Biot.
\newblock Theory of propagation of elastic waves in a fluid-saturated porous
  solid. i. low frequency range.
\newblock {\em The Journal of the acoustical Society of america},
  28(2):168--178, 1956.

\bibitem{Fellah2004ultrasonic}
Zine El Abiddine Fellah, Jean-Yves Chapelon, Sylvain Berger, Walter Lauriks, and Claude Depollier.
\newblock Ultrasonic wave propagation in human cancellous bone: Application of
  Biot theory.
\newblock {\em The Journal of the Acoustical Society of America},
  116(1):61--73, 2004.

\bibitem{bociu2020nonlinear}
Lorena Bociu and Justin~T. Webster.
\newblock Nonlinear quasi-static poroelasticity.
\newblock {\em arXiv preprint arXiv:2011.12356}, 2020.

\bibitem{allaire1991homogenization1}
Gr{\'e}goire Allaire.
\newblock Homogenization of the Navier-Stokes equations in open sets perforated
  with tiny holes I. abstract framework, a volume distribution of holes.
\newblock {\em Archive for Rational Mechanics and Analysis}, 113(3):209--259,
  1991.

\bibitem{allaire1991homogenization2}
Gr{\'e}goire Allaire.
\newblock Homogenization of the Navier-Stokes equations in open sets perforated
  with tiny holes II: Non-critical sizes of the holes for a volume distribution
  and a surface distribution of holes.
\newblock {\em Archive for rational mechanics and analysis}, 113(3):261--298,
  1991.

\bibitem{brinkman1949calculation}
Hendrik~C. Brinkman.
\newblock A calculation of the viscous force exerted by a flowing fluid on a
  dense swarm of particles.
\newblock {\em Flow, Turbulence and Combustion}, 1(1):27--34, 1949.

\bibitem{brinkman1952viscosity}
Hendrik~C. Brinkman.
\newblock The viscosity of concentrated suspensions and solutions.
\newblock {\em The Journal of Chemical Physics}, 20(4):571--571, 1952.

\bibitem{srinivasan2014thermodynamic}
Shriram Srinivasan and Kumbakonam~R. Rajagopal.
\newblock A thermodynamic basis for the derivation of the Darcy, Forchheimer
  and Brinkman models for flows through porous media and their generalizations.
\newblock {\em International Journal of Non-Linear Mechanics}, 58:162--166,
  2014.

\bibitem{lacave2016vanishing}
Christophe Lacave and Anna~L. Mazzucato.
\newblock The vanishing viscosity limit in the presence of a porous medium.
\newblock {\em Mathematische Annalen}, 365(3):1527--1557, 2016.

\bibitem{FGBPu2021}
Fran\c{c}ois Gay-Balmaz and Vakhtang Putkaradze.
\newblock Variational geometric approach to the thermodynamics of porous media.
\newblock {\em Zeitschrift f\"ur Angewandte Mathematik und Mechanik (ZAMM),
  under consideration}, 2021.

\bibitem{geeves1999structural}
Michael~A. Geeves and Kenneth~C. Holmes.
\newblock Structural mechanism of muscle contraction.
\newblock {\em Annual review of biochemistry}, 68(1):687--728, 1999.

\bibitem{gay2014exact}
Fran{\c{c}}ois Gay-Balmaz and Vakhtang Putkaradze.
\newblock Exact geometric theory for flexible, fluid-conducting tubes.
\newblock {\em Comptes Rendus M{\'e}canique}, 342(2):79--84, 2014.

\bibitem{gay2015flexible}
Fran{\c{c}}ois Gay-Balmaz and Vakhtang Putkaradze.
\newblock On flexible tubes conveying fluid: geometric nonlinear theory,
  stability and dynamics.
\newblock {\em Journal of Nonlinear Science}, 25(4):889--936, 2015.

\bibitem{gay2018stability}
Fran{\c{c}}ois Gay-Balmaz, Dimitri Georgievskii, and Vakhtang Putkaradze.
\newblock Stability of helical tubes conveying fluid.
\newblock {\em Journal of Fluids and Structures}, 78:146--174, 2018.

\bibitem{heshmati2019vibration}
Mahmood Heshmati, Yaser Amini, and Farhang Daneshmand.
\newblock Vibration and instability analysis of closed-cell poroelastic pipes
  conveying fluid.
\newblock {\em European Journal of Mechanics-A/Solids}, 73:356--365, 2019.

\bibitem{bukac2015effects}
Martina Bukac, Ivan Yotov, Rana Zakerzadeh, and Paolo Zunino.
\newblock Effects of poroelasticity on fluid-structure interaction in arteries:
  A computational sensitivity study.
\newblock In {\em Modeling the heart and the circulatory system}, pages
  197--220. Springer, 2015.

\bibitem{gay2017variational}
Fran{\c{c}}ois Gay-Balmaz.
\newblock A variational derivation of the thermodynamics of a moist atmosphere
  with rain process and its pseudoincompressible approximation.
\newblock {\em Geophysical and Astrophysical Fluid Dynamics}, 113, 2019.

\bibitem{gay2017lagrangian}
Fran{\c{c}}ois Gay-Balmaz and Hiroaki Yoshimura.
\newblock A {L}agrangian variational formulation for nonequilibrium
  thermodynamics. {P}art {I}: discrete systems.
\newblock {\em Journal of Geometry and Physics}, 111:169--193, 2017.

\bibitem{gay2017lagrangian2}
Fran{\c{c}}ois Gay-Balmaz and Hiroaki Yoshimura.
\newblock A {L}agrangian variational formulation for nonequilibrium
  thermodynamics. {P}art {II}: continuum systems.
\newblock {\em Journal of Geometry and Physics}, 111:194--212, 2017.

\bibitem{GBYo2018}
Fran{\c{c}}ois Gay-Balmaz and Hiroaki Yoshimura.
\newblock A variational formulation of nonequilibrium thermodynamics for
  discrete open systems with mass and heat transfer.
\newblock {\em Entropy}, 20, 2018.

\bibitem{GBYo2019}
Fran{\c{c}}ois Gay-Balmaz and Hiroaki Yoshimura.
\newblock From {L}agrangian mechanics to nonequilibrium thermodynamics: a
  variational perspective.
\newblock {\em Entropy}, 21, 2019.

\bibitem{WeMa1997}
Jeffrey~M. Wendlandt and Jerrold~E. Marsden.
\newblock Mechanical integrators derived from a discrete variational principle.
\newblock {\em Physica D}, 106:223--246, 1997.

\bibitem{MaWe2001}
Jerrold~E. Marsden and Matthew West.
\newblock Discrete mechanics and variational integrators.
\newblock {\em Acta Numerica}, pages 1--158, 2001.

\bibitem{gay2016variational}
Fran{\c{c}}ois Gay-Balmaz and Vakhtang Putkaradze.
\newblock Variational discretizations for the dynamics of fluid-conveying
  flexible tubes.
\newblock {\em Comptes Rendus M{\'e}canique}, 344(11-12):769--775, 2016.

\bibitem{FGBPu2020}
Fran{\c{c}}ois Gay-Balmaz and Vakhtang Putkaradze.
\newblock {\em Variational Methods for Fluid-Structure Interactions, in
  Handbook of Variational Methods for Nonlinear Geometric Data}, pages
  175--205.
\newblock Springer, 2020.

\end{thebibliography}

\end{document}